\documentclass[amsmath,amssymb]{revtex4}

\usepackage{epsfig}
\usepackage{graphicx}
\usepackage{dcolumn}
\usepackage{bm}

\usepackage{color}

\usepackage{tikz} 
\usetikzlibrary{matrix,arrows}

\begin{document}

\title{Effective statistical physics of Anosov systems}

\author{Steven Huntsman}%
\email{sh@eqnets.com}
\affiliation{%
Physics Department, Naval Postgraduate School, Monterey, California
}%
\affiliation{%
Equilibrium Networks, Alexandria, Virginia
}%

\date{\today}

\begin{abstract}
We present evidence indicating that Anosov systems can be endowed with a unique physically reasonable effective temperature. Results for the two paradigmatic Anosov systems (i.e., the cat map and the geodesic flow on a surface of constant negative curvature) are used to justify a proposal for extending Ruelle's thermodynamical formalism into a comprehensive theory of statistical physics for nonequilibrium steady states satisfying the Gallavotti-Cohen chaotic hypothesis.
\end{abstract}

\maketitle

\tableofcontents

\section{\label{sec:introduction} Introduction}

The \emph{chaotic hypothesis} of Gallavotti and Cohen \cite{GC} is that 
\begin{quote} for the purpose of studying macroscopic properties, the time evolution map [$T$] of a many-particle system can be regarded as a mixing Anosov map \cite{Gallavotti1}. \end{quote}
The principal consequence of this hypothesis to date is the \emph{fluctuation theorem} for deterministic reversible dynamics \cite{GC,Gentile}. Like its analogue for stochastic dynamics \cite{Kurchan1998,LebowitzS}, the fluctuation theorem provides a quantitative relation between entropy production values of equal magnitude and opposite signs through consideration of the time-reversed dynamics. These fluctuation relations apply even to singular systems such as particle systems obeying Lennard-Jones potentials \cite{BGGZ} and fit into the mathematical framework of large-deviations theory \cite{JQQ}. They have been shown to generalize both the Onsager reciprocity relations and the Green-Kubo formulae linking fluxes and transport coefficients; as such they play a fundamental role in statistical physics \cite{Gallavotti1}.

In this paper we explore the chaotic hypothesis from a different perspective, presenting evidence indicating that Anosov systems can be endowed with an essentially unique effective temperature and a corresponding energy function that can be calculated directly in terms of the ergodic statistics of the time evolution. In the context of Anosov systems this prescription is entirely dynamical and draws on the SRB measure of the system, highlighting its physical relevance \cite{Cohen1,Cohen2}. On the other hand, the present considerations inform the choice of an appropriate timescale for computing the effective temperature and the scaling behavior of the effective temperature with respect to the number of states.

Results for the two paradigmatic Anosov systems (i.e., the cat map and the geodesic flow on a surface of constant negative curvature) are used to justify a proposed framework extending Ruelle's thermodynamical formalism into a comprehensive theory of statistical physics for nonequilibrium steady states satisfying the chaotic hypothesis. The key results are a series of analytical and numerical calculations identifying nontrivial limiting behavior of the effective temperature under successive physically motivated refinements of phase space and culminating in sections \ref{sec:greedy}, \ref{sec:catflow}, and \ref{sec:geodesic2}. These combine to suggest that a generalized variational principle (likely to be recognizable as minimizing the effective free energy in some way; see also appendix \ref{sec:variational}) singles out a preferred effective temperature and concomitant effective energy function.

The paper is organized as follows. Background on Anosov systems and SRB measures is given in section \ref{sec:chaotic}. Section \ref{sec:gedanken} serves to couch the construction of Markov partitions in a physical context and describes how to obtain a partition-independent energy function before discussing the role of the mixing time as the preferred characteristic time for an Anosov system.

In section \ref{sec:temperature} we briefly describe the unique effective inverse temperature of a finite system with stationary probabilities $p_j$ and characteristic time $t_\infty$ that is both consistent with equilibrium statistical physics and physical constraints on scaling behavior. Up to a fixed choice of scale, it is given by $\beta = t_\infty \lVert p \rVert \sqrt{ \lVert \gamma \lVert^{2} + 1 }$, where $\gamma_k = \frac{1}{n} \sum_{j=1}^n \log p_j - \log p_k$. By using this and closing the Gibbs relation $p_k = Z^{-1}e^{-\beta E_k}$, it is possible to use the idiom of equilibrium statistical physics to describe the behavior of nonequilibrium steady states. Most of the paper will deal with understanding this approach in the context of a physically reasonable application to Anosov systems.

Section \ref{sec:catmap} introduces the Arnol'd-Avez cat map and the classical Markov partition of Adler and Weiss before illustrating the first indication of nontrivial limiting behavior for the effective temperature. In section \ref{sec:nontriviallimiting} a general explanation for this behavior is given; section \ref{sec:nontriviality} illustrates why it is nontrivial by way of an example. The implications of this limiting behavior for constraining the detailed form of the effective temperature in a somewhat unexpected way are discussed in section \ref{sec:implication}.

The issue of uniformity of Markov partitions is discussed in section \ref{sec:uniformity} and serves to provide background for a physically motivated class of refinements. Section \ref{sec:greedy} continues in this vein by introducing a method of obtaining the most uniform possible refinements of Markov partitions  w/r/t the ambient Riemannian measure and identifies an apparently unique effective temperature for the cat map on this basis. The role of ensembles and the thermodynamical limit is discussed in section \ref{sec:ensembles} to provide a broader context for the ideas presented in this paper before section \ref{sec:proposal} outlines the essentials of a proposal for extending Ruelle's thermodynamical formalism to a complete theory of statistical physics for nonequilibrium steady states in the context of the chaotic hypothesis. The so-called Ulam method for approximating SRB measures via piecewise approximations on generic (non-Markov) partitions is briefly touched on in section \ref{sec:ulam}, along with its relevance for approximating the mixing time.

In section \ref{sec:catflow} a continuous-time version of the cat map is introduced and related to a Hamiltonian system before introducing the cat flow as the simplest Anosov flow and relating its effective statistical physics to that of the cat map, providing an indication that Anosov flows also appear to enjoy a uniquely determined effective temperature. The subject of Anosov flows is continued with the introduction of the paradigmatic example of geodesic flow on a surface of constant negative curvature in section \ref{sec:geodesic} and culminates in the numerical demonstration of a well-behaved limit for the effective temperature in section \ref{sec:geodesic2}. 

Sections \ref{sec:rugh} and \ref{sec:fdtemp} discuss the effective temperature focused on in this paper in relation to the dynamical temperature introduced by Rugh \cite{Rugh} and the fluctuation-dissipation (effective) temperature \cite{CKP}, respectively. After the conclusion appendices are provided on a technical lemma, the so-called variational principle, and a single Glauber-Ising spin as a basis for comparison of the fluctuation-dissipation temperature with the effective temperature at the center of our considerations.

\section{\label{sec:chaotic} Background on dynamical systems}

In this section we provide definitions and notation for the necessary mathematical background, following (e.g.) \cite{JQQ,Bowen2,KH,Chernov}. For an overview of these concepts from the point of view of statistical physics see \S 9 of \cite{Gallavotti1}. (A physical concept more generally familiar than the chaotic hypothesis but still incorporating much of the mathematical infrastructure introduced immediately below is furnished by Lyapunov exponents. \footnote{Let $T: M \rightarrow M$ be a diffeomorphism and let $\mu$ be an $T$-invariant measure. By Oseledets' multiplicative ergodic theorem (see, e.g., \cite{Walters}), the tangent space decomposes as $T_x M = \bigoplus_{k=1}^\ell T_x^{(k)}$, where the Lyapunov exponents $\lambda_k := \lim_{n \uparrow \infty} n^{-1} \log \lVert D_x T^n v^{(k)} \rVert$ are well defined $\mu$-almost everywhere for $v^{(k)} \in T_x^{(k)}$. For an Anosov diffeomorphism the Lyapunov exponents are constant and the hyperbolic structure is manifested as $E_x^s = \bigoplus_{k:\lambda_k < 0} T_x^{(k)}$ and $E_x^u = \bigoplus_{k:\lambda_k > 0} T_x^{(k)}$.})

A \emph{diffeomorphism} between Riemannian manifolds $M$ and $N$ is a map $T: M \rightarrow N$ such that both $T$ and $T^{-1}$ are smooth. Typically we will write $Tx$ in place of $T(x)$. A closed subset $M_0$ of $M$ is \emph{hyperbolic} if $TM_0 = M_0$ and the bundle of tangent spaces $\{T_xM\}_{x \in M_0}$ admits a decomposition $T_xM = E_x^s \oplus E_x^u$ into \emph{stable} and \emph{unstable} subspaces varying continuously w/r/t $x$ and such that a) the decomposition is invariant: i.e., the derivative satisfies $D_x T(E_x^{(s,u)}) = E_{Tx}^{(s,u)}$ and b) there exist $C > 0$ and $0 < \lambda < 1$ such that $\lVert D_x T^{(n,-n)} v^{(s,u)} \rVert \le C\lambda^n \lVert v^{(s,u)} \rVert$ for all $v^{(s,u)} \in E_x^{(s,u)}$ and $n \ge 0$. Note that hyperbolicity does not depend on a particular choice of Riemannian metric, but only on the existence of a suitable one. 

\begin{figure}[htbp]
\includegraphics[trim = 20mm 35mm 20mm 25mm, clip, width=10cm,keepaspectratio]{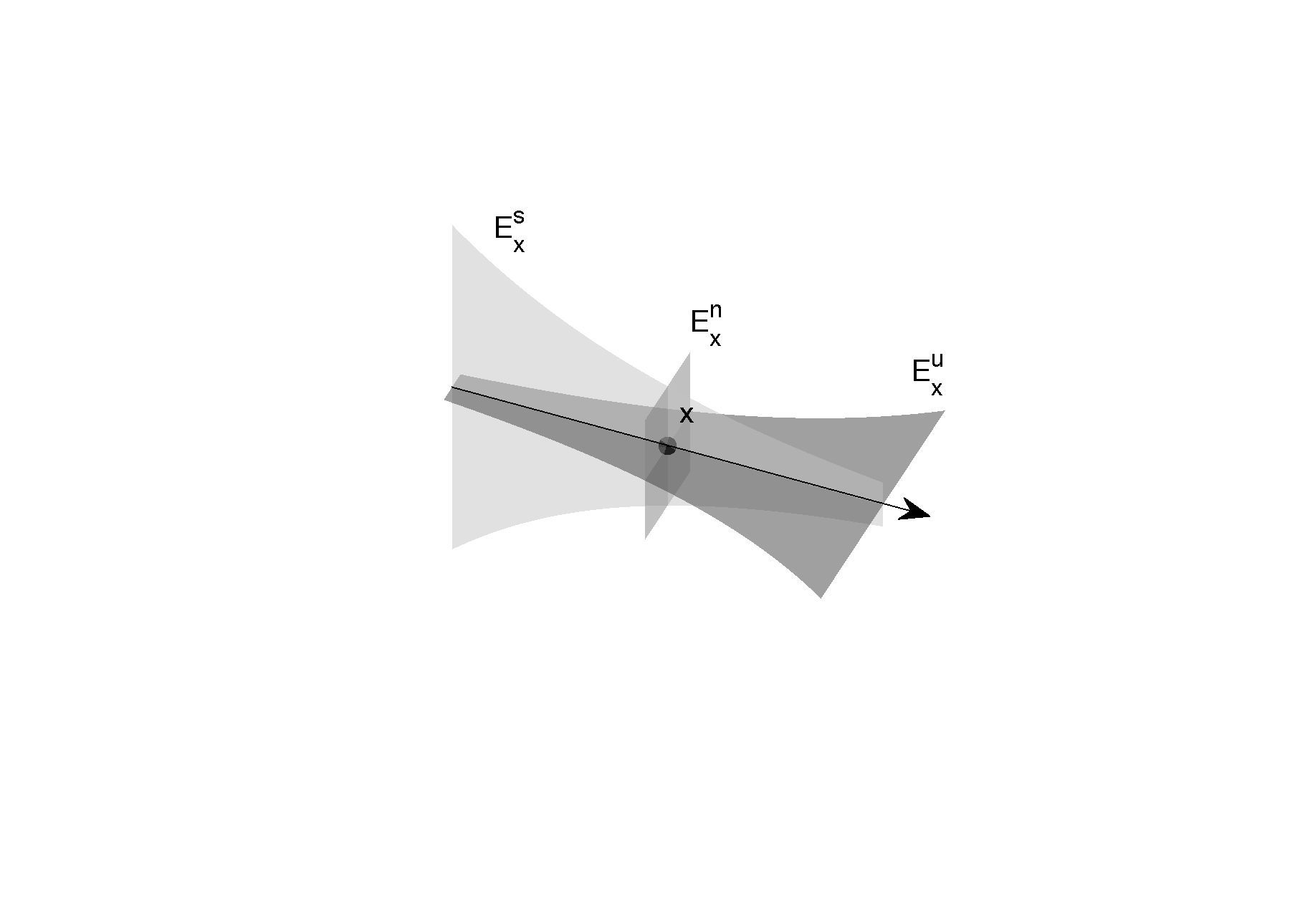}
\caption{ \label{fig:anosovflow} Schematic local geometry of a three-dimensional Anosov flow.}
\end{figure}

A point $x \in M$ is \emph{nonwandering} iff for all $U \subset M$ with $x \in U$, there exists $n > 0$ such that $T^nU \cap U \ne \varnothing$. Denote the set of nonwandering points by $\Omega(T)$. $T$ is \emph{Axiom A} if $\Omega(T)$ is hyperbolic and equals the closure of the set of periodic points. $T$ is \emph{Anosov} if $M$ itself is hyperbolic. An \emph{Anosov flow} is defined similarly, but with an invariant decomposition of the tangent bundle of the form $T_xM = E_x^n \oplus E_x^s \oplus E_x^u$, where the \emph{neutral} direction of the flow corresponds to a one-dimensional family of subspaces $E_x^n$. (See figure \ref{fig:anosovflow}.) Thus the construction of Anosov diffeomorphisms and flows require at least two and three dimensions, respectively.

If $\mu$ is a $T$-invariant probability measure on $M$, then $T$ is \emph{mixing} if $\mu(X \cap T^{-k}Y) \rightarrow \mu(X) \mu(Y)$ for measurable sets $X$ and $Y$. While mixing is a very strong statistical property of a measure-preserving dynamical system, in the special case of Anosov diffeomorphisms it appears to reduce to a technical assumption for practical purposes, and one that is satisfied in all known cases when $M$ is connected. (There are Anosov flows that are not mixing, and we shall actually encounter the simplest such flow in this paper; however, the assumption of mixing still appears to be technical in nature.) Typically we will denote the normalized Riemannian volume on $M$ by $\nu$. The normalized \emph{Liouville measure} on the unit tangent bundle is given by a product of the normalized Riemannian volumes on tangent spheres $UT_xM := \{v \in T_xM: \lVert v \rVert = 1\}$ and on $M$. A dynamical approach to the Liouville measure adapted to the present context is given in \cite{Gallavotti2b}.

The chaotic hypothesis was motivated by Ruelle's elucidation of a thermodynamical formalism for Axiom A systems that provides a great deal of explanatory power for the study of nonequilibrium steady states \cite{Ruelle1}. In particular, the theory for mixing Anosov systems gives the existence of a \emph{unique} $T$-invariant probability measure $\mu_{SRB}$, called the Sinai-Ruelle-Bowen or \emph{SRB measure} \cite{Young} and satisfying $\lim_{k \uparrow \infty} k^{-1} \sum_{j=0}^k F(T^j x) = \int_M F \ d \mu_{SRB}$ for $\nu$-almost all $x$ when $F$ is continuous. \footnote{In general, for a $C^2$ diffeomorphism $T$ with at least one Lyapunov exponent that is a.e. positive w/r/t $\nu$, a SRB measure is a $T$-invariant measure whose conditional measures on unstable manifolds are absolutely continuous w/r/t the relative Riemannian volume.} 

Because in general $\mu_{SRB}$ is singular w/r/t $\nu$, a naive density of the form $p = d\mu_{SRB}/d\nu$ typically fails to be well-defined. Consequently, a naive generalization of the formula for the inverse effective temperature in which sums are replaced by integrals w/r/t $\nu$ also typically fails to be well-defined. Much of this paper is implicitly concerned with overcoming obstacles raised by this fact, which is specifically addressed in more detail in section \ref{sec:implication}.

The existence of a unique SRB measure generalizing the microcanonical ensemble, along with the chaotic hypothesis and the associated fluctuation theorem, serves to indicate the relevance of the theory of Anosov systems to statistical physics. The expressiveness of the theory is due chiefly to the consequences of a local product structure on $M$ inherited from the decomposition into stable and unstable tangent spaces. A brief sketch will serve to introduce the basic constructions of interest in this regard. 

Let $T$ be an Anosov diffeomorphism. For $x \in M$ and $U \subset M$, let $W^{(s,u)}(x)$ respectively denote the global stable and unstable manifolds containing $x$, let $W^{(s,u)}_\varepsilon(x)$ denote local stable and unstable manifolds (defined respectively as the sets $\{y \in M : d(T^{(n,-n)}x,T^{(n,-n)}y) \le \varepsilon, \forall n \ge 0\}$), and let $W^{(s,u)}(x,U)$ denote the intersection of $U$ and local stable and unstable manifolds that minimally stretch across $U$. A subset $R$ of $M$ is called a \emph{rectangle} if its diameter is sufficiently small that for any $x, y \in R$ there exists $\varepsilon$ such that $[x,y] := W^{s}_\varepsilon(x) \cap W^{u}_\varepsilon(y)$ consists of a single point also contained in $R$. The local product structure of a rectangle $R$ is illuminated by taking $z \in R$: it happens that $R = \{[x,y]: x \in W^u(z,R), y \in W^s(z,R)\}$ and there is a unique representation of the form $x = [x^u,x^s]$, where $x^{(s,u)} \in W^{(s,u)}(z,R)$.

A collection of subsets $\mathcal{R} = \{R_1,\dots, R_n\}$ is called a \emph{partition} of $M$ if the $R_j$ have pairwise disjoint interiors and $\bigcup_j R_j = M$. If the set $R_j$ equals the closure $\overline{\mbox{int} R_j}$ of its interior it is called \emph{proper}. We shall generally not concern ourselves with ambiguities associated with the boundary $\partial \mathcal{R}$, as it has (Riemannian and SRB) measure zero and thus no physical significance. With this in mind, write $\mathcal{R}(x)$ for an element of $\mathcal{R}$ containing $x$ ($\mathcal{R}(x)$ is almost surely unique). A partition $\mathcal{R} = \{R_1,\dots, R_n\}$ of $M$ into proper rectangles is called \emph{Markov} if 
\begin{equation}
\label{eq:markovpartition}
TW^s(x,R_j) \subset W^s(Tx,R_k); \quad W^u(Tx,R_k) \subset TW^u(x,R_j)
\end{equation}
for $x \in \mbox{int} R_j \cap T^{-1} \mbox{int} R_k$. It can be shown that $M$ admits Markov partitions of arbitrarily small diameter. 

The Markov criteria \eqref{eq:markovpartition} means that the forward and inverse images of each rectangle respectively stretch across the unstable and stable directions in such a way that the stable and unstable boundaries of these images are contained within the stable and unstable boundaries of other rectangles. Markov partitions can be regarded as coarse-grained phase space cells well-suited for describing (a suitable time discretization of) physical dynamics, and particularly from the point of view of statistical physics. A suitable diameter for physical coarse-graining is one for which nontrivial intersections of the form $\mbox{int} R_j \cap T^{-1} \mbox{int} R_k$ are connected and for which the physically interesting observables are approximately or effectively constant on the partition elements.

A similar notion of \emph{Markov sections} applies to Anosov flows. Details can be found in \cite{Chernov}.

The key property of Markov partitions for the ergodic theory of Anosov systems is their correspondence to simple and natural symbolic dynamical encodings. Given a partition $\mathcal{R} = \{R_1,\dots, R_n\}$ of $M$, the entries $a(\mathcal{R})_{jk}$ of the corresponding \emph{transition matrix} (not to be confused with a stochastic matrix, although see section \ref{sec:ulam} for a related construction in this vein) are equal to 0 or 1 if $\mbox{int}R_j \cap T^{-1} \mbox{int} R_k$ is respectively empty or nonempty. The dynamics of $T$ w/r/t a Markov partition $\mathcal{R}$ is captured by the \emph{subshift of finite type} or \emph{topological Markov chain}
\begin{equation}
\label{eq:subshiftoffinitetype}
\Sigma_{a(\mathcal{R})} := \left \{s \in \{1,\dots,n\}^\mathbb{Z} : a(\mathcal{R})_{s_j,s_{j+1}} \equiv 1 \right\}.
\end{equation}
For $s \in \Sigma_{a(\mathcal{R})}$, the set $\pi(s) := \bigcap_{k \in \mathbb{Z}} T^{-k} R_{s_k}$ consists of a single point and $\pi \circ \ell = T \circ \pi$, where $\ell$ denotes the left shift operator on $\Sigma_{a(\mathcal{R})}$ that sends $s_j$ to $s_{j+1}$. That is, points in $M$ and symbolic sequences in $\Sigma_{a(\mathcal{R})}$ are in a well-behaved one-to-one correspondence (except on a set of measure zero corresponding to the images of the partition boundary) called a \emph{topological conjugacy}. \footnote{Two maps $f: X \rightarrow X$ and $g : Y \rightarrow Y$ are called \emph{topologically conjugate} if there is some $h : X \rightarrow Y$ such that $f = h \circ g \circ h^{-1}$. In this case, $f^n = h \circ g^n \circ h^{-1}$ as well. Intuitively, the notion of topological conjugacy can be regarded as equivalence w/r/t a change of coordinates.} 

This correspondence--and in particular the simple description of the symbolic sequences encoding the dynamics--is the payoff for the framework described above. A Markov partition (which can be realized in settings more general than Anosov or even Axiom A systems) enables the reformulation of dynamics in terms of a one-dimensional spin model \cite{BeckS} with short-range interactions in which the permitted spin configurations are specified by the transition matrix (a/k/a a nearest neighbor ``hard core'' interaction). In this context the SRB measure of the original system yields a Gibbs measure of the corresponding spin system. 

This point of view immediately leads to nontrivial consequences which further serve to indicate the physical relevance of the Markov description of dynamics: for example, a unique SRB measure corresponds to the absence of phase transitions in one-dimensional short-ranged spin models \cite{Gallavotti1}. By the same token, it suggests the construction of $d$-dimensional lattices of coupled maps corresponding to $(d+1)$-dimensional spin systems capable of exhibiting phase transitions \cite{CF}. However realizing such phenomena in the Anosov context will probably require either strong coupling and projections onto a subsystem, or perhaps both (see section \ref{sec:ensembles} for details). 

In this work we focus on aspects of the two paradigmatic Anosov systems: the Arnol'd-Avez cat map and the geodesic flow on a surface of constant negative curvature.

\section{\label{sec:gedanken} A \emph{Gedankenexperiment}}

Before proceeding with particular systems, however, we shall make some general observations about the effective statistical physics of finite systems, and in particular the ``experimental'' character of the framework. Towards that end, consider the following simple classical \emph{Gedankenexperiment} on a system represented by a mixing Anosov diffeomorphism $T$.

The experimenter's access to the system (including the underlying manifold $M$) is limited to an ``oracle'' that accepts input $x$ and an integer $k$ and returns $T^k x$ with a resolution of $\varepsilon$. That is, the experimenter can evolve the system perfectly, but she can only distinguish points (including initial conditions) that are at least a distance $\varepsilon$ apart. 

Assuming henceforth that $M$ is compact and finite-dimensional, the experimenter can approximate a (sufficiently) regular partition $\mathcal{X} = \{X_i\}$ of $M$ into rectangular microcells of nearly equal Riemannian measure. She can also use the information gained from the oracle to approximate a suitable Markov partition $\mathcal{R}$ to a degree of accuracy limited only by her patience (as measured by the number of evaluations $x \rightarrow Tx$ she performs) and $\varepsilon$ by adapting the original construction of Markov partitions by Sinai (see \cite{Chernov} for details). For reasons that will become clear, it will make sense to consider both a microscopic regular partition and a mesoscopic or coarse-grained Markov partition simultaneously. 

The procedure to approximate the Markov partition begins with an initial partition $\mathcal{R}_0$ into smooth connected (approximate) rectangles constructed by considering the action of $T$ on suitable local stable and unstable manifolds through a set of points such that balls of small radius around them cover $M$. In order to form an (approximate) Markov partition, the stable and unstable boundaries $\partial^{(s,u)} \mathcal{R}_0$ (which are defined in the obvious way) are perturbed along the unstable direction along with relatively small expansions or contractions in such a way as to preserve the rectangular property. A new partition $\mathcal{R}_1$ is defined so that $T \partial^s \mathcal{R}_1 \subset \partial^s \mathcal{R}_0$ and $T^{-1} \partial^u \mathcal{R}_1 \subset \partial^u \mathcal{R}_0$. Now it turns out that the conditions $T \partial^s \mathcal{R} \subset \partial^s \mathcal{R}$ and $T^{-1} \partial^u \mathcal{R} \subset \partial^u \mathcal{R}$ imply that $\mathcal{R}$ is Markov. By an iteration of the procedure just outlined, therefore, a sequence $\{\mathcal{R}_k\}$ of partitions can in principle be constructed that converges exponentially quickly to a Markov partition.

Because our experimenter is careful, she will have taken pains to establish that she has approximated $\mathcal{R}$ at the best possible resolution $\varepsilon$ (which we assume is comparable to the diameter of $\mathcal{X}$ and small compared to the diameter of the coarse Markov partition $\mathcal{R}$), so that in the overwhelming number of cases she can reliably obtain the partition elements $\{\mathcal{R}(T^k x_0)\}$ corresponding to the initial condition $x_0$, and in most cases even the elements $\{\mathcal{X}(T^k x_0)\}$ (and at worst one of the neighboring microcells, which will be functionally equivalent). Because the experimenter has no direct knowledge of $T$, she cannot compute anything from first principles, but instead must rely only on symbolic data of this form. 

Since Markov partitions are not unique, the experimenter will typically consider $\mathcal{X}$ to provide a more physically fundamental discretization than $\mathcal{R}$. Indeed, as we shall see in section \ref{sec:ulam}, her best (but still not unique) strategy for constructing a suitable $\mathcal{R}$ will be to take its diameter as large as possible while keeping the SRB measures of elements of $\mathcal{X}$ that intersect $R_j$ indistinguishable (up to a desired tolerance, which introduces a free but untroublesome parameter). It is nevertheless the case that Markov partitions are physically relevant, and not least for the reasons outlined in section \ref{sec:chaotic}, though it is the case that any \emph{particular} Markov partition cannot be considered physically relevant in isolation. \footnote{Another more exotic reason one might expect Markov partitions to have physical relevance, especially in concert with a microscopic partition $\mathcal{X}$, is because of the relationship between periodic orbits (which correspond to periodic sequences of the form $\{\mathcal{R}(T^k x_0)\}$ for any $\mathcal{R}$) and the density of states in semiclassical quantum theory \cite{Gutzwiller}.}

Write $p_i^{(k)}(x) := k^{-1} \sum_{\ell = 0}^{k-1} 1_{X_i}(T^\ell x)$ for the empirical probability distribution. After some characteristic timescale that is independent of the initial condition, the experimenter will almost surely be able to conclude that $\sum_{i:X_i \cap R_j \ne \varnothing} p_i^{(k)}(x_0)$ is converging to some value, which she will (be able to approximately) identify as $\mu_{SRB}(R_j)$. At this time she stops (paying attention to) the evolution since as we shall see below this is a sufficiently long interval to extract an appropriate $t_\infty$.

Under the physically reasonable assumptions above, she can obtain the probabilities and effective energies of the microcells straightforwardly. For the moment, write $r$ and $x$ for indices corresponding to $\mathcal{R}$ and $\mathcal{X}$, respectively. Assume that $\nu_x \equiv \nu(X_x) \equiv 1/\lvert \mathcal{X} \rvert$ and $\tilde \nu_r \equiv \nu(R_r) = N_r/\lvert \mathcal{X} \rvert$, and write $\mu_x \equiv \mu_{SRB}(X_x) = e^{-\gamma_x}/Z$ and $\tilde \mu_r \equiv  \mu_{SRB}(R_r) = e^{-\tilde \gamma_r}/Z$. Now provided that $\mathcal{R}$ is (as it should be) small enough so that any interesting observables are practically constant on its rectangles, we have that $\mu_x/\nu_x \approx \tilde \mu_{r(x)}/\tilde \nu_{r(x)}$, where $X_x \cap R_{r(x)} \ne \varnothing$: equivalently, $\gamma_x \approx \tilde \gamma_{r(x)} + \log N_{r(x)}$. We can use this equality to define 
\begin{equation}
\gamma_{r(x)} := \tilde \gamma_{r(x)} + \log N_{r(x)}
\end{equation} 
and after writing $E := \beta^{-1}\gamma$ it follows that $E_{r(x)} \approx E_x$, provided that $\beta$ does not depend on $\mathcal{R}$ or $\mathcal{X}$ (as is the case for the situations of interest in this paper). Therefore nothing significant is lost by restricting attention to a Markov partition in general, and given $\beta$ we can in principle reconstruct the corresponding effective energy function to an accuracy limited only by $\varepsilon$. (Note also that if $\sum_r \tilde \gamma_r = 0$, then $\sum_r \gamma_r = \log \lvert \mathcal{X} \rvert$, so we might in this event consider $\gamma'_r := \gamma_r - \log \lvert \mathcal{X} \rvert / \log \lvert \mathcal{R} \rvert$, which again sums to zero.)

In the case where $\mu_{SRB} = \nu$ (which is the case with the systems considered explicitly in this paper) $\gamma_x = 0$. However, since the more interesting case $\mu_{SRB} \ne \nu$ is presently infeasible to examine in any detail, we do not bother with any such rescaling in the remainder of this paper (though we remark here that it appears to be ultimately irrelevant for the computation of the physically preferred inverse effective temperature). In particular, we will exhibit typical Anosov systems for which the effective temperature and spectral density (i.e., the energy levels, with only relative multiplicities accounted for) appear to have unique well-defined limits as a function of certain types of partition refinements. This result will inform most of our discussion throughout.

The \emph{a priori} identification of an appropriate characteristic time $t_\infty$ is perhaps the most difficult conceptual issue the experimenter faces. One reason for this is that several parameters that can be interpreted as timescales to consider arise rather naturally (though some may depend on the partition). Among these timescales are the inverse topological entropy (see appendix \ref{sec:variational}), the number of elements in the partition 
\footnote{Let $T$ be an ergodic transformation of a probability space $(M,\Omega, \mathbb{P})$ endowed with a finite partition $\mathcal{R} = \{R_1,\dots,R_n\}$, where the $R_j$ are a.s. disjoint and have positive probability. For $x \in R_k$ define $\tau(x) := \inf \{\ell>0:T^\ell x \in R_k\}$. The Kac lemma (see, e.g. \cite{Baez-Duarte,Saussol}) gives that $\int_{R_k} \tau(x) \ d\mathbb{P}(x) = 1$. Now $\int_X \tau(x) \ d\mathbb{P}(x) = \sum_k \int_{R_k} \tau(x) \ d\mathbb{P}(x) = n$, or equivalently $\mathbb{E}\tau = n$. In words, the average first return time is simply the number of elements of the partition. 
}, 
or the number of timesteps necessary for any two elements of $\mathcal{R}$ to communicate, i.e. the least integer $k$ s.t. $a(\mathcal{R})^k$ has strictly positive entries \cite{BFG}. Like all of the candidates mentioned except the inverse topological entropy, a Poincar\'e or recurrence time is evidently \emph{not} an appropriate choice in the present context because it is far from intrinsic. To see this note that a particular discretization scale of the phase space is necessary for a Poincar\'e time to be defined, but for a system such as the geodesic flow on a surface of constant negative curvature there is no single preferred finite discretization scale by the remarks above on effective energies. 

However there is another particular timescale that appears to be entirely appropriate for the present purposes, namely the so-called mixing time of the system. The relaxation time is generally similar and may also be considered.
\footnote{
An analogous mixing time also exists for Markov processes. Define the \emph{total variation distance} between two probability measures $\mathbb{P}$ and $\mathbb{P}'$ as $\lVert \mathbb{P}-\mathbb{P}' \rVert_{TV} := \sup_X \lvert \mathbb{P}(X) - \mathbb{P}'(X) \rvert = \frac{1}{2} \sum_x \lvert \mathbb{P}(x) - \mathbb{P}'(x) \rvert$, where the last equality can be taken as a lemma. If $P$ is the transition matrix for a Markov chain with stationary distribution $p$ and $\delta > 0$, set $d(\ell) := \sup_j \lVert P^\ell(j,\cdot) - p \rVert_{TV}$. The standard convergence proof for Markov chains shows that $d(\ell)$ is bounded by a decaying exponential, which motivates the definition of \emph{mixing time} as $t_{mix}(\delta) := \inf \left \{\ell: d(\ell) \le \delta \right \}$. The \emph{relaxation time} is given by the inverse absolute spectral gap, viz. $t_{rel} := (1-\lambda_*)^{-1}$, where $\lambda_* = \sup \left \{ \lvert \lambda \rvert: \lambda \in \mbox{spec}(P) \land \lambda \ne 1 \right \}$. If $P$ is reversible, irreducible, and aperiodic, then the two timescales are closely related: viz. $(t_{rel}-1) \cdot \log(1/2\delta) \le t_{mix}(\delta) \le t_{rel} \cdot \log(1/\delta \inf_j p_j)$ \cite{LPW}. For the Ising model with Glauber dynamics, both $t_{mix} := t_{mix}(1/4) \ge t_{mix}(\delta)/\lceil \log_2 \delta^{-1} \rceil$ and $t_{rel}$ depend on the inverse temperature, and in fact these timescales can exhibit sharp transitions depending on the temperature. In general however one expects the temperature dependence to be monotonic as a function of a characteristic timescale; indeed, the effective inverse temperature $\beta$ and associated timescale $t_\infty$ scale linearly with each other \cite{Huntsman}. 
} 
See also section \ref{sec:ulam} for obtaining such a $t_\infty$ via the Ulam method.
By replacing sets with characteristic functions and making an obvious generalization of scope, the property of mixing can be considered from the point of view of the time correlation function $C_{F,G}(k) := \int F \cdot (G \circ T^k) \ d \mu - \int F \ d \mu \cdot \int G \ d \mu$. Ruelle \cite{Ruelle0} and Sinai \cite{Sinai2} showed that $C_{F,G}(k)$ decays exponentially if $T$ is Axiom A with dense unstable manifolds and a connected attractor (these criteria are conjectured to be satisfied for generic Anosov diffeomorphisms when $M$ is connected; see also \cite{Young0} for an extension to partially hyperbolic systems). For more detailed results in the particular case of hyperbolic toral autmorphisms (i.e., generalized cat maps discussed in section \ref{sec:catmap}), see \cite{BSTV,BriniS}.

An analogous result for Anosov flows remained unproven for more than two decades until Chernov \cite{Chernov2} utilized what amount to Ulam approximations (see section \ref{sec:ulam}) in the process of demonstrating decay bounded by an exponential in $\sqrt{t}$ for the time correlation function for (e.g.) geodesic flows on compact surfaces of variable negative curvature. The application of the Ulam method in this context rests on the construction of a Markov partition of small diameter (which can be done by refining any initial Markov partition by images under $T$) and subsequent approximation of $\mu$ on a rectangle by a product measure with accuracy controlled (essentially) by the diameter of the small partition \cite{Chernov3}. These results were later improved using spectral methods by Dolgopyat \cite{Dolgopyat} and Liverani \cite{Liverani} to show exponential decay of $C_{F,G}$ for the cases of clear physical relevance. 

It is natural to expect on purely physical grounds that the time correlation (which tends to zero iff $\mu$ is mixing) will decay exponentially. Although there are examples of Axiom A flows with arbitrarily slow decay of correlations, this expectation is nevertheless valid for the known physically relevant examples and conjectured to be valid for arbitrary mixing Anosov flows (it is known that the decay is superpolynomial in this case) \cite{Chernov}. Besides accounting for cases of obvious interest here, this circle of results indicates that we can use the decay rate or \emph{mixing time} of the time correlation function--which depends only on the Anosov diffeomorphism or flow in all known cases--as a natural dynamically intrinsic timescale that is independent of the Markov partition, unlike most otherwise plausible candidates for $t_\infty$. The approximation of the mixing time is briefly touched upon in section \ref{sec:ulam}.

So now our experimenter has not only a natural timescale $t_\infty$ but also for each initial condition a symbolic sequence of (approximate) partition elements, and she can form from the frequencies of symbols an empirical probability associated with each partition element. This is a gigantic amount of information, and it is only natural to assume that if $T$ represented the dynamics of some equilibrium system, then our experimenter could in principle go beyond the construction of an (approximate) SRB measure to provide a full characterization of the Gibbsian statistical physics, including an effective temperature and an effective energy spectrum. Taking the idea that ``there is no conceptual difference between stationary states in equilibrium and out of equilibrium'' \cite{Gallavotti2b} to its natural conclusion, the experimenter should be able to provide an effective temperature and energies in either case. 

This idea was anticipated in \cite{Gallavotti1b}, where the rates of expansion and contraction under the action of $T$ were said to ``provide an `energy function' that assigns relative probabilistic weights to the coarse grained cells,'' and in fact was followed up with a proposal to define an effective temperature of a thermostat keeping the system in a stationary nonequilibrium state by $\dot W/ \dot S$, where $\dot W$ is the work rate of external forces on the system and $\dot S$ is the entropy production rate (see also \cite{GC2}). However, the proposal of \cite{Gallavotti1b} still requires reference to a predefined energy function of some sort in order to define a sensible notion of work rate.

\section{\label{sec:temperature} The effective temperature}

There is a quick derivation of the Gibbs distribution for a finite system from the basic postulate that \emph{the probability of a state depends only on its energy}. This approach has the benefit of motivating the construction of the effective temperature presented later in the section. While this approach does not motivate the introduction of entropy, the standard information-theoretic motivation provides an adequate remedy. We sketch the derivation here and note that it can be made rigorous without substantial difficulty. 

The key observation is that energy is only defined up to an additive constant. This and the basic postulate imply that
\begin{equation}
\mathbb{P}(E_k) = \frac{f(E_k)}{\sum_j f(E_j)} = \frac{f(E_k + \varepsilon)}{\sum_j f(E_j + \varepsilon)}
\end{equation}
for some function $f$ and $\varepsilon$ arbitrary. Define
\begin{equation}
g_E(\varepsilon) := \frac{\sum_j f(E_j + \varepsilon)}{\sum_j f(E_j)}
\end{equation}
and note that $ g_E(0) = 1$ by definition. It follows that
\begin{equation}
\mathbb{P}(E_k) = \frac{f(E_k)}{\sum_j f(E_j + \varepsilon)} g_E(\varepsilon) = \frac{f(E_k + \varepsilon)}{\sum_j f(E_j + \varepsilon)}.
\end{equation}
Therefore $f(E_k) \cdot g_E(\varepsilon) = f(E_k + \varepsilon)$, implying that $f(E_k + \varepsilon) - f(E_k) = (g_E(\varepsilon) - 1) \cdot f(E_k)$. In turn, since $ g_E(0) = 1$, we obtain $f'(E_k) = g'_E(0) \cdot f(E_k)$. 

As a result,
\begin{equation}
f(E_k) = C \exp (g'_E(0)E_k).
\end{equation}
Setting, without loss of generality, $\beta_o := -g'_E(0)$ and $C \equiv 1$ produces the Gibbs distribution so long as the ``ordinary'' temperature is \emph{defined} via $T := 1/k_B\beta_o.$ Note also that $g_E(\varepsilon) = \exp(-\beta_o \varepsilon)$ so that $g_E \equiv g$, as required for the self-consistency of the argument. Although the derivation here is only appropriate for a fixed $\beta_o$, this amounts to the canonical ensemble.

Besides the naturalness of using an invariance principle, there is a good reason to treat the notion of temperature as central to the derivation of the Gibbs distribution in lieu of entropy. Namely, entropies can be introduced, defined, and applied in much broader contexts than temperature, precisely because of the abstract and general nature of the entropy concept. Temperature, on the other hand, has no obvious information-theoretical interpretation, and often its physical interpretation is nontrivial. It is nevertheless generally expected that the kinetic temperature (which serves as a convenient operational parameter in general physical situations by considering weak coupling) can be bijectively associated with the ``true'' temperature, though they may not be equal \cite{MD}. The present paper aims to broaden the scope of applicability of (effective) temperature but does not attempt to supply any information-theoretical interpretation.

With the preceding arguments in mind, we pause briefly to consider some general implications of a widely applicable effective temperature before reviewing its derivation. A stationary (or sufficiently slowly varying) physical system with $n < \infty$ states is typically described in terms of its state energies and inverse temperature whenever possible. While this description greatly aids the theorist armed with a Hamiltonian in making predictions about the statistical behavior of simple systems in equilibrium, in practice many systems of interest are too complex or far from equilibrium to admit easy characterization in terms of their microscopic dynamics. They ``are governed by emergent rules. This means, in practice, that if you are locked in a room with the system Hamiltonian, you can't figure the rules out in the absence of experiment'' \cite{LaughlinP}. 

Such situations are remedied by the identification of effective theories. With a physically reasonable effective temperature determined in terms of experimentally accessible data, an experimenter can close the Gibbs relation to arrive at an effective energy function. This procedure appears to offer the prospect of formulating good effective physical theories for many systems beyond the scope of present methods. The present paper will develop this technique in the context of the two paradigmatic Anosov systems and also indicate how it can be applied more generally, including to manifestly nonequilibrium systems.

We now give a brief summary of the derivation presented in \cite{Huntsman,Huntsman2} and drawing on earlier work in \cite{FordH,Ford}. 

The behavior of a stationary system with $n < \infty$ states and characteristic timescale (here, a mixing or relaxation time) $t_\infty$ and occupation probabilities $p_j$ may equivalently be described in terms of $t := t_\infty p \equiv (t_1,\dots,t_n)$. Write $H = (E_1,\dots,E_n,\beta^{-1})$ for a tuple of notional state energies appended with an effective temperature. Now the typical goal of equilibrium statistical physics is to produce predictions along the lines $H \mapsto p$ using the Gibbs relation $p_j = Z^{-1}e^{-\beta E_j}$. While this map is obviously not invertible and $t_\infty$ does not enter the usual equilibrium picture explicitly, as we shall sketch below there is an essentially unique physically reasonable bijection between $t$ and $H$, and the map $t \mapsto H$ is also of interest for purposes of characterization.

To construct this map, first (w/l/o/g) reset the zero point of energy so that $\sum_j E_j = 0$ (with a trivial adjustment, we can later set $\sum_j E_j$ to any constant, e.g. $n \beta^{-1}$; note also that this does not place any meaningful constraint on the effective internal energy $\sum_j p_j E_j$) and invoke the Gibbs relation to obtain 
\begin{equation}
\label{eq:betaE}
\gamma_k := \beta E_k = \frac{1}{n} \sum_{j=1}^n \log p_j - \log p_k.
\end{equation}
Note that $\beta H = (\gamma_1,\dots,\gamma_n,1)$, so we can compute $\lVert \beta H \rVert = \sqrt{\lVert \gamma \rVert^2 + 1}$. If we knew $\lVert H \rVert$, we could obtain $\beta \equiv \beta(t) = \sqrt{\lVert \gamma \rVert^2 + 1}/\lVert H \rVert$ (as well as $H$ itself) trivially. We will accomplish this below using simple symmetry and physical considerations.

\begin{figure}[htbp]
\includegraphics[trim = 10mm 20mm 10mm 25mm, clip, width=10cm,keepaspectratio]{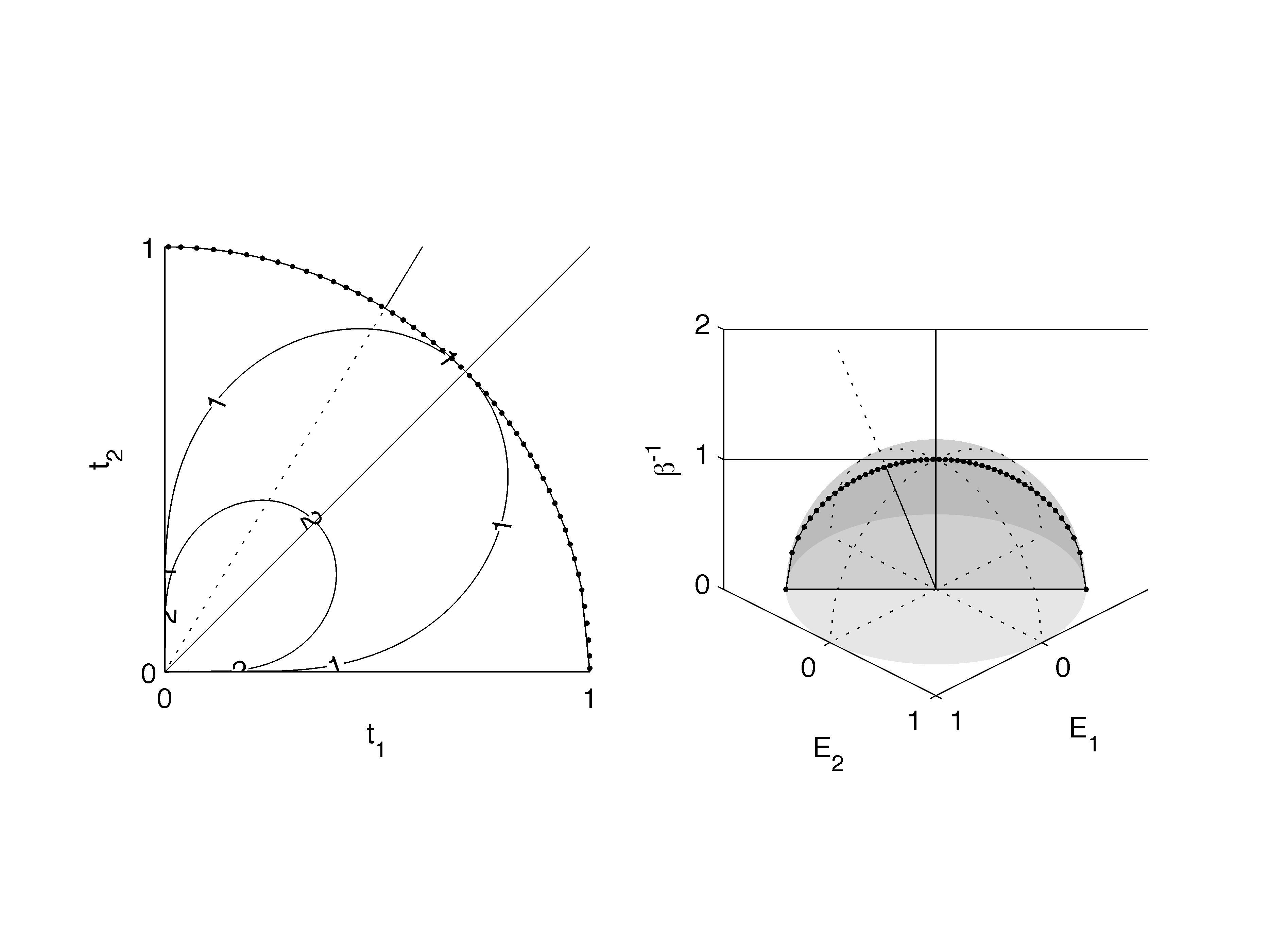}
\caption{ \label{fig:t2H} Geometry of the bijection $t \leftrightarrow H$ for two states. Contours of $\beta^{-1} = 1,2$ are shown in both coordinate systems.}
\end{figure}

A dilation $t \mapsto C t$ (equivalently, $t_\infty \mapsto C t_\infty$) leaves $p$ invariant, so by \eqref{eq:betaE} it must also leave $\gamma$ invariant. Under the \emph{Ansatz} $\beta \equiv \beta(t)$, this dilation effects the transformation
\begin{equation}
H  = \frac{1}{\beta(t)}(\gamma,1) \mapsto \frac{1}{\beta(C t)}(\gamma,1)
\end{equation}
which is itself a dilation by $\beta(t)/\beta(C t)$. This is equivalent to the observation that $p$ is constant on rays in both $t$- and $H$-coordinates, where $t_k/\sum_j t_j = p_k = e^{-\beta E_k}/\sum_j e^{-\beta E_j}$. By the same token, under mild differentiability assumptions on the map between $t$ and $H$ we have that $dt = d\lVert t \rVert e_r^{(t)} \iff dH = d\lVert H \rVert e_r^{(H)}$, where $e_r^{(\cdot)}$ denotes a unit radial vector. That is, rays and orthants of spheres in $t$-coordinates map to rays and hemispheres in $H$-coordinates, and conversely (see figure \ref{fig:t2H}).

The observations above show that $\lVert H(t) \rVert = \lVert H(\lVert t \rVert \hat 1) \rVert = 1/\beta(\lVert t \rVert \hat 1)$, where $\hat 1 := (1,\dots,1)/\sqrt{n}$. Moreover, 
\begin{equation}
\beta(t) = \beta(\lVert t \rVert \hat 1) \cdot \sqrt{\lVert \gamma \rVert^2 + 1}.
\end{equation}
Meanwhile, a multitude of physical considerations detailed in \cite{Huntsman,Huntsman2} (e.g., dimensional analysis, scaling for ideal gas systems, extended canonical transformations, etc.) require that $\beta$ scales as $t_\infty$, so that (up to an overall constant)
\begin{equation}
\label{eq:temperature}
\beta(t) = \lVert t \rVert \sqrt{ \lVert \gamma \rVert^2 + 1 }  = t_\infty \lVert p \rVert \sqrt{ \lVert \gamma \rVert^2 + 1 } 
\end{equation}
where $\lVert \gamma \rVert^2$ is computed using \eqref{eq:betaE}.

The case of a two-state system serves to illustrate the basic features of the bijection between $t$ and $H$ depicted in figure \ref{fig:t2H}. 
A few lines of algebra yield
\begin{equation}
\beta = \left [ (t_1^2 + t_2^2) \left( \frac{1}{2} \log^2 \frac{t_1}{t_2} + 1 \right) \right]^{1/2}; \quad E_1 = -\frac{1}{2\beta} \log \frac{t_1}{t_2}; \quad E_2 = -E_1.
\end{equation}
In the other direction, we enforce $E_j \mapsto E_j - (E_1 + E_2)/2$, so that again $E_2 = -E_1$ and
\begin{equation}
t_\infty = \beta \left [ \frac{\cosh 2 \gamma_1}{2\cosh^2 \gamma_1} \left ( 2\gamma_1^2 + 1 \right ) \right ]^{-1/2}; \quad t_1 = t_\infty \frac{e^{-\gamma_1}}{e^{-\gamma_1}+e^{\gamma_1}}; \quad t_2 = t_\infty \frac{e^{\gamma_1}}{e^{-\gamma_1}+e^{\gamma_1}}.
\end{equation}

With (\ref{eq:temperature}) and the Gibbs relation to obtain the energies, an experimenter has everything she needs in order to provide an effective description of the system in the idiom of equilibrium statistical physics. The principal difficulties associated with the practical application of \eqref{eq:temperature} are generally the identification of an appropriate coarse-grained partition of phase space and timescale. Even in the case of a single-spin system, where the partition is not an issue, and a proper choice of timescale exists, this choice is still not completely obvious (see appendix \ref{sec:glauber}). Provided however that the partition and timescale chosen for an actual equilibrium physical system are appropriate, this effective framework will (re)capture the essential features of the usual statistical physics. More importantly, however, it allows equilibrium and steady-state (or even sufficiently slowly-varying) nonequilibrium systems to be treated on the same footing.

The central theme of this paper is that given $t_\infty$ of the sort mentioned in section \ref{sec:gedanken}, there appears to be a uniquely determined choice of effective inverse temperature for Anosov systems based on physically reasonable considerations. The particulars bear close similarity with the well known variational principle relating the topological pressure and entropy of dynamical systems (see appendix \ref{sec:variational}).

\section{\label{sec:catmap} A simple Markov partition for the cat map}

Let $A \in \pm SL(d,\mathbb{Z})$, i.e., $A$ is an integral $d \times d$ matrix with determinant $\pm 1$. If $A$ has no eigenvalues of modulus 1, then $A$ has both stable and unstable eigenspaces, $A\mathbb{Z}^d = \mathbb{Z}^d$ and $A$ determines an invertible map from the torus $\mathbb{T}^d \cong \mathbb{R}^d/\mathbb{Z}^d$ to itself. For these reasons such maps are called \emph{hyperbolic toral automorphisms}; they provide the simplest examples of Anosov diffeomorphisms. 

The \emph{Arnol'd-Avez cat map} $T_A$ is the two-dimensional hyperbolic toral automorphism defined by $A =\left(\begin{smallmatrix}
  2 & 1 \\
  1 & 1
\end{smallmatrix} \right)$. That is, $T_Ax = Ax \mbox{ mod } 1$, where the modulus is taken componentwise. Henceforth we shall restrict ourselves to this choice of $A$ unless stated otherwise. The relative simplicity of the cat map (or of more general two-dimensional hyperbolic toral automorphisms) provides fertile ground for exploring the effective statistical physics of Anosov systems. For example, the SRB measure is just (the pushforward of) Lebesgue measure, which also serves as a probability measure. 

\begin{figure}[htbp]
\includegraphics[trim = 0mm 50mm 0mm 50mm, clip, width=10cm,keepaspectratio]{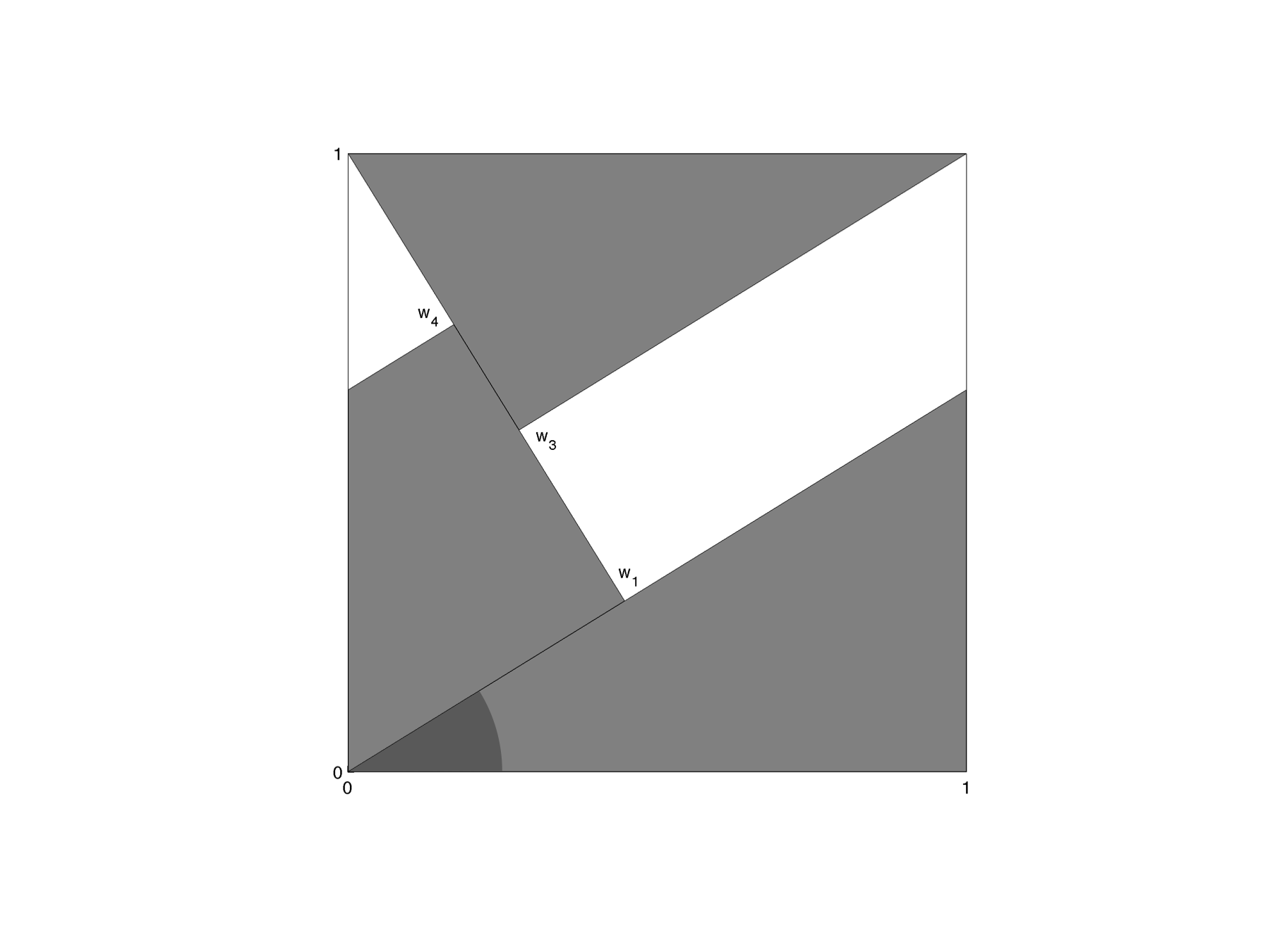}
\caption{ \label{fig:twoelementpartition} A partition $\mathcal{R}_0(\theta)$ of $\mathbb{T}^2$. The shaded angle $\theta$ is in $(0, \pi/4)$.}
\end{figure}

With this in mind, we describe a Markov partition for the cat map originally constructed by Adler and Weiss. For $0 < \theta < \pi/4$ consider the partition $\mathcal{R}_0(\theta)$ of $\mathbb{T}^2$ indicated in figure \ref{fig:twoelementpartition}.
Writing $c := \cos \theta$, $s := \sin \theta$, $e_+ := (c, s)^*$ and $e_- := (s, -c)^*$, it is easy to see that the nontrivial vertices of both rectangles in $\mathcal{R}_0(\theta)$ are $w_1 = se_+$, $w_3 = e_2 + se_-$, and $w_4 = (c + s)e_+ - e_1$. Moreover, $c^2 = \frac{1}{3-\phi}$ and $s^2 = \frac{2-\phi}{3-\phi}$, where $\phi := \frac{1+\sqrt{5}}{2}$ is the golden ratio. 

Write $\lambda_\pm = \phi^{\pm 2}$. For $\theta_A := \cot^{-1} \phi$ it is easy to verify that $e_+$ and $e_-$ are respectively the normalized unstable and stable eigenvectors corresponding to $A$; the eigenvalues are $\lambda_\pm$, and their logarithms are the Lyapunov exponents. Thus like $\mathcal{R}_0(\theta_A)$, the rectangles in the partition $T_A \mathcal{R}_0(\theta_A)$ of $\mathbb{T}^2$ have edges parallel to the $x_\pm$-directions; more specifically, $T_A w_1 \sim s \lambda_+ e_+$, $T_A w_3 \sim s \lambda_- e_-$, and $T_A w_4 \sim (c+s)\lambda_+ e_+$, where $\sim$ indicates equivalence under the projection $\pi: \mathbb{R}^2 \rightarrow \mathbb{T}^2$. 

\begin{figure}[htbp]
\includegraphics[width=10cm,keepaspectratio]{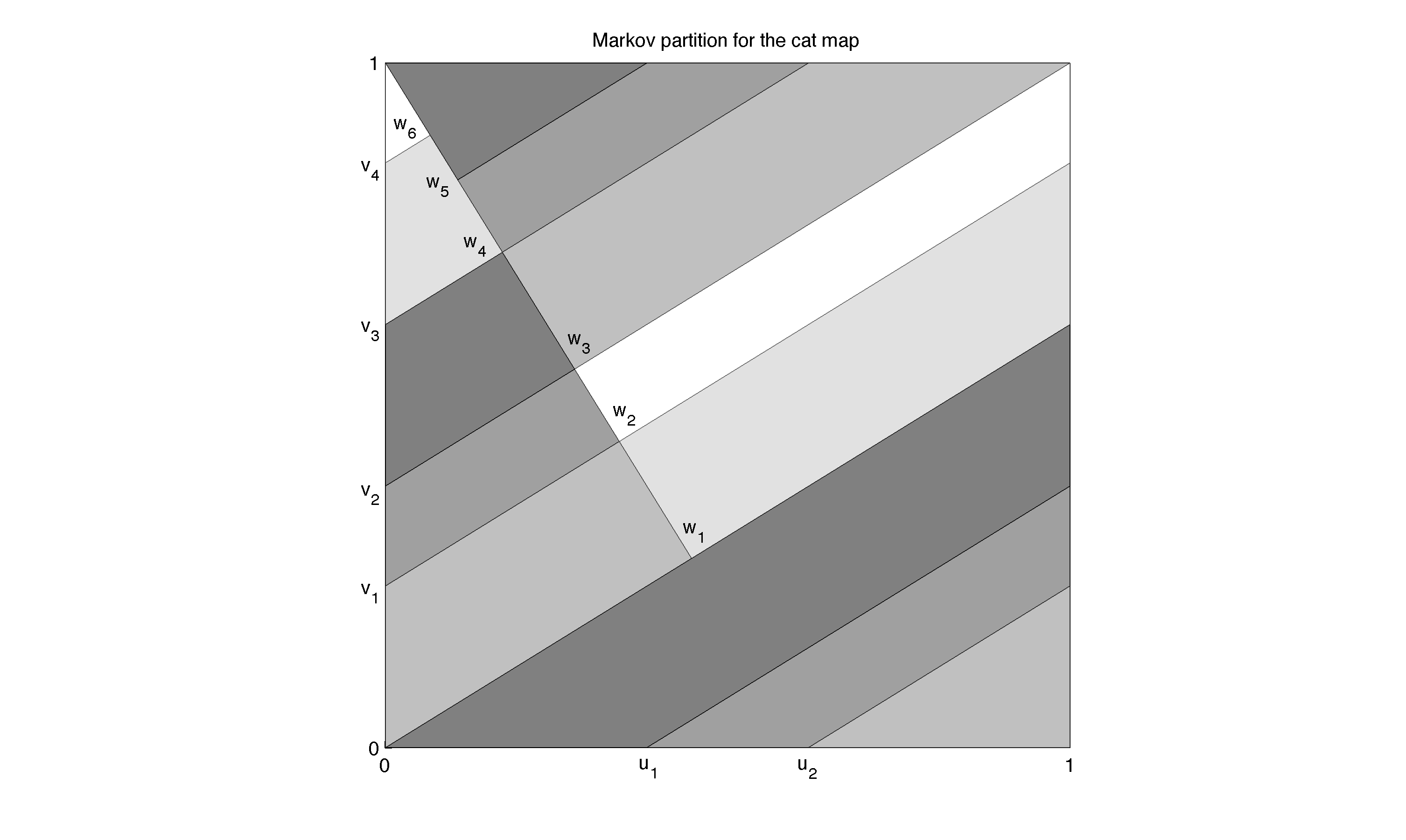}
\caption{ \label{fig:markovpartition} $\mathcal{R}_A =  \mathcal{R}_0(\theta_A) \lor T_A\mathcal{R}_0(\theta_A)$ is a Markov partition for the cat map $T_A$. The constituent rectangles are labeled in order of their original encounter by starting at the origin and proceeding counterclockwise around the unit square, so that the first rectangle is the darkest and the fifth is the lightest.}
\end{figure}

It follows that $\mathcal{R}_A =  \mathcal{R}_0(\theta_A) \lor T_A\mathcal{R}_0(\theta_A)$ is a Markov partition for $T_A$, where the join of partitions (i.e., the partition formed by intersections) is indicated by the symbol $\lor$. Indeed, elementary topological considerations along with the equalities $\lambda_+s = c + s$ and $\lambda_-c = c - s$ suffice to establish that $\mathcal{R}_A$ is as shown in figure \ref{fig:markovpartition}.
In particular, the lengths of the resulting five rectangles in the contracting ($x_-$) direction take on only two values: call the greater length $a$ and the smaller length $b$. It follows that $a = 2s - c$ and $b = 2c - 3s$, and from here all the points $u_*$, $v_*$, and $w_*$ indicated in figure \ref{fig:markovpartition} can be calculated straightforwardly. 

\begin{figure}[htbp]
\includegraphics[trim = 10mm 40mm 10mm 30mm, clip, width=10cm,keepaspectratio]{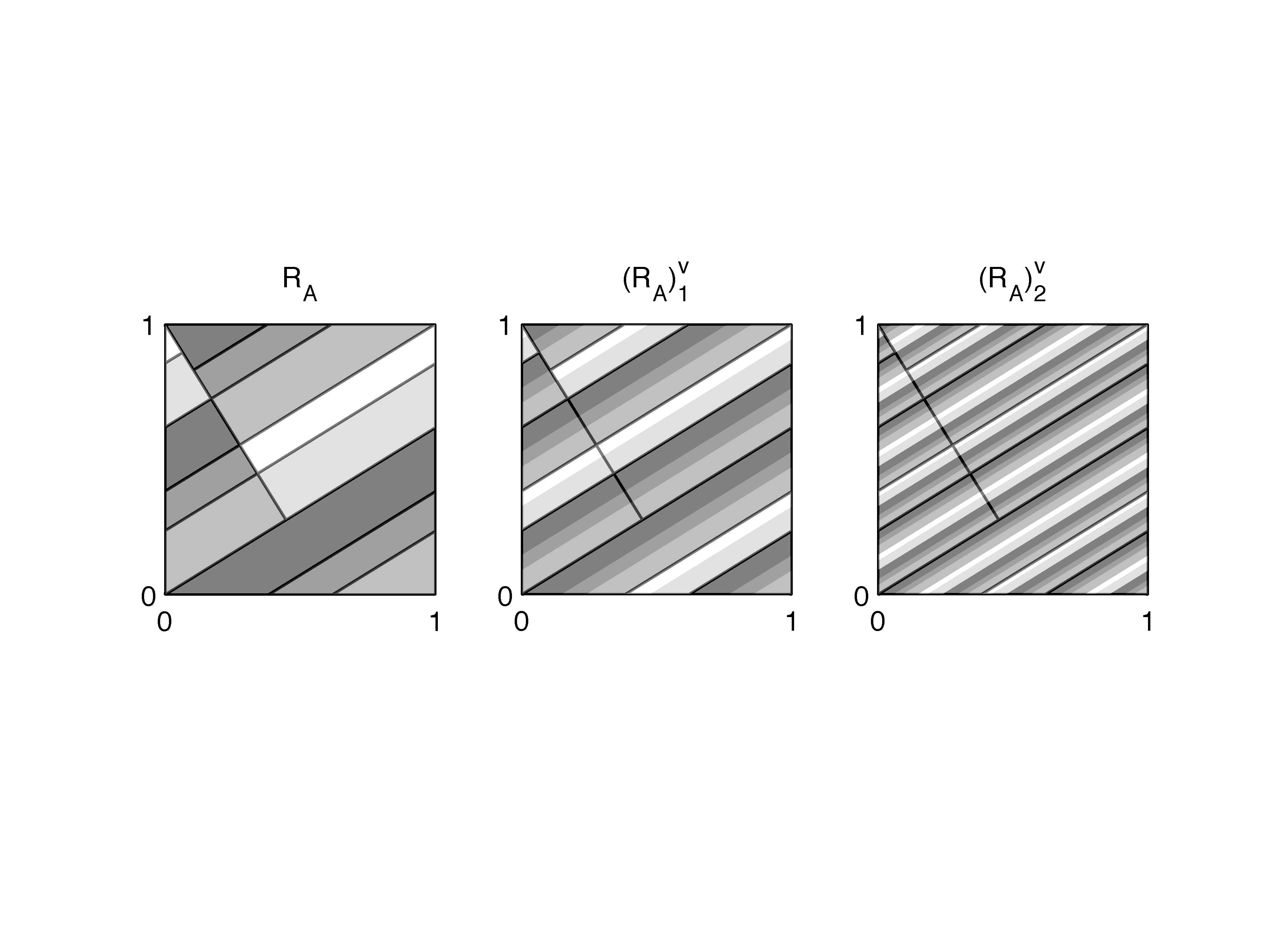}
\caption{ \label{fig:awrefinements} $(\mathcal{R}_A)^\lor_m \equiv \bigvee_{j=0}^{m} T_A^j \mathcal{R}$ for (L) $m = 0$, (C) $m = 1$, and (R) $m = 2$.}
\end{figure}

Note that if $\mathcal{R}$ is a Markov partition for $T$ then so is $\bigvee_{j=-m'}^{m} T^j \mathcal{R}$. Bearing this in mind, it is a straightforward (if slightly tedious) combinatorial exercise to show that a partition of the form $\bigvee_{j=-m'}^{m} T_A^j \mathcal{R}_A$ always contains $n_0 := F_{2(m+m')+1}$, $n_1 := 2F_{2(m+m')+2}$, and $n_2 := F_{2(m+m')+3}$ rectangles of relative measure $1$, $\phi$, and $\phi^2$, respectively, where $F_k = (\phi^k -(-\phi)^{-k})/\sqrt{5}$ is the $k$th Fibonacci number (see figure \ref{fig:awrefinements}).
The corresponding probabilities are $1/\Sigma$, $\phi/\Sigma$, and $\phi^2/\Sigma$, where $\Sigma := n_0 + n_1 \phi + n_2 \phi^2$. Now $\lVert p \rVert^2 = (n_0 + n_1 \phi^2 + n_2 \phi^4)/\Sigma^2$, and substitution yields
\begin{equation}
\lVert p \rVert^2 = \frac{ F_{2(m+m')+1}+2F_{2(m+m')+2}\phi^2 + F_{2(m+m')+3}\phi^4 }{ (F_{2(m+m')+1}+2F_{2(m+m')+2}\phi + F_{2(m+m')+3}\phi^2)^2 } \sim \frac{\phi^{2(m+m')+1}(1+2\phi^3 + \phi^6)/\sqrt{5}}{\phi^{4(m+m')+2}(1+2\phi^2 + \phi^4)^2/5}.
\end{equation}
Reducing powers of $\phi$ using $\phi^2 = \phi + 1$ leads to the asymptotic formula 
\begin{equation}
\lVert p \rVert^2 \sim \frac{4 \sqrt{5}}{25 \phi} \phi^{-2(m+m')} =: C_1 \phi^{-2(m+m')}
\end{equation}
with $C_1 \approx 0.2211$.

Meanwhile, with $n \equiv n_1 + n_2 + n_3$, we can see that $n^{-1} \sum_j \log p_j = (n_0/n) \log (1/\Sigma) + (n_1/n) \log (\phi/\Sigma) + (n_2/n) \log (\phi^2/\Sigma) =  (n_1/n) \log \phi + (2n_2/n) \log \phi - \log \Sigma$. When we form $\gamma_k = n^{-1} \sum_j \log p_j - \log p_k$ the contributions from $\log \Sigma$ cancel, so that $\gamma$ has $n_0$ entries equal to $\log \phi \cdot (n_1 + 2n_2)/n$, $n_1$ entries equal to $\log \phi \cdot (n_1 + 2n_2)/n - \log \phi = \log \phi \cdot (n_2-n_0)/n$, and $n_2$ entries equal to $\log \phi \cdot (n_1 + 2n_2)/n - 2\log \phi = -\log \phi \cdot (2n_0 + n_1)/n$. Substituting for the $n_j$ gives that $\gamma$ has $F_{2(m+m')+1}$ entries equal to $2\log \phi \cdot F_{2(m+m')+4}/F_{2(m+m')+5}$, $2F_{2(m+m')+2}$ entries equal to $\log \phi \cdot F_{2(m+m')+2}/F_{2(m+m')+5}$, and $F_{2(m+m')+3}$ entries equal to $-2 \log \phi \cdot F_{2(m+m')+3}/F_{2(m+m')+5}$.

It follows that 
\begin{equation}
\lVert \gamma \rVert^2 = \left( \frac{ \log \phi}{F_{2(m+m')+5}} \right )^2 \cdot (F_{2(m+m')+1} \cdot 4F_{2(m+m')+4}^2 + 2F_{2(m+m')+2} \cdot F_{2(m+m')+2}^2 + F_{2(m+m')+3} \cdot 4F_{2(m+m')+3}^2) 
\end{equation}
and using $F_\ell \sim \phi^\ell/\sqrt{5}$ gives
\begin{equation}
\lVert \gamma \rVert^2 \sim \frac{ \log^2 \phi}{\sqrt{5}} \cdot (16\phi+10) \phi^{2(m+m')-4} =: C_2 \phi^{2(m+m')}
\end{equation}
where $C_2 \approx 0.5422$.

These approximations above are very good; note also that although the number of states increases exponentially, the density of spectra (i.e., not only the values of $E$ but also their relative multiplicities) converges.

With this in hand, we can quickly establish the behavior of $\beta/t_\infty$ as a function of the class of Markov partitions induced by $\mathcal{R}_A$. To a very good approximation we have that for $m+m' >> 1$
\begin{equation}
\frac{\beta}{t_\infty} = \lVert p \rVert \sqrt{ \lVert \gamma \lVert^{2} + 1 } \approx \lVert p \rVert \cdot \lVert \gamma \lVert = \sqrt{C_1 C_2} \approx 0.3463.
\end{equation}
So the dependence of the effective temperature on the partition of the form described above is essentially nonexistent once its diameter is sufficiently small. As we shall see, this behavior holds more generally and yet is nontrivial.

\section{\label{sec:nontriviallimiting} Limiting behavior for $\beta/t_\infty$ under refinements of the Markov partition by its images}

Given a Markov partition $\mathcal{R}$ for $T_A$, let $z_+$ and $z_- \equiv z_-^{(0)}$ denote the extents of the corresponding rectangles in the unstable and stable directions, respectively. By the Markov property, the number $B^{(m)}_{jk}$ of (connected components of) rectangles in $\mathcal{R}^\lor_m := \bigvee_{j=0}^m T_A^j \mathcal{R}$ that are contained within the $j$th rectangle of $\mathcal{R}$ and have extent $z^{(m)}_{-,k}$ in the stable direction is a well-defined integer. (Note that when considering SRB measures of rectangles in $\bigvee_{j=-m'}^{m} T^j \mathcal{R}$ it suffices to set $m' = 0$ by the $T$-invariance of $\mu_{SRB}$. We exploit this here and henceforth for notational convenience more than anything else.)

Let $\alpha_{kj}$ be the number of times that the interior of $T_A R_k$ crosses the interior of $R_j$ in the stable direction--i.e., $\alpha$ is the so-called \emph{Markov matrix} of $\mathcal{R}$, and its spectrum consists of the absolute value of the spectrum of $A$ along with zeros and roots of unity \cite{Snavely}.
\footnote{Note that by taking suitable partition refinements we can force the Markov and transition matrices to coincide.}
Now $B^{(m)}_{k\ell}$ is the number of rectangles in $T_A\mathcal{R}^\lor_m \cap T_A R_k$ with stable extent $\lambda_- z^{(m)}_{-,\ell} = z^{(m+1)}_{-,\ell}$. It follows that $\sum_k \alpha_{kj} B^{(m)}_{k\ell}$ gives the number of rectangles in $T_A \mathcal{R}^\lor_m \cap R_j = \mathcal{R}^\lor_{m+1} \cap R_j$ with stable extent $z^{(m+1)}_{-,\ell}$: that is, we have $B^{(m+1)} = \alpha^*B^{(m)}$.

The matrix $B^{(m)}$ has a left inverse, and furthermore $z_-^{(m)} := B^{(m)} \backslash z_-^{(0)}$ gives the corresponding extents of rectangles of $\mathcal{R}^\lor_m$ in the stable direction. This gives us everything we need in order to compute $\beta/t_\infty$ for this family of partitions. Because the SRB measure is Lebesgue measure, it admits a product decomposition along the stable and unstable directions. There are $B^{(m)}_{jk}$ rectangles of measure $\zeta^{(m)}_{jk} := z_{+,j} z^{(m)}_{-,k}$, so 
\begin{equation}
\lVert p \rVert^2 = \sum_{j,k} B^{(m)}_{jk} (z_{+,j} z^{(m)}_{-,k})^2
\end{equation}
and
\begin{equation}
\lVert \gamma \rVert^2 = \sum_{r,s} B^{(m)}_{rs} \left( \frac{1}{\sum_{t,u}B^{(m)}_{t,u}} \sum_{j,k} B^{(m)}_{jk} \log \frac{z_{+,j} z^{(m)}_{-,k}}{z_{+,r} z^{(m)}_{-,s}} \right)^2.
\end{equation}

The useful aspect of this construction is that $B^{(m)}$ and $z_-^{(m)}$ can generally be computed without too much difficulty and only slight tedium. For $\mathcal{R}_A$, we have by visual inspection that $z_+ = c(\phi,\phi,\phi,1,1)^*$ and that 
\begin{equation}
\begin{pmatrix}
  3 & 2 \\
  2 & 1
\end{pmatrix} \begin{pmatrix}
  z_{-,1} \\
  z_{-,2}
\end{pmatrix} = \begin{pmatrix}
  c \\
  s
\end{pmatrix},
\end{equation}
so $z_{-,1} = 2s-c = b\phi$, where $b: = z_{-,2} = 2c-3s$. Moreover $z_- \equiv z_-^{(0)} = b\cdot(\phi,1,\phi,\phi,1)^*$ and again by inspection
\begin{equation}
B^{(0)} = \begin{pmatrix}
  0 & 1 \\
  1 & 0 \\
  0 & 1 \\
  0 & 1 \\
  1 & 0 \\
\end{pmatrix}.
\end{equation}

Since (as visual inspection of figure \ref{fig:greedyaw} makes clear)
\begin{equation}
\alpha = a(\mathcal{R}_A) = \begin{pmatrix} 
1 & 0 & 1 & 1 & 0 \\ 
1 & 0 & 1 & 1 & 0 \\ 
1 & 0 & 1 & 1 & 0 \\ 
0 & 1 & 0 & 0 & 1 \\ 
0 & 1 & 0 & 0 & 1
\end{pmatrix}
\end{equation}
it is easy to check that for $m > 0$
\begin{equation}
B^{(m)} = \alpha^*B^{(m-1)} = \begin{pmatrix}
  F_{2m} & F_{2m+1} \\
  F_{2m-1} & F_{2m} \\
  F_{2m} & F_{2m+1} \\
  F_{2m} & F_{2m+1} \\
  F_{2m-1} & F_{2m} \\
\end{pmatrix} \sim \frac{\phi^{2m}}{\sqrt{5}} \begin{pmatrix}
  1 & \phi \\
  \phi^{-1} & 1 \\
  1 & \phi \\
  1 & \phi \\
  \phi^{-1} & 1 \\
\end{pmatrix}
\end{equation}
where as usual $F_\ell = (\phi^\ell - (-\phi)^{-\ell})/\sqrt{5}$ is the $\ell$th Fibonacci number.

Now we note that the inverse of
\begin{equation}
\begin{pmatrix}
  F_{2m} & F_{2m+1} \\
  F_{2m-1} & F_{2m} \\
\end{pmatrix}
\end{equation}
is 
\begin{equation}
\begin{pmatrix}
  -F_{2m} & F_{2m+1} \\
  F_{2m-1} & -F_{2m} \\
\end{pmatrix}.
\end{equation}
Using this and the row redundancy of $B^{(m)}$ and $z_-$ we can easily compute $z_-^{(m)} = B^{(m)} \backslash z_-$, obtaining 
\begin{equation}
z_-^{(m)} = b \phi^{-2m} \begin{pmatrix}
  1 \\
  \phi \\
\end{pmatrix}.
\end{equation}

Thus the matrix $\zeta^{(m)}$ with entries $\zeta^{(m)}_{jk} := z_{+,j}z^{(m)}_{-,k}$ is 
\begin{equation}
\zeta^{(m)} = cb \phi^{-2m}\begin{pmatrix}
  \phi & \phi^2 \\
  \phi & \phi^2 \\
  \phi & \phi^2 \\
  1 & \phi \\
  1 & \phi \\
\end{pmatrix}
\end{equation}
and
\begin{equation}
\lVert p \rVert^2 = \sum_{jk} B^{(m)}_{jk} \left ( \zeta^{(m)}_{jk} \right )^2 \sim \frac{\phi^{2m}}{\sqrt{5}}  (cb \phi^{-2m})^2 \cdot \left ( \phi^2 + \phi^5 + \phi + \phi^4 + \phi^2 + \phi^5 + 1 + \phi^3 + \phi^{-1} + \phi^2 \right )
\end{equation}
where the sum in parentheses on the RHS is in lexicographic order w/r/t matrix entries. Simplifying this using $\phi^2 = \phi + 1$ leads to
\begin{equation}
\lVert p \rVert^2 \sim \frac{4c^2b^2}{\sqrt{5}}  (5\phi+3) \phi^{-2m} = C_1 \phi^{-2m}
\end{equation}
with $C_1 \approx 0.2211$ as before.

By the same token, we have 
\begin{equation}
\lVert \gamma \rVert^2 = \sum_{r,s} B^{(m)}_{rs} \left( \frac{X}{\sum_{t,u}B^{(m)}_{t,u}} \sum_{j,k} \frac{B^{(m)}_{jk}}{X} \log \frac{\zeta^{(m)}_{jk}}{Y} - \log \frac{\zeta^{(m)}_{rs}}{Y} \right)^2
\end{equation}
for any scalars $X, Y$. The choices $X = \phi^{2m}/\sqrt{5}$, $Y = cb \phi^{-2m}$ are most convenient and lead to 
\begin{equation}
\sum_{j,k} \frac{B^{(m)}_{jk}}{X} \log \frac{\zeta^{(m)}_{jk}}{Y} \sim \log \phi \cdot (1 + 2\phi + \phi^{-1} + 2 + 1 + 2\phi + 0 + \phi + 0 + 1) = \log \phi \cdot (6\phi  + 4).
\end{equation}

Furthermore, $\sum_{j,k} B^{(m)}_{jk}/X = 5\phi + 3$, so if we note that $\frac{6\phi + 4}{5\phi+3} \log \phi = 2(\phi - 1) \log \phi$ then
\begin{equation}
\lVert \gamma \rVert^2 \sim \frac{\log^2 \phi}{\sqrt{5}}  \cdot \left ( \Delta_3 + \phi\Delta_4 + \phi^{-1}\Delta_3 + \Delta_4 + \Delta_3 +\phi\Delta_4 + \Delta_2 + \phi\Delta_3 + \phi^{-1}\Delta_2 + \Delta_3 \right ) \cdot \phi^{2m}
\end{equation}
where $\Delta_\ell := (2\phi-\ell)^2 = 4(1-\ell)\phi + 5\ell -2$.
Now
\begin{equation}
\lVert \gamma \rVert^2 \sim \frac{\log^2 \phi }{\sqrt{5}} \cdot \left ( \phi [-4\phi+8] + [2\phi+2][-8\phi+13] + [2\phi+1][-12\phi+20] \right ) \cdot \phi^{2m} = \frac{2(\phi+1)}{\sqrt{5}} \log^2 \phi \cdot \phi^{2m} = C_2 \phi^{2m}
\end{equation}
where as we have already seen $C_2 \approx 0.5422$.

In summary, this reproduces the calculation of the previous section: $\beta/t_\infty \sim \sqrt{C_1 C_2} \approx 0.3463$.

\begin{figure}[htbp]
\includegraphics[width=8cm,keepaspectratio]{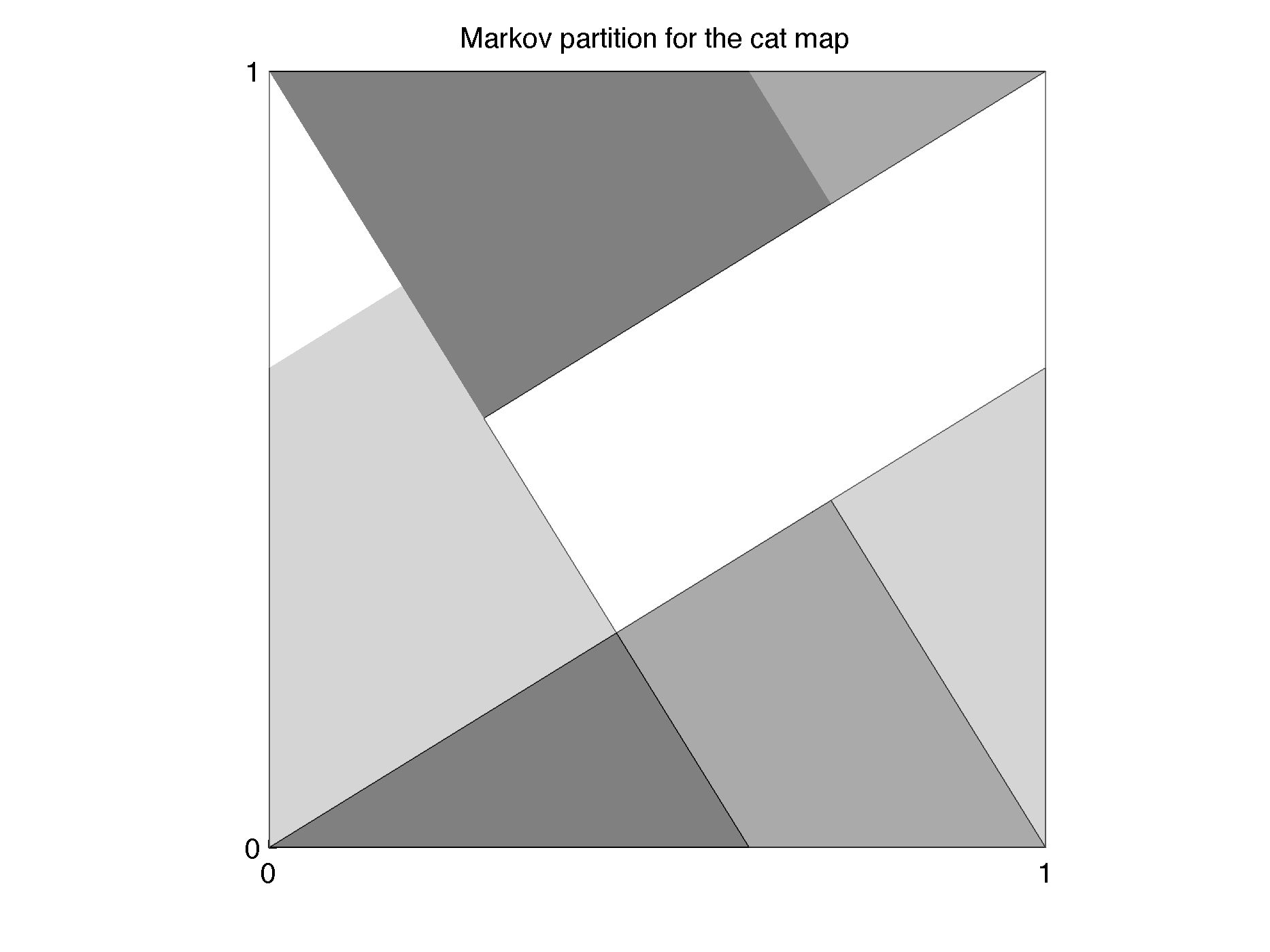}
\caption{ \label{fig:alternatecatpartition} Another Markov partition $\mathcal{R}'_A$ for $T_A$. For concreteness, we label the darkest rectangle $R'_1$, the next darkest rectangle $R'_2$, the next darkest rectangle $R'_3$, and the white rectangle $R'_4$.}
\end{figure}

Another Markov partition $\mathcal{R}'_A$ for $T_A$ is indicated in figure \ref{fig:alternatecatpartition}.
The calculation for $\mathcal{R}'_A$ is carried out along entirely similar lines. We have $z_+ = s(1,\phi^{-1},1,\phi)^*$ and $z_- = s(1,1,1,\phi^{-1})^*$. Visual inspection of figure \ref{fig:greedych} shows that
\begin{equation}
B^{(0)} = \begin{pmatrix}
  0 & 1 \\
  0 & 1 \\
  0 & 1 \\
  1 & 0 \\
\end{pmatrix}
\end{equation}
and
\begin{equation}
\alpha = a(\mathcal{R}'_A) = \begin{pmatrix} 
1 & 1 & 1 & 0 \\ 
1 & 1 & 0 & 0 \\ 
0 & 0 & 1 & 1 \\ 
1 & 1 & 1 & 1
\end{pmatrix},
\end{equation}
so that
for $m > 0$
\begin{equation}
B^{(m)} = \begin{pmatrix}
  F_{2m} & F_{2m+1} \\
  F_{2m} & F_{2m+1} \\
  F_{2m} & F_{2m+1} \\
  F_{2m-1} & F_{2m} \\
\end{pmatrix} \sim \frac{\phi^{2m}}{\sqrt{5}} \begin{pmatrix}
  1 & \phi \\
  1 & \phi \\
  1 & \phi \\
  \phi^{-1} & 1 \\
\end{pmatrix}.
\end{equation}

We can compute $z_-^{(m)} = B^{(m)} \backslash z_-$ just as for $\mathcal{R}_A$, obtaining 
\begin{equation}
z_-^{(m)} = s \phi^{-2m-1} \begin{pmatrix}
  1 \\
  \phi \\
\end{pmatrix}.
\end{equation}
Thus the matrix $\zeta^{(m)}$ is 
\begin{equation}
\zeta^{(m)} = s^2 \phi^{-2m-1}\begin{pmatrix}
  1 & \phi \\
  \phi^{-1} & 1 \\
  1 & \phi \\
  \phi & \phi^2 \\
\end{pmatrix}
\end{equation}
and
\begin{equation}
\lVert p \rVert^2 = \sum_{jk} B^{(m)}_{jk} \left ( \zeta^{(m)}_{jk} \right )^2 \sim \frac{\phi^{2m}}{\sqrt{5}}  (s^2 \phi^{-2m-1})^2 \cdot \left ( 1+ \phi^3 + \phi^{-2} + \phi + 1 + \phi^3 + \phi + \phi^4 \right )
\end{equation}
which simplifies to
\begin{equation}
\lVert p \rVert^2 \sim \frac{s^4}{\sqrt{5}} \phi^{-3} (9\phi^2 + 6\phi + \phi^{-1}) \phi^{-2m} = \frac{8s^4}{\sqrt{5}} \phi^{-2m} =: C'_1 \phi^{-2m}
\end{equation}
with $C'_1 \approx 0.2733$.

Meanwhile, taking $X = \phi^{2m}/\sqrt{5}$, $Y = s^2 \phi^{-2m-1}$ leads to 
\begin{equation}
\sum_{j,k} \frac{B^{(m)}_{jk}}{X} \log \frac{\zeta^{(m)}_{jk}}{Y} \sim \log \phi \cdot ( 0 + \phi - 1 + 0 + 0 + \phi + \phi^{-1} + 2 ) = 3 \phi \log \phi.
\end{equation}
Similarly, $\sum_{j,k} B^{(m)}_{jk}/X = 3\phi + 4 + \phi^{-1} = 4\phi + 3$. Writing $\Delta'_\ell := \frac{3\phi}{4\phi+3}-\ell$, we obtain
\begin{equation}
\lVert \gamma \rVert^2 \sim \frac{\phi^{2m}}{\sqrt{5}} \log^2 \phi \cdot \left ( \Delta'_0 + \phi \Delta'_1 + \Delta'_{-1} +\phi \Delta'_0 + \Delta'_0 + \phi \Delta'_1 + \phi^{-1}\Delta'_1 + \Delta'_2 \right )
\end{equation}
Now 
\begin{equation}
\lVert \gamma \rVert^2 \sim \frac{\phi^{2m}}{\sqrt{5}}  \log^2 \phi \cdot \left ( \left[ \frac{3\phi}{4\phi+3} \right]^2[4\phi+3] -\frac{3\phi}{4\phi+3}6\phi + 3\phi +4 \right ) = \frac{  \log^2 \phi }{\sqrt{5}}  \cdot \frac{ 28\phi + 15 }{ 4\phi +3 } \cdot \phi^{2m} =: C'_2 \phi^{2m}
\end{equation}
with $C'_2 \approx 0.6593$.

So for the family of partitions induced by $\mathcal{R}'_A$ we have that $\beta/t_\infty \sim \sqrt{C'_1 C'_2} \approx 0.4245 \ne 0.3463$: the limiting values of $\beta/t_\infty$ differ for the partition families induced by $\mathcal{R}_A$ and $\mathcal{R}'_A$.

However, another initial partition gives the same result as for $\mathcal{R}_A$. For the two-element Markov partition $\mathcal{R}''_A$ from example 1.4 of \cite{Snavely} and depicted in figure \ref{fig:greedysn} below, we have $z_+ = (s,c)^*$ and $z_- = (s,c)^*$. From figure \ref{fig:greedysn} we see that
\begin{equation}
B^{(0)} = \begin{pmatrix}
  1 & 0 \\
  0 & 1 \\
\end{pmatrix}
\end{equation}
and
\begin{equation}
\alpha = \begin{pmatrix} 
1 & 1 \\ 
1 & 2 \\ 
\end{pmatrix},
\end{equation}
so that for $m > 0$
\begin{equation}
B^{(m)} = \begin{pmatrix}
  F_{2m-1} & F_{2m} \\
  F_{2m} & F_{2m+1} \\
\end{pmatrix} \sim \frac{\phi^{2m}}{\sqrt{5}} \begin{pmatrix}
  \phi^{-1} & 1 \\
  1 & \phi \\
\end{pmatrix}.
\end{equation}
We have 
\begin{equation}
z_-^{(m)} = (B^{(m)})^{-1}z_- = \begin{pmatrix}
  F_{2m+1} & -F_{2m} \\
  -F_{2m} & F_{2m-1} \\
\end{pmatrix} \begin{pmatrix}
  s \\
  c \\
\end{pmatrix} = s \phi^{-2m} \begin{pmatrix}
  1 \\
  \phi \\
\end{pmatrix}
\end{equation}
so that 
\begin{equation}
\zeta^{(m)} = s^2 \phi^{-2m}\begin{pmatrix}
  1 & \phi \\
  \phi & \phi^2 \\
\end{pmatrix}
\end{equation}
and
\begin{equation}
\lVert p \rVert^2 = \sum_{jk} B^{(m)}_{jk} \left ( \zeta^{(m)}_{jk} \right )^2 \sim \frac{\phi^{2m}}{\sqrt{5}}  (s^2 \phi^{-2m})^2 \cdot \left ( \phi^{-1} + \phi^2 + \phi^2 + \phi^5 \right ) = \frac{4s^4}{\sqrt{5}}   \cdot \left ( 2\phi + 1 \right )\phi^{-2m} =: C''_1 \phi^{-2m}
\end{equation}
with $C''_1 \approx 0.5789$.

Continuing, with $X = \phi^{2m}/\sqrt{5}$, $Y = s^2 \phi^{-2m}$ we have 
\begin{equation}
\sum_{j,k} \frac{B^{(m)}_{jk}}{X} \log \frac{\zeta^{(m)}_{jk}}{Y} \sim \log \phi \cdot ( 0 + 1 + 1 + 2\phi ) = 2\phi^2 \log \phi
\end{equation}
and $\sum_{j,k} B^{(m)}_{jk}/X = \phi + 2 + \phi^{-1} = \phi^3$. Since $\phi^{-3} \cdot 2\phi^2 \log \phi = 2(\phi - 1) \log \phi$,  
\begin{equation}
\lVert \gamma \rVert^2 \sim \frac{\phi^{2m}}{\sqrt{5}}  \log^2 \phi \cdot \left ( \phi^{-1} [2(\phi-1)]^2 + [2(\phi-1)-1]^2 + [2(\phi-1)-1]^2 + \phi[2(\phi-1)-2]^2 \right )
\end{equation}
which simplifies to 
\begin{equation}
\lVert \gamma \rVert^2 \sim \frac{\phi^{2m}}{\sqrt{5}}  \log^2 \phi \cdot \left ( \phi^{-1} [-4\phi+8] + [-8\phi+13] + [-8\phi+13] + \phi[-12\phi+20] \right ) = \frac{2}{\sqrt{5}}  \log^2 \phi \cdot \phi^{2m} =: C''_2 \phi^{2m}
\end{equation}
with $C''_2 \approx 0.2071$. 

Therefore for $\mathcal{R}''_A$, we have $\beta/t_\infty \sim \sqrt{C''_1 C''_2} \approx 0.3463$. Note that $\sqrt{C''_1 C''_2} = \sqrt{C_1 C_2} \ne \sqrt{C'_1 C'_2} $. This suggests that differences between limiting values of $\beta/t_\infty$ among initial Markov partitions has a subtle origin.

The key implication can be loosely put that while $\beta/t_\infty$ may have a complex dependence on the ``shape'' of the Markov partition used, it does not depend substantially on the ``scale'' of that partition. This holds in greater generality, as we proceed to sketch below.

For a generic two-dimensional hyperbolic toral automorphism, the rectangles in a Markov partition are not rectangular in the Euclidean sense, but instead are (unions of connected) parallelograms since the defining matrix $A \in \pm SL(2, \mathbb{Z})$ need not be symmetric. Nevertheless, the key elements of the scheme outlined above remain effectively unchanged, provided that one first defines a defect angle as the difference between the angles of the two eigendirections less $\pi/2$ and then multiplies $z_-^{(m)}$ by the cosine of the defect angle in the calculations for $\lVert p \rVert^2$ and $\lVert \gamma \rVert^2$. Because this cosine contributes only a multiplicative constant for $\lVert p \rVert^2$ and does not contribute at all to $\lVert \gamma \rVert^2$, it does not affect the qualitative picture. 

In general, by linearity and the Markov property $B^{(m)}$ will satisfy a linear recurrence relation with integral coefficients. For $A = \left(\begin{smallmatrix}
  2 & 1 \\
  1 & 1
\end{smallmatrix} \right)$ the recurrence relation is $B^{(m)} = 3B^{(m-1)}-B^{(m-2)}$, whereas for $A = \left(\begin{smallmatrix}
  3 & 2 \\
  1 & 1
\end{smallmatrix} \right)$ the recurrence relation turns out to be $B^{(m)} = 4B^{(m-1)}-B^{(m-2)}$. (Note that unlike the cat map, the second example does not have orthogonal stable and unstable directions.)
The recurrence relation itself does not depend on the particular Markov partition, but only on the defining matrix $A \in \pm SL(2,\mathbb{Z})$ and the initial condition.

A sketch of this is as follows. If $A \in SL(2,\mathbb{Z})$ then its characteristic polynomial is $\chi_A(y) = y^2 - \mbox{Tr}A \cdot y + 1$. By the Cayley-Hamilton theorem, $\chi_A(A) = 0$, so that powers of $A$ satisfy a second-order linear recurrence relation with integral coefficients. Moreover, the spectral radius of $\alpha$ satisfies $\rho(\alpha) = \rho(A)$ for any Markov partition $\mathcal{R}$ by Theorem 2.2 of \cite{Snavely}.
It follows that $B^{(m+1)} = \alpha^*B^{(m)}$ satisfies (up to the initial condition) the same recurrence relation as the powers of $A$, viz. $B^{(m)} = \mbox{Tr}A \cdot B^{(m-1)} - B^{(m-2)}$. 

More importantly for our purposes, and simply by virtue of the fact that $B^{(m)}$ satisfies a linear recurrence relation, it has a well-defined scaling behavior that (by construction) is the inverse of that for $z_-^{(m)}$. Because $\lVert p \rVert^2$ scales as $z_-^{(m)}$ and $\lVert \gamma \rVert^2$ scales as $B^{(m)}$, the two terms counterbalance each other and so $\beta/t_\infty$ converges to a finite nonzero value as $m \uparrow \infty$. It follows similarly that the density of spectra (i.e., the effective energy values counting only relative multiplicities) converges to a finite set.

The linear substructure of hyperbolic toral automorphisms appears to be the fundamental reason for this phenomenon. Although even in three dimensions Markov partitions for hyperbolic toral automorphisms are necessarily fractal \cite{KH}, it is reasonable to conjecture that some weaker form of this phenomenon should continue to hold in higher dimensions, since linearity is still preserved. Moreover, because every Anosov diffeomorphism of the torus $\mathbb{T}^d$ is topologically conjugate to a hyperbolic toral automorphism and these maps play such a central role in the theory of Anosov diffeomorphisms, the significance of the above phenomenon should not be underestimated from the point of view of statistical physics in general, and of the chaotic hypothesis in particular.

That said, and although there is a sense in which SRB measures for Anosov systems are almost product measures \cite{BPS}, it is far from clear that the analogue of the recurrence relation here should necessarily enjoy the same linearity properties. Indeed, we shall see in section \ref{sec:geodesic} that the situation can be considerably more complicated in general, although consideration of physically reasonable refinements of Markov partitions still appear to yield unique limits for $\beta$. 

Finally, while the results of this and the preceding section serve to motivate the relevance and applicability of effective statistical physics to the Ruelle program, they should not be misinterpreted as fundamental (apart from the asymptotic independence of $\beta/t_\infty$ w/r/t scale). Their import derives from the indirect evidence they provide both to support and to further constrain the functional form \eqref{eq:temperature} of the effective temperature, and also for the impetus they provide for the results in later sections. Specifically, these results suggest that neither $\beta$ nor $t_\infty$ should have any explicit dependence on the number of states, which in turn lends additional credence to the already-mentioned identification of $t_\infty$ with the mixing time for Anosov systems and motivates the approach taken in section \ref{sec:greedy}.

\section{\label{sec:nontriviality} Nontriviality of limiting behavior}

The balancing act detailed above in sections \ref{sec:catmap} and \ref{sec:nontriviallimiting} is far from trivial. To see this, first let $q \in (0, 1)$. Consider a family of partitions of the unit interval defined inductively by setting $\mathcal{Y}^{(0)} := \{[0,1]\}$ and forming $\mathcal{Y}^{(m+1)}$ by subdividing the intervals in $\mathcal{Y}^{(m)}$ into subintervals of relative (Lebesgue) measure $q$ and $1-q$. 

It is not hard to see that the corresponding probability tuple $p^{(m)}$ has $\binom{m}{r}$ entries equal to $q^r (1-q)^{m-r}$, from which it follows that 
\begin{equation}
\lVert p^{(m)} \rVert^2 = \left(q^2 + (1-q)^2\right)^m.
\end{equation} 
Similarly, 
\begin{equation}
2^{-m} \sum_j \log p_j^{(m)} = m \left( \frac{1}{2} \log \frac{q}{1-q} + \log (1-q) \right),
\end{equation}
where we have used $\sum_r \binom{m}{r} = 2^m$ and $\sum_r \binom{m}{r} r = m2^{m-1}$. Now $\gamma^{(m)}$ has $\binom{m}{r}$ entries equal to $\left( \frac{m}{2} - r \right) \log \frac{q}{1-q}$, so 
\begin{equation}
\lVert \gamma^{(m)} \rVert^2 = \log^2 \frac{q}{1-q} \cdot \sum_{r = 0}^m \binom{m}{r} \left( \frac{m}{2} - r \right)^2 = \log^2 \frac{q}{1-q} \cdot m 2^{m-2},
\end{equation}
where we have also used $\sum_r \binom{m}{r} r^2 = m(m+1)2^{m-2}$. The net result is that $\beta/t_\infty$ cannot have a finite nonzero limit in this situation: indeed, $\beta/t_\infty$ tends to infinity as $m$ increases unless $q = 1/2$, in which case $\beta/t_\infty$ tends to zero.

Viewed another way, the existence of a nontrivial limit requires that $\lVert \gamma \rVert^2 \equiv -n \log^2 Z + \lVert \log p \rVert^2$ and $\lVert p \rVert^2$ have precisely inverse scaling behavior (note that $\log Z \equiv -\frac{1}{n} \sum_j \log p_j$ and that $n$ has its own scaling behavior). Even if the entries of $e^{\kappa m} \cdot p$ are bounded by polynomials in $m$, this cannot be expected to hold in any generality. In this context the fact that $\beta/t_\infty$ converges in the manner detailed earlier can be seen to take on a special significance, particularly in light of the relevance of Anosov systems and Markov partitions for nonequilibrium steady states.

\section{\label{sec:implication} An implication for the detailed form of the effective temperature}

The derivation of the form for $\beta$ actually leaves open the possibility that $\beta = f(n) \cdot t_\infty \lVert p \rVert \sqrt{\lVert \gamma \rVert^2 +1}$, where $n = \dim p$. For $p$ uniform this simplifies to $\beta = f(n) \cdot t_\infty / \sqrt{n}$, so it might therefore be imagined that an appropriate choice of $f$ other than unity would be $f(n) = \sqrt{n}$. 

Another slightly more detailed argument in the same direction is as follows. Consider a Hamiltonian system with phase space measure $\Gamma$ and such that $\Gamma(E) := \int_{H \le E} \ d\Gamma < \infty$. If the system is ergodic on constant-energy levels then an expression for the inverse temperature that is particularly well-suited for low-dimensional systems is $\beta_B(E) = \partial_E \log \Gamma(E)$, and up to an additive constant the concomitant entropy is given by $S = \log \Gamma(E)$ \cite{Berdichevsky}. (In the usual expressions, the volume $\Gamma(E)$ is replaced by the relative volume $\partial_E \Gamma$.) If now the system is mixing and a constant-energy level $H = E$ is decomposed into cells of equal measure, then the inverse effective temperature simplifies to $\beta = t_\infty n^{-1/2}$, where $n$ is the number of cells. This would seem to suggest that a factor $n^{-1/2}$ should be absorbed into the expression for the effective temperature, so that $t_\infty$ would be proportional to the relative volume of the energy level, which is just $\partial_E \Gamma$. In such an event we would have $\beta \equiv C \partial_E \Gamma$ and $\beta_B = \Gamma(E)^{-1} \partial_E \Gamma$.

Based on these considerations, scaling according to $f(n) = \sqrt{n}$ might seem to have the advantage of a well-defined microcanonical limit and a clear correspondence with established low-dimensional statistical physics, but as we have seen above that is actually not the case in at least one example where the SRB measure (which we recall generalizes the microcanonical measure) is uniform. Later we shall encounter another example in the geodesic flow on a surface of constant negative curvature: see section \ref{sec:geodesic2}.

Indeed, consider the continuous version of the formula for $\beta$: for this to make sense it is necessary that $\mu_{SRB}$ be absolutely continuous w/r/t $\nu$, which amounts to saying that the probability density $p$ of the SRB measure w/r/t $\nu$ is well-defined and determined by the equality $\int_X d\mu_{SRB} = \int_X p \ d\nu$ for all $\nu$-measurable sets $X$. Although for hyperbolic toral automorphisms and geodesic flows this is the case (indeed, here $p = 1$), in general such a $p$ does not exist: a SRB measure is generally not absolutely continuous w/r/t $\nu$ (in fact absolute continuity of the conditional measure on unstable manifolds alone is an equivalent criterion for \emph{defining} SRB measures \cite{Young}). For example, deterministic dissipative systems will typically contract phase space, so the concomitant SRB measure will necessarily be supported on a (typically dense) set of zero Riemannian measure \cite{Gallavotti1}. At the same time, the restriction of $\mu_{SRB}$ to a $\sigma$-algebra generated by any given set of finite microcells produces a well-defined density on microcells. This observation is related to the Ulam method (see section \ref{sec:ulam}). This makes it clear that in practice a discretization is always required on physical grounds for nonequilibrium steady states (and is usually invoked in equilibrium as well).

So while considerations of the microcanonical ensemble might be taken to suggest that $t_\infty$ should be scaled so as to absorb a ``missing'' factor of $\sqrt{n}$ in a putative continuous limit, the detailed examination of Anosov systems as motivated by the chaotic hypothesis casts doubt on this idea. 


Finally, there is a preferred (characteristic) scale for a Markovian coarse graining: namely, just small enough in order for any physically interesting observables to be well-defined and effectively constant on rectangles. Meanwhile, a secondary fine-graining also arises naturally in the consideration of entropy production in nonequilibrium steady states; its characteristic scale determines the microscopic (local, quasi-ergodic) description of dynamics via a relationship between the mesoscopic coarse-graining scale in concert with the Lyapunov spectrum and a concomitant time scale \cite{Gallavotti1b}. These considerations provide further support to the idea that an effective inverse temperature should be (at least somewhat) insensitive to the scales of phase space discretizations. (See also section \ref{sec:ulam}.)

As it happens, our own considerations naturally and felicitously lead to a class of microscopic Markov partitions as phase space discretizations that appear to obviate scaling issues of the sort discussed here and that furthermore also have physically reasonable uniformity properties. It is to this aspect of our framework that we now turn.

\section{\label{sec:uniformity} Uniformity of partitions}

The variation in the measures $\mu_{SRB}(R_j)$ for $R_j \in \mathcal{R}$ is responsible for the variation in coarse-grained state energies. However, any reasonable fine-graining of phase space should be approximately uniform w/r/t the Riemannian (or associated Liouville) measure. In the examples considered explicitly in this paper, $\mu_{SRB} = \nu$. This results in a degree of tension as one expects the dynamics to contribute to an effective temperature in some way other than through $t_\infty$ even for these simple cases, a possibility that uniform SRB measures of rectangles would preclude. But as we shall see in the sequel and thereafter, this tension also suggests a physically reasonable method for coarse-graining that is at the heart of our results.

Recall that the Anosov property is independent of the particular choice of Riemannian metric. By locally perturbing the metric inside of rectangles, the concomitant Riemannian measure can be made to satisfy $\nu(R_j) = \lvert \mathcal{R} \rvert^{-1}$ \cite{MathOverflow4}. However, such a construction is far from natural, and is physically unmotivated. Moreover, the considerable evidence that ``geodesically conjugate'' Riemannian manifolds of negative curvature are isometric casts doubt on the existence of an analogous approach for geodesic flows \cite{EHS,Eberlein}. More to the point, the Riemannian/Liouville measure associated with such a modified metric--which in the case of a surface of negative curvature is also the SRB measure--will not be invariant under the original geodesic flow \cite{BGGZ}.

It is therefore appropriate to briefly discuss the sense in which we may restrict consideration to a ``natural'' Riemannian measure. While not every Anosov system will preserve a natural Riemannian measure, the prototypical ones will: for hyperbolic toral automorphisms, this is just the (pushforward of) Lebesgue measure, and for the geodesic flow on a surface of constant negative curvature it will be the Liouville measure. More generally, so-called \emph{conservative} diffeomorphisms preserve a natural Riemannian measure (and diffeomorphisms in general preserve an equivalence class of measures) \cite{Wilkinson}. A wide class of conservative diffeomorphisms is furnished by Hamiltonian systems, and it is natural to couch the otherwise implicit notion of a ``natural'' Riemannian measure in this context (we will briefly touch upon the related context of Hamiltonian systems as they pertain to nonequilibrium statistical physics below).

Bearing this in mind, it appears unlikely that a Markov partition can typically be constructed such that (either) a ``natural'' Riemannian (or the SRB) measure is uniform on rectangles \cite{Gallavotti2b}.
\footnote{
Although in general Markov partitions will not have such strong regularity properties, the so-called baker map (which is not Anosov) has SRB measure equal to Lebesgue measure on the unit square and admits Markov partitions consisting of regular grids of squares \cite{LaCourS}. 
}
In the cases considered explicitly here, viz. the cat map and the (Poincar\'e map for the) geodesic flow on a surface of constant negative curvature, the measures of rectangles will often be controlled in a weaker sense. Suppose that $\mu$ is a measure of maximal entropy (see, e.g., \cite{Walters} for background), as is the case for the examples mentioned here. \footnote{This line of reasoning actually provides a demonstration of the well-known fact that the SRB measure for a hyperbolic toral automorphism (i.e., the pushforward of Lebesgue measure) is a measure of maximal entropy. By the same scaling argument as employed in section \ref{sec:nontriviallimiting}, we have that $\log(z_+ z_-^{(m)*}) = -Cm1 + o(m)$, where $C$ is a positive constant and $1$ (here) denotes a matrix of all ones. Since $\sum_{j,k} B^{(m)}_{jk}z_{+,j}z_{-,k}^{(m)} \equiv 1$, we have that $- \frac{1}{Cm} \sum_{j,k} B^{(m)}_{jk}z_{+,j}z_{-,k}^{(m)} \log (z_{+,j}z_{-,k}^{(m)}) \rightarrow 1$, and moreover that $C$ equals both the Kolmogorov-Sinai and topological entropy.} 
Then the \emph{Kolmogorov-Sinai} or \emph{metric entropy} equals the \emph{topological entropy} \footnote{
Here we are implicitly using the Adler-Konheim-McAndrew formulation of topological entropy, which can however be shown equivalent to the Bowen-Dinaburg formulation touched on in appendix \ref{sec:variational}.
}; i.e.
\begin{equation}
- \frac{1}{\log \lvert \mathcal{R}^\vee_m \rvert} \sum_{R \in \mathcal{R}^\vee_m} \mu(R) \log \mu(R) \rightarrow 1 
\end{equation}
where as before we write $\mathcal{R}^\vee_m := \bigvee_{j=0}^m T^j \mathcal{R}$. It follows in this event that the ratios of terms of the form $\log \mu(R)$ tend to unity. However, this only implies that the differences between measures of rectangles are subexponential, and suggests that the utility of entropy for analytically evaluating the uniformity of individual partitions is limited in practice.

Basically, it appears unlikely that Markov partitions for Anosov systems can be generically constructed so that the Riemannian measures of rectangles are precisely uniform, though virtually by default the measures of rectangles will differ only by polynomial factors. This deficit, which might appear to be a mere nuisiance, in fact suggests an avenue whereby a significant conceptual link between effective and traditional equilibrium statistical physics can be forged.

\section{\label{sec:greedy} Greedy refinements of Markov partitions}

Define $\mathcal{R} \vee_T R_j := \left \{ R_k \right \}_{k \ne j} \cup \left \{ T^{-1} \left(TR_j\cap R_k\right) : a_{jk} \ne 0 \right\}$. We have the following

\textsc{Lemma.} $\mathcal{R} \vee_T R_j$ is Markov. 

\textsc{Proof.} See appendix \ref{sec:lemmaapp}. A heuristic sketch reduces to the definition of a Markov partition: namely, the images of rectangles in $T(\mathcal{R} \vee_T R_j)$ will stretch across rectangles in $\mathcal{R} \vee_T R_j$ in the unstable direction, and rectangles in $\mathcal{R} \vee_T R_j$ will stretch across the images of rectangles in $T(\mathcal{R} \vee_T R_j)$ in the stable direction. $\Box$

\textsc{Corollary.} If $\mathcal{S}$ is a refinement of $\mathcal{R}$ and $\mathcal{R} \vee_T R_j$ is a refinement of $\mathcal{S}$, then $\mathcal{S}$ is Markov. In this event call $\mathcal{S}$ a \emph{quasi-minimal refinement} of $\mathcal{R}$. $\Box$

The generalized refinement $\mathcal{R} \vee_{T^t} R_j := \{R_k\}_{k \ne j} \cup \{T^{-t}(T^tR_j \cap R_k)\}$ fails to be Markov for $t$ generic, since (e.g.) the analogue of iv) in the proof fails to hold. This obstruction precludes arbitrarily detailed control over the refinement of Markov partitions. However, the residual control offered by quasi-minimal refinements is still of considerable interest, as we proceed to illustrate.

Call a quasi-minimal refinement \emph{minimal} if it is not a nontrivial refinement of any nontrivial quasi-minimal refinement. Set $\mathcal{R}_0 := \mathcal{R}$. For $k \ge 0$, let $\mathcal{R}_{k+1}$ be a minimal refinement of $\mathcal{R}_k$ of maximal $\nu$-entropy. (An alternative extremal principle could turn out be more appropriate in general, e.g. minimizing the effective free energy.) These \emph{greedy} refinements are asymptotically unique, in the sense that any two sequences of greedy refinements of an initial partition must share infinitely many elements. 

\begin{figure}[htbp]
\includegraphics[trim = 0mm 0mm 0mm 0mm, clip, width=8cm,keepaspectratio]{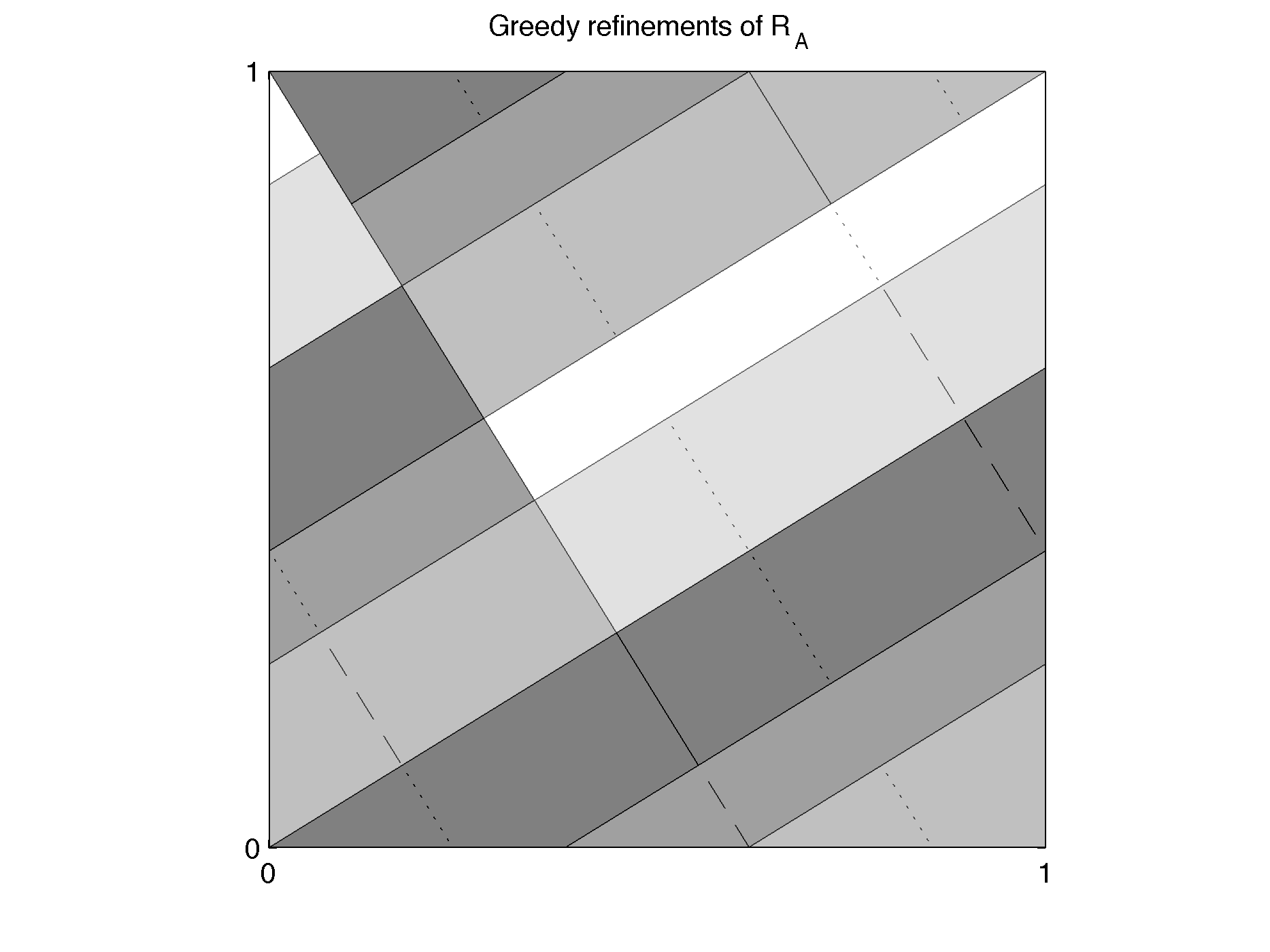}
\caption{ \label{fig:greedyawpartitions} Markov partitions $\mathcal{R}_{A,1}$, $\mathcal{R}_{A,2}$ and $\mathcal{R}_{A,3}$ obtained by the first three rounds of greedy refinements of $\mathcal{R}_A$, indicated respectively by subsequently considering the solid, dashed and dotted line segments.}
\end{figure}

\begin{figure}[htbp]
\includegraphics[trim = 10mm 50mm 10mm 50mm, clip, width=15cm,keepaspectratio]{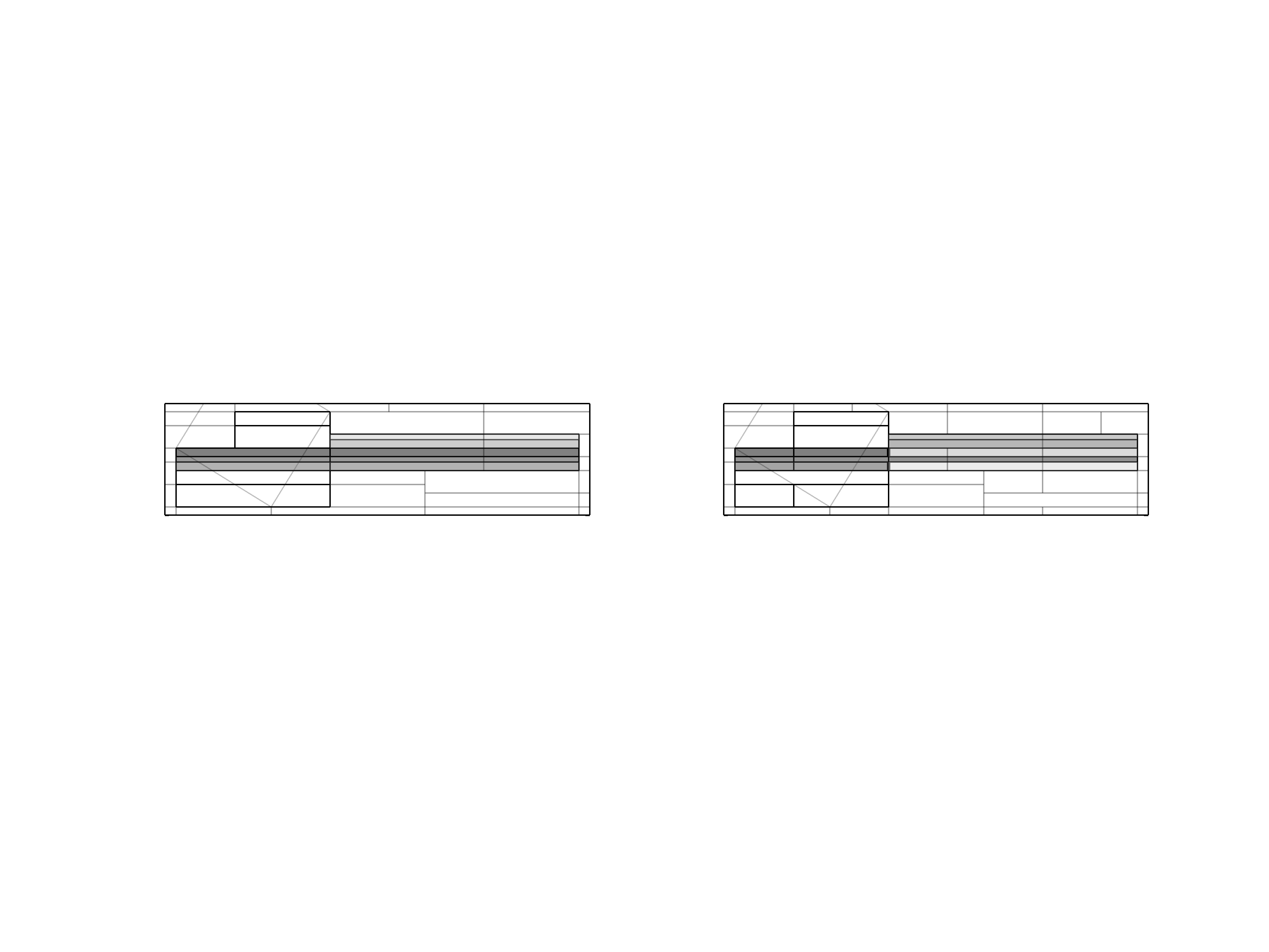}
\caption{ \label{fig:greedyaw} Depiction of greedy refinements in eigencoordinates of $T_A$. A portion of the unit (Cartesian) square and translates of the current partition are shown along with the action of $T_A$ on the current partition. New boundary elements introduced by greedy partitioning are shown in bold. (L) $\mathcal{R}_{A,0}$. (R) $\mathcal{R}_{A,1}$.}
\end{figure}

It can be shown that a sequence of greedy refinements of $\mathcal{R}_A$ contains a subsequence of refinements that are local maxima of the normalized entropy $S_\nu(\mathcal{R})/\log \lvert \mathcal{R} \rvert$ (see, e.g., figure \ref{fig:greedyg234} for an example of this phenomenon in a more complicated setting). The rectangles in the elements $\mathcal{R}_{A,m}$ of this subsequence have relative measures $1$ and $\phi$ and respective multiplicities $L_{m+1}$ and $L_{m+2}$. Here the \emph{Lucas numbers} are given by $L_{m+2} = L_{m+1} + L_m$ with $L_1 = 1$ and $L_2 = 3$, so that $L_m$ is the closest integer to $\phi^m$. (See figures \ref{fig:greedyawpartitions} and \ref{fig:greedyaw}. Note also that the recurrence relations for the Fibonacci and Lucas numbers differ only in their initial conditions.)

It follows that 
\begin{equation}
\lVert p \rVert^2 = \frac{L_{m+2}\phi^2 + L_{m+1}}{(L_{m+2}\phi + L_{m+1})^2} \approx \frac{2}{5} \phi^{-m-1}
\end{equation}
and
\begin{equation}
\lVert \gamma \rVert^2 = \frac{L_{m+2} L_{m+1}^2 + L_{m+1} L_{m+2}^2}{L_{m+3}^2} \log^2 \phi \approx \phi^m \log^2 \phi,
\end{equation}
so that for $\mathcal{R}_{A,m}$
\begin{equation}
\beta/t_\infty \rightarrow \sqrt{\frac{2}{5\phi}} \log \phi \approx 0.2393.
\end{equation}
Besides producing another family of partitions--by a different mechanism--for which $\beta/t_\infty$ converges to a finite value, this also results in the lowest asymptotic value seen thus far. This is to be expected because of the uniformity of the partitions so obtained, which serves not only to maximize the entropy but also to minimize $\beta/t_\infty$.

\begin{figure}[htbp]
\includegraphics[trim = 10mm 50mm 10mm 50mm, clip, width=15cm,keepaspectratio]{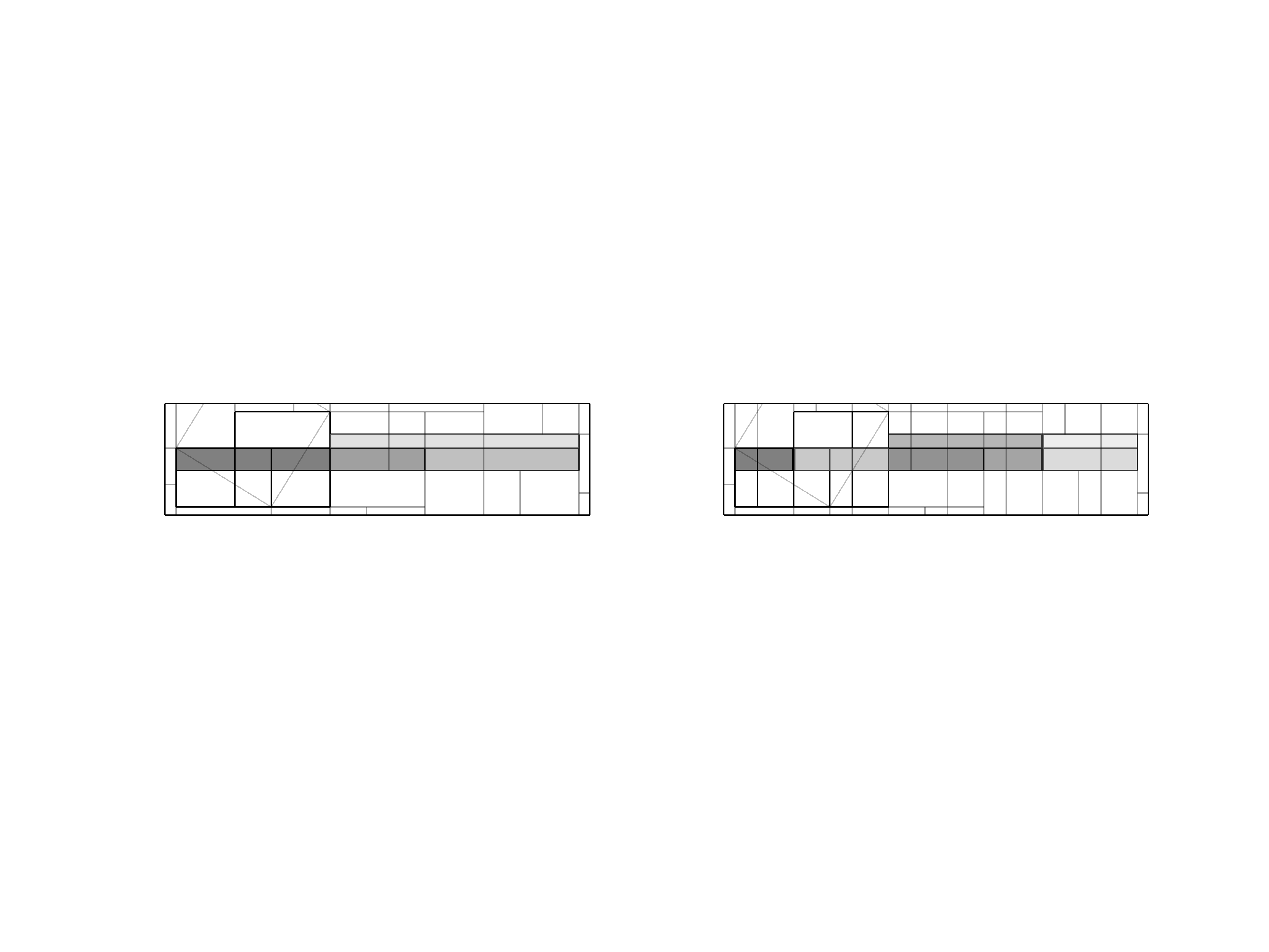}
\caption{ \label{fig:greedych} (L) $\mathcal{R}'_{A,0}$. (R) $\mathcal{R}'_{A,1}$. Note that the result of a round of greedy refinements is not unique, though the value of the resulting entropy is.}
\end{figure}

Remarkably, we obtain precisely the same results when starting with $\mathcal{R}'_A$ instead. To see this, first consider the $\mathcal{R}_{A,m}$. $\mathcal{R}_A$ has five rectangles: one with relative measure 1, two with relative measure $\phi$, and two with relative measure $\phi^2$. We summarize this in the obvious shorthand $(1\times 1, \phi \times 2, \phi^2 \times 2)_{@\mathcal{R}_A}$. The greedy refinement $\mathcal{R}_{A,1}$ is obtained by dividing each of the two rectangles of relative measure $\phi^2$ into two subrectangles of relative (w/r/t $\mathcal{R}_A$) measure 1 and $\phi$, leading to $(1\times 3, \phi \times 4)_{@\mathcal{R}_{A,1}}$. To get $\mathcal{R}_{A,2}$, each of the four rectangles of relative measure $\phi$ is further decomposed, and so on. (See figures \ref{fig:greedyawpartitions} and \ref{fig:greedyaw}.) Meanwhile, we have $(1\times 1, \phi \times 3)_{@\mathcal{R}'_A}$. Subdividing the three rectangles of relative measure $\phi$ produces $\mathcal{R}'_{A,1}$, for which we have $(1\times 3, \phi \times 4)_{@\mathcal{R}'_{A,1}}$: see figure \ref{fig:greedych}. Therefore we see not only how the Lucas numbers come about, but more importantly that the measures of greedy refinements exhibit a stabilization property leading to the equality of the limits for $\beta/t_\infty$. 

\begin{figure}[htbp]
\includegraphics[trim = 10mm 65mm 10mm 10mm, clip, width=15cm,keepaspectratio]{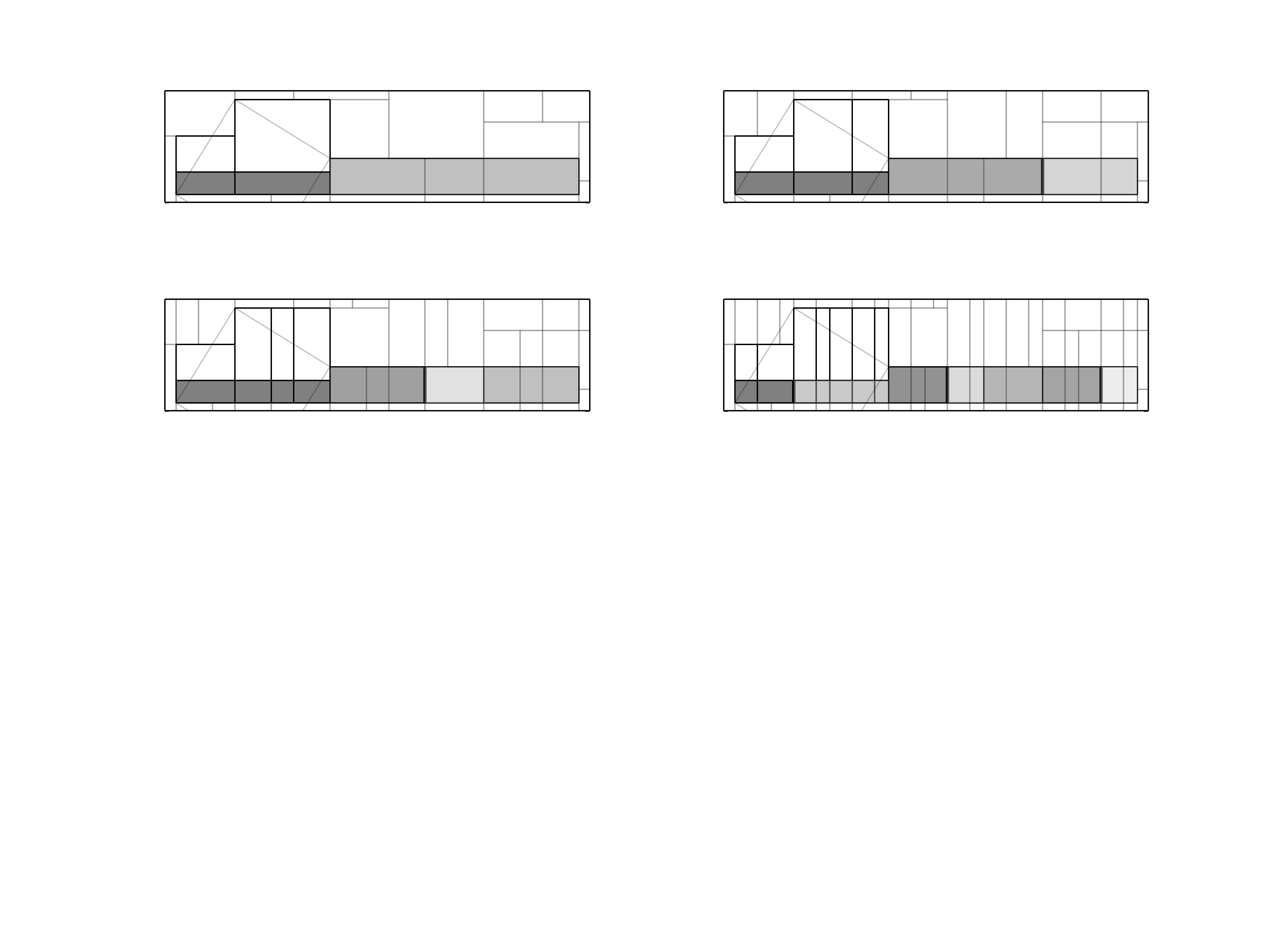}
\caption{ \label{fig:greedysn} Upper left: $\mathcal{R}''_{A,0}$. Upper right: $\mathcal{R}''_{A,1}$. Lower left: $\mathcal{R}''_{A,2}$. Lower right: $\mathcal{R}''_{A,3}$.}
\end{figure}

Moreover, the two-element Markov partition from example 1.4 of \cite{Snavely} (see figure \ref{fig:greedysn}) also yields the same result. For convenience we denote this partition by $\mathcal{R}''_A$ and we start with $(1\times 1, \phi^2 \times 1)_{@\mathcal{R}''_A}$. The first greedy refinement leads to $(1\times 2, \phi \times 1)_{@\mathcal{R}''_{A,1}}$; the second greedy refinement leads to $(1\times 1, \phi \times 3)_{@\mathcal{R}''_{A,2}}$; the third round of three greedy refinements leads to  $(1\times 3, \phi \times 4)_{@\mathcal{R}''_{A,3}}$, and so on.

In light of these equalities it is natural to conjecture that this limiting behavior is universal for the cat map (as well as for hyperbolic toral automorphisms in two dimensions and even more generally in higher dimensions and for constant-ceiling suspensions): i.e., that greedy refinements of any Markov partition $\mathcal{R}$ produce a (sub)sequence of partitions whose measures eventually coincide with those obtained in the same way from any other Markov partition $\mathcal{R}'$. We further conjecture that the limit inferior of $\beta/t_\infty$ over greedy refinements with diameter tending to zero is the physically preferred inverse effective temperature.

To understand why this is plausible in the special case of $T_A$, suppose we have for an initial partition $\mathcal{R}$ that $(1 \times c_0, \phi \times c_1, \dots \phi^u \times c_u)_{@\mathcal{R}}$. We expect that each of the rectangles of relative measure $\phi^v$ can be greedily refined into a rectangle of relative measure $\phi^{v-1}$ and $\phi^{v-2}$; by repeating this, we expect to eventually obtain a refinement $\mathcal{R}'$ s.t. $(1 \times \tilde c_0, \phi \times \tilde c_1)_{@\mathcal{R}'}$, with $\tilde c_0 = c_0 + \sum_{m = 1}^{u-2} F_m c_{m+1} + F_{u-2}c_u$ and $\tilde c_1 = c_1 + \sum_{m=2}^{u-1} F_m c_m + F_{u-1}c_u$. (Note that this greedy refinement $\mathcal{R}'$ can itself be greedily refined.) So the idea is that the allowable initial relative measure multiplicities $\{c_m\}$ appear to be constrained by the Markov condition in such a way that greedy refinements always lead to a subsequence for which the relative measures are independent of the initial condition $\{c_m\}$.

Although this phenomenon does not appear to admit a straightforward generalization to arbitrary Anosov systems, supporting evidence in the context of the geodesic flow on a surface of constant negative curvature (see section \ref{sec:geodesic2}) and the variational principle (see appendix \ref{sec:variational}) do provide a reasonable basis to assume that its essence holds more generally. This also leads to a far-reaching conjecture about the effective statistical physics of nonequilibrium steady states couched in the context of SRB measures for Anosov systems. Before we summarize this proposal, however, we will briefly turn to the role of effective and actual ensembles.

\section{\label{sec:ensembles} Ensembles}

The effective temperature leads to an effective canonical ensemble: in the case of Anosov systems, an effective state space is a Markovian coarse-graining of the underlying configuration space. However this underlying configuration space can also support an actual microcanonical ensemble (recall that the SRB measure generalizes the microcanonical ensemble), for example in the case of geodesic flow on a surface of constant negative curvature discussed in sections \ref{sec:geodesic} and \ref{sec:geodesic2}. While the key to resolving this disconnect has already been described in section \ref{sec:gedanken}, the mechanisms by which effective and actual canonical ensembles can be brought into closer correspondence will be addressed here.

In general, from the point of view of effective statistical physics no fundamental distinction is drawn between equilibrium and nonequilibrium steady states, provided that a suitable characteristic timescale $t_\infty$ is given. However, systems producing nonequilibrium steady states must necessarily be thermostatted in some way \cite{MD,Klages}. A proposal of Gallavotti \cite{Gallavotti1} (see also, e.g., \cite{ES,Gallavotti3,BGGZ,GP1,GP2}) suggests that dynamics in the thermodynamical limit should be insensitive to the details of the thermostat (i.e., the concomitant SRB measures should tend to the same limit). By extension, this ``ensemble equivalence'' proposal holds that nonequilibrium ensembles enjoy a greater latitude in their definition than equilibrium ensembles. 

A mechanism of particular interest to the case of Anosov systems is provided by the so-called Gaussian isokinetic (GIK) thermostat, which preserves the Anosov property. 
\footnote{As a dynamical system, the GIK thermostat including an electric field $E$ is defined by $v = \dot{x}$ and $\nabla_v v = E - v \langle E, v \rangle/\langle v,v \rangle$. Here $\nabla$ denotes the covariant derivative. On a surface of constant negative curvature this is a (dissipative, reversible) Anosov system \cite{BGM,PW}. Physically, the electric field that $E$ represents creates currents circulating around the holes in the surface the particle is restricted to. The thermostat itself maintains the kinetic energy of the system under the influence of $E$. Unless $E$ has a global potential, the Anosov GIK thermostat is dissipative \cite{DP1} and (near equilibrium) has positive entropy production \cite{DP2}. For a phenomenological analogue (i.e., a dissipative reversible perturbation) for the cat map, see \cite{BGM}.} 
The basic idea of the Gaussian thermostat for an autonomous ODE $\dot x = F(x)$ is to project $F(x)$ onto the tangent space $T_x\Sigma$ of a specified constraint manifold $\Sigma$; thus Gaussian thermostats can be employed to fix not only the kinetic or internal energy, but also a field or current \cite{MD}. The GIK and Gaussian isoenergetic (GIE) thermostats have been shown to be formally equivalent in the thermodynamical limit \cite{Ruelle2}. Thus both the GIK and GIE correspond to the microcanonical ensemble (although spatial degrees of freedom are canonically distributed, in most cases of interest this is trivial \cite{MD}). 

However, assuming the validity of the ensemble equivalence proposal and of the chaotic hypothesis, we may simultaneously consider a non-Gaussian thermostat such as the Nos\'e-Hoover (NH) thermostat \cite{Nose1,Nose2,Hoover} and retain effective Anosov properties in the thermodynamical limit: the chaotic hypothesis has been shown to be applicable to both Gaussian and NH thermostats \cite{BGGZ}. Unlike the GIK or GIE thermostats, the NH thermostat corresponds to the canonical ensemble in equilibrium, and maintains the internal energy in nonequilibrium. Both Gaussian and NH thermostats enjoy the seemingly contradictory properties of determinism, dissipativity, and time-reversibility; however, the NH thermostat can be regarded as a generalized Gaussian thermostat in certain contexts. Indeed, the NH thermostat itself admits generalization, providing considerable freedom for the construction of a concrete deterministic reservoir that yields the canonical distribution in many cases, of which the so-called Lorentz gas provides a particularly well-studied example \cite{Klages}. 

While the dynamics of a thermostatted system appear to depend on the details of the thermostat, thermodynamical properties appear not to. That is, the ensemble equivalence proposal is supported by evidence, but may not be applicable to equivalence of chaotic properties (such as the oft-noted equality between phase space contraction and entropy production rates). In particular, transport properties can vary as a function of the thermostat (though this does not seem to be the case with Gaussian or NH thermostats). However, it is conceivable that such variability is due to the presence of only a small number of degrees of freedom, and that the ensemble equivalence proposal holds in full generality in the thermodynamical limit \cite{Klages}.

Even though Gaussian and NH thermostats can be realized in a Hamiltonian framework, in general projecting out thermostat/reservoir degrees of freedom will transform Hamiltonian dynamics into non-Hamiltonian dynamics \cite{Klages}. Indeed, it has been argued that the study of nonequilibrium processes requires the consideration of either non-Hamiltonian or infinite systems \cite{Ruelle1b}.

Having discussed thermostats, we therefore briefly turn to the complementary topics of weak coupling and infinite systems. A natural system to consider the role of ensembles and the thermodynamical limit in the present context is a coupled lattice of Anosov diffeomorphisms, particularly as studied in \cite{BonettoKL,BFG}. Aside from the broad range of physical applications of generic \emph{coupled map lattices} \cite{CF}, coupled lattices of Anosov diffeomorphisms have recently been employed in concert with renormalization group analyses in attempts to understand the origins of diffusion from deterministic microdynamics with local conservation laws \cite{Kupiainen}.

To tie these considerations into our framework in a more concrete manner through a relevant class of examples, write $\mathcal{T} := (\mathbb{T}^2)^{\mathbb{Z}^d}$ and $\mathcal{T}_N := (\mathbb{T}^2)^{V_N^d}$, where $V \equiv V_N^d := \{-N,\dots,N\}^d$. The metric on $\mathcal{T}$ is given by a weighted sum of the metrics on each copy of $\mathbb{T}^2$, with the weights rapidly decreasing on $\mathbb{Z}^d$. Let $\mathcal{A} : \mathcal{T} \rightarrow \mathcal{T}$ be given by $(\mathcal{A}x)_\omega = T_A x_\omega$ for $\omega \in \mathbb{Z}^d$. That is, $\mathcal{A}$ is the Cartesian product of copies of $T_A$ indexed by elements of $\mathbb{Z}^d$. Let $\mathcal{A}_N$ be the restriction of $\mathcal{A}$ to $\mathcal{T}_N$. If $g: \mathcal{T} \rightarrow \mathbb{T}^2$ is well-behaved (e.g., analytic and depending only on tori in some finite neighborhood of the origin) and $\varepsilon$ small, we set $(\mathcal{A}_\varepsilon x)_\omega := T_A x_\omega + \varepsilon g(\tau_{-\omega} x)$, where $(\tau_{-\omega} x)_\zeta := x_{-\omega + \zeta}$. That is, $\mathcal{A}_\varepsilon$ is obtained from $\mathcal{A}$ by adding a weak $\mathbb{Z}^d$-invariant coupling. Considering periodic boundary conditions on $\mathcal{T}_N$ yields a concomitant map $\mathcal{A}_{N,\varepsilon}$.

In \cite{BonettoKL,BFG} it was shown that the SRB measure corresponding to $\mathcal{A}_\varepsilon$ exists and is well behaved for $\varepsilon$ sufficiently small, and that it equals the weak limit of the SRB measures of $\mathcal{A}_{N,\varepsilon}$. Moreover, if the coupling is not degenerate in a certain technical sense, then the projected SRB measures of $\mathcal{A}_{N,\varepsilon}$ and $\mathcal{A}_\varepsilon$ are absolutely continuous w/r/t Lebesgue measure (which is the projection of the SRB measures of $\mathcal{A}_N$ and $\mathcal{A}$ onto $\mathbb{T}^2$). In practical terms this means that the projected SRB measures of $\mathcal{A}_{N,\varepsilon}$ and $\mathcal{A}_\varepsilon$ can be given in terms of probability densities w/r/t the projected SRB measures of $\mathcal{A}_N$ and $\mathcal{A}$.

Before concluding this section, we will briefly discuss the key technical observation at the root of these constructions. It is easily seen that the Cartesian product of Markov partitions is again a Markov partition for the associated product map, and as we shall sketch, a small perturbation of an Anosov diffeomorphism is again Anosov and has a Markov partition that can be simply characterized. 

A basic property of an Anosov 
diffeomorphism $T$ is \emph{structural stability}, i.e., the property that any $T'$ sufficiently close to $T$ in the natural metric is topologically conjugate to $T$. Let $T'$ be a small smooth perturbation of $T$ and $U$ the corresponding topological conjugacy, viz. $U^{-1}T'U = T$. It turns out that if $\mathcal{R}$ is a Markov partition for $T$ then
\begin{equation}
UW^s_{(T)}(x,R_j) = W^s_{(T')}(Ux,UR_j)
\end{equation}
and similarly for the unstable manifolds.

To see this, let $x' \in W^s_{(T)}(x,R_j)$, i.e. $x' \in R_j$ and $d(T^nx,T^nx') \downarrow 0$. Then $Ux' \in UR_j$ and $d(U^{-1}T'^n(Ux),U^{-1}T'^n(Ux')) \downarrow 0$. By continuity, $d(T'^n(Ux),T'^n(Ux')) \downarrow 0$. So $Ux' \in W^s_{(T')}(Ux,UR_j)$. Now $UW^s_{(T)}(x,R_j) = W^s_{(T')}(Ux,UR_j)$ implies that 
\begin{equation}
T'UW^s_{(T)}(x,R_j) = T'W^s_{(T')}(Ux,UR_j) \subset UW^s_{(T)}(U^{-1}T'Ux,R_k) = W^s_{(T')}(T'Ux,UR_k).
\end{equation}
As a consequence of this and the analogous result for unstable manifolds, $U\mathcal{R}$ is a Markov partition for $T'$. 

In particular, and without delving into the issue of explicitly constructing the topological conjugacy if it is not already given (but see, e.g. \cite{MathOverflow3} for general considerations along these lines and \cite{BFG} for an explicit treatment in the case of a coupled cat map lattice), a Markov partition for $T_A$ can therefore be used to obtain a Markov partition for $\mathcal{A}_\varepsilon$. Since as we have seen the ``thermodynamical limit" of finite lattices of coupled cat maps is well-behaved, we can in principle consider the dynamics on a reduced space (such as $\mathcal{T}_N$) from the point of view of the (actual) canonical ensemble. 

The details of an analogous construction for Anosov flows are presently unknown, though it is reasonable to anticipate that one should exist. The paradigmatic example of an Anosov flow (also considered later here) is the geodesic (Hamiltonian) flow on a surface of constant negative curvature. The primary obstacle is that a nonzero coupling destroys the invariance of the individual uncoupled Hamiltonians, thereby precluding uniform hyperbolicity \cite{ABF,Kupiainen}. 

In summary, while the application of effective statistical physics to infinite systems in general requires care, considerations of both the thermodynamical limit and of projected thermostatted dynamics appear to be tenable in the present context. In the same spirit, while some care is clearly required in discussing the role of effective and actual ensembles, in the presence of a weak external coupling or thermostatting the conceptual distinction between effective and actual ensembles can reasonably be expected to range from subtle to nonexistent.

On this basis our framework can be viewed as a proposal for a significant extension of the program initiated by Ruelle for a general theory of nonequilibrium statistical physics. We now turn to the explicit statement of this proposal before corroborating it in the other paradigmatic--and more physically relevant--example of an Anosov system.

\section{\label{sec:proposal} The proposal in a nutshell}

In the preceding sections we have couched the chaotic hypothesis in the context of effective statistical physics. On the basis of results obtained above (and corroborated in section \ref{sec:geodesic2}) for greedy refinements, we have outlined the key elements of a proposal for a comprehensive framework for nonequilibrium statistical physics that simultaneously incorporates and extends the formalism originally introduced by Ruelle and subsequently refined by Gallavotti and coworkers. Besides introducing several new ideas in this paper, we have also provided evidence to support a nascent theory of statistical physics that is truly intrinsic: i.e., that provides a general framework in which not only entropy but also concepts like (effective) temperature and energy can be understood simply in terms of raw temporal information about the dynamics, and without reference to a predetermined Hamiltonian (while bearing in mind that if a Hamiltonian exists, then it can be reconstructed using this framework). 

The proposal, in a nutshell, is that

\begin{quote}
classical physical systems in either equilibrium or nonequilibrium steady states and with a well-defined characteristic timescale $t_\infty$ can be described using the idiom of equilibrium statistical physics. For many-particle systems the chaotic hypothesis is assumed to hold. For Anosov-like systems $t_\infty$ is the mixing or relaxation time, and there is a conjecturally unique and well-defined inverse effective temperature given by the infimum over initial Markov partitions of the limits inferior of the inverse effective temperatures over greedy refinements. This inverse effective temperature may be used to (re)construct an effective energy function by a suitable application of the Gibbs relation. A generalization of the variational principle (see appendix \ref{sec:variational}) is conjectured to hold in which the effective free energy is minimized.
\end{quote}
Note that the proposed characterization of the preferred inverse effective temperature may be stronger than necessary, since for the cat map the initial Markov partitions that we have examined all lead to the same result. 

It is reasonable to speculate that a relationship effectively of the sort indicated by the following commutative diagram might hold:
\[ 
\begin{tikzpicture}[descr/.style={fill=white}] 
\matrix(m)[matrix of math nodes, row sep=3em, column sep=2.8em, 
text height=2.5ex, text depth=0.25ex] 
{\mathcal{R}& &\mathcal{R}'\\&\mathcal{R} \lor \mathcal{R}'\\&\mathcal{G}(\mathcal{R} \lor \mathcal{R}')\\}; 
\path[dashed,->,font=\scriptsize](m-1-1) edge node[descr] {$\mathcal{G}$?} (m-3-2); 
\path[->,font=\scriptsize](m-1-1) edge node[descr] {$\lor$} (m-2-2); 
\path[->,font=\scriptsize](m-1-3) edge node[descr] {$\lor$} (m-2-2); 
\path[dashed,->,font=\scriptsize](m-1-3) edge node[descr] {$\mathcal{G}$?} (m-3-2);
\path[->,font=\scriptsize](m-2-2) edge node[descr] {$\mathcal{G}$} (m-3-2);
\end{tikzpicture} 
\] 
Here $\mathcal{R}$ and $\mathcal{R}'$ are arbitrary Markov partitions, the notation $\mathcal{G}$ indicates greedy refinement, and the dashed arrows are the speculative part. If some greedy refinement $\mathcal{G}(\mathcal{R} \lor \mathcal{R}')$ of $\mathcal{R} \lor \mathcal{R}'$ were also a greedy refinement of both $\mathcal{R}$ and $\mathcal{R}'$, then the process of taking greedy refinements would be asymptotically unique. This in turn would provide a strong structural framework in which limits of the sort we have considered could be dealt with in great generality. 

Establishing a relationship along such lines would likely involve the space of all Markov partitions, and in particular the realization of this space as a contractible simplicial complex \cite{Nakahara,Wagoner,BadoianW,Zizza}. Indeed, ordered triples of the form $(\mathcal{R},\mathcal{R} \lor_T R_j,\mathcal{R} \lor T^{-1}\mathcal{R})$ correspond to certain triangles in this simplicial complex, as do more general triples in which the middle entry is replaced by the result of a suitably constrained sequence of minimal refinements. In this setting the aim is to construct two sequences of edges, each corresponding to greedy refinements of an initial Markov partition, that eventually coincide.

Recent work aimed at characterizing associated spaces of Markov partitions for a generic two-dimensional hyperbolic toral automorphism \cite{AnosovKK,Klimenko,SiemaszkoW} could serve as a basis for developing the necessary ideas in a concrete setting. This is of particular importance from the physical point of view since a complete theory of greedy or even minimal refinements might require the incorporation of abstract constructions from algebraic topology and algebraic K-theory \cite{Magurn,Wagoner,Zizza}. 

A few points bear enumeration:
\begin{itemize}

\item While the detailed results obtained here for prototypical Anosov diffeomorphisms (viz., two-dimensional hyperbolic toral automorphisms and, as discussed in section \ref{sec:geodesic2}, the geodesic flow on a surface of constant negative curvature) are not by themselves obviously fundamental and we do not presently have anything resembling a mathematical proof of the sort ultimately required, our analysis of Markov partitions in the light of the chaotic hypothesis provides strong motivation for the applicability of our framework for effective statistical physics to generic nonequilibrium steady states, incorporating and extending the Ruelle program. This analysis also suggests additional nontrivial constraints on the form of the effective temperature.

\item The relevant objects and quantities involved can all be obtained from dynamics alone, yet without external reference to a Hamiltonian. The existence of a dynamical basis for framing the effective temperature in relation to a broad class of systems of interest for statistical physics suggests its physical relevance \cite{Cohen1,Cohen2}. The important aspect of its application is to start with Markov partitions of $M$, and obtain after greedy refinements the concomitant effective temperatures, then to identify the appropriate extremal limiting effective temperature and its concomitant effective energy function, then select any Markov partition such that the effective energies can be treated as constant on its rectangles and apply our extension of the Ruelle framework. 

\item The simultaneous use of greedy--and therefore more nearly regular--microscopic and mesoscopic Markov partitions can be done self-consistently in the presence of a reservoir or thermostat, or in the thermodynamical limit. It turns out in fact that such considerations lead to estimates on the mixing time of an Anosov system, which is a physically natural choice for $t_\infty$ (see sections \ref{sec:gedanken} and \ref{sec:ulam}). This means that in the event that the system has a known Hamiltonian, the approach in this paper will (at least up to an overall timescale, if not manifestly) facilitate the recovery of traditional equilibrium statistical physics. It is for this reason that we can plausibly claim that our proposal suggests an extension of the Ruelle program to a potentially complete theory of statistical physics for nonequilibrium steady states.

\end{itemize}

Now that the basic ideas have been sketched, we will briefly detour to discuss the Ulam method before continuing with continuous-time systems.

\section{\label{sec:ulam} The Ulam method}

While coarse-graining Anosov systems via Markov partitions is entirely natural, a simultaneous fine-graining is both anticipated and necessary from the point of view of statistical physics \cite{Gallavotti1,Gallavotti1b}. It is reasonable to expect that the SRB measure should be obtainable from a fairly generic fine-graining and that Markovity should be in some sense a technical convenience in this regard. The present section aims to illustrate how this is likely to be the case while also proposing a recipe for approximating the mixing time, which as we pointed out in section \ref{sec:gedanken} is a physically natural choice for $t_\infty$.

In order to explain how this can be, some background is necessary. The \emph{Perron-Frobenius} or \emph{transfer operator} $\mathcal{P}$ (not to be confused with a partition) is given by $\int_X \mathcal{P}f = \int_{T^{-1}X} f$ \cite{BeckS}. The SRB measure is given by an eigenvector of $\mathcal{P}$ with unit eigenvalue; the relaxation time (which can generally be expected to be related to the mixing time) is given by the second largest eigenvalue of $\mathcal{P}$. The Ulam conjecture \cite{Ulam} is that the key spectral properties of the Perron-Frobenius operator can be obtained by a finite-dimensional approximation. Although open in its general form, the Ulam conjecture has been established for important special cases and supporting numerical work performed in others (a recent survey is \cite{DLZ}).

The Ulam approximation (see also \S 17.5 of \cite{BeckS}) is a so-called Galerkin method, and starts from a (typically non-Markov) partition $\mathcal{S}_m = \{S_1^{(m)},\dots,S_{n_m}^{(m)}\}$ and the Riemannian measure $\nu$. The matrix $P_{jk}^{(m)}[\mathcal{S}] := \nu(S_j \cap T^{-1}S_k)/\nu(S_j)$ is clearly stochastic, and as such admits a corresponding invariant distribution $\Pi^{(m)} = \Pi^{(m)}P^{(m)}$. If one takes Markov partitions $\mathcal{R}_m$ with the diameters of rectangles tending to zero, the corresponding sequence of probability measures $\mathbb{P}_m(X) := \sum_{j=1}^{n_m} \Pi_j^{(m)} \nu(X \cap R_j^{(m)})$ converges weakly to the SRB measure for two-dimensional Anosov systems \cite{Froyland0}. More generally for Markov partitions for two-dimensional Anosov systems, the isolated spectra of $P^{(m)}$ approximate the isolated spectrum of the Perron-Frobenius operator for the projection of $T$ onto the unstable boundary of $\mathcal{R}$. This result can be used to obtain good bounds on the mixing time, which we have identified as a natural choice for $t_\infty$ \cite{Froyland1}.

Though a satisfactory explanation of the numerical effectiveness of the Ulam method does not exist at present \cite{BlankKL}, it succeeds in considerable generality for approximating SRB measures \cite{Froyland2,Froyland3,DJ}. It should be noted that the success of the Ulam method in approximating SRB measures does not entail success in approximating the corresponding Perron-Frobenius operator itself: indeed, for non-Markov partitions the Ulam approximation can fail to provide a reasonable approximation of the Perron-Frobenius operator of the cat map. However it has been argued that such failures are due to a lack of subunit isolated spectra of the Perron-Frobenius operator, and that such a situation should preclude the expectation of any reasonable spectral approximation in the first place \cite{Froyland4}. 

In short, the Ulam method appears (but has not yet been proven) to provide precisely the tool required in order to dynamically obtain the SRB measure in a physically relevant way without reliance on Markov partitions, as well as for providing (bounds on) a suitable $t_\infty$.

This is especially noteworthy when one considers that not only is the construction of ($\mu_{SRB}$-) regular Markov partitions (or even a sequence of Markov partitions such that the Riemannian measures of rectangles tend to equality) unlikely, but more remarkably that the boundaries of Markov partitions are generically (and even in the case of hyperbolic toral automorphisms in more than two dimensions are always) fractal \cite{KH}, so that Markov partitions are intrinsically difficult to construct precisely. This latter fact suggests that in practice approximations of the SRB measure will generally be necessary in any event.

\section{\label{sec:catflow} The continuous cat map and the cat flow}

Having already discussed Anosov diffeomorphisms, we shall proceed to introduce Anosov flows. As a preliminary, we first briefly touch upon a continuous version of the cat map that informs the simplest Anosov flow. 

Iterates of $T_A$ correspond to unit-frequency stroboscopic projections onto $\mathbb{T}^2$ of trajectories on $\mathbb{R}^2$ generated by the Hamiltonian $H_A(\xi,p) := K(p^2 - \xi^2 + \xi p)$, where $K = \sinh^{-1}(\sqrt{5}/2)/\sqrt{5}$. To see this, consider for the moment the more general Hamiltonian of the form $H(\xi,p) =  \frac{C}{2}p^2 -\frac{D}{2}\xi^2 + E \xi p$. Now Hamilton's equations are $\dot x = \mathcal{L}x$, where $x = \left(\begin{smallmatrix}
  \xi \\
  p
\end{smallmatrix} \right)$ and the Liouvillian is $\mathcal{L} = \left(\begin{smallmatrix}
  E & C \\
  D & -E
\end{smallmatrix} \right)$. Writing $L := \sqrt{E^2 + CD}$, we have that 
\begin{equation}
\exp \mathcal{L} = \begin{pmatrix}
  \cosh L + \frac{E}{L} \sinh L & \frac{C}{L} \sinh L \\
  \frac{D}{L} \sinh L & \cosh L - \frac{E}{L} \sinh L \\
\end{pmatrix}.
\end{equation}
Now a bit of elementary algebra leads to $\log \left(\begin{smallmatrix}
  2 & 1 \\
  1 & 1
\end{smallmatrix} \right) = K\left(\begin{smallmatrix}
  1 & 2 \\
  2 & -1
\end{smallmatrix} \right) \equiv \mathcal{L}_A$, so the corresponding Hamiltonian is $H_A$.

\begin{figure}[htbp]
\includegraphics[width=8cm,keepaspectratio]{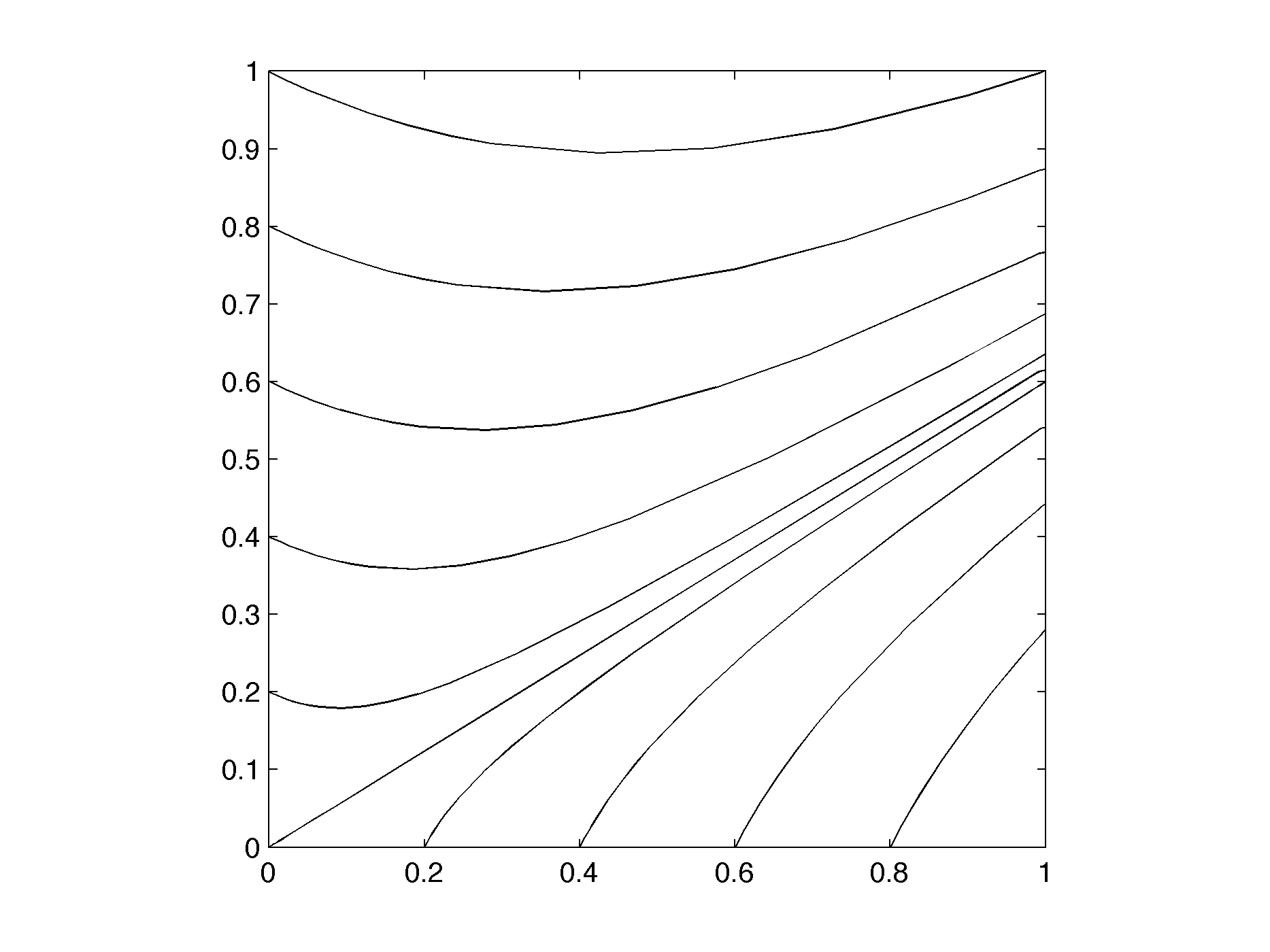}
\caption{ \label{fig:naivecatflow} The naive cat flow.}
\end{figure}

The restriction of the associated Hamiltonian flow to $[0,1]^2$ is shown in figure \ref{fig:naivecatflow}. Projecting this restricted flow onto the torus results in a singular flow, called the \emph{naive cat flow}. (Note that the projection of $(0,1/2)$ is a repeller.)
On the other hand, projecting the Hamiltonian flow on $\mathbb{R}^2$ onto the torus does not result in a flow at all, but it does produce a continuous-time evolution $x(t) = \pi(e^{t\mathcal{L}}x(0))$ called the \emph{continuous cat map} because of the equality $x(\ell) = T_A^\ell x(0)$.

The cat map can be used to construct perhaps the simplest Anosov flow; the continuous cat map can help visualize this flow. Take the product $\mathbb{T}^2 \times \mathbb{R}$ and introduce the equivalence relation $(T_Ax, z) \sim (x, z + 1)$. Write $\mathcal{M} := \mathbb{T}^2 \times \mathbb{R} / \sim$ and consider the flow on $\mathcal{M}$ generated by the vector field $e_z$. This ``suspension'' of the cat map is called the \emph{cat flow}. It is an Anosov flow for $\mathcal{M}$ \cite{CG,AK} with Anosov metric
\begin{equation}
ds^2 = \lambda_+^{2z} dx_+^2 + \lambda_-^{2z} dx_-^2 + dz^2.
\end{equation}

Given a Markov partition $\mathcal{R} = \{R_1,\dots, R_n\}$ for the cat map and $m \ge 3$, consider the sets $R_{jk} := R_j \times \{ \frac{k}{m} - j\varepsilon \}$, where $1 \le k \le m$ and $\varepsilon < \frac{1}{mn}$. The section $\mathcal{R}' := \{R_{jk}\}_{j,k}$ is readily seen to be a proper section for the cat flow \cite{Mathoverflow2}. The Poincar\'e map for $\mathcal{R}'$ sends $R_{jk}$ to $R_{j,k+1}$ for $1 \le k \le m-1$, and because $\mathcal{R}$ is a Markov partition for the cat map it follows that $\mathcal{R}'$ is a Markov section for the cat flow. Theorems apply for Markov sections of Anosov flows that are similar to those for Markov partitions of Anosov diffeomorphisms \cite{Bowen} and as such the present situation may be analyzed in a manner essentially similar to that of previous sections. 

Indeed, for the characteristic times $t(x) = (t_1(x),\dots, t_n(x)) \equiv t_\infty p(x)$ w/r/t $\mathcal{R}$ of a finite segment of the orbit of $x$ under the cat map $T_A$, the corresponding set of return times for the cat flow w/r/t the Markov section $\mathcal{R}'$ tends to $(t/m)^{\oplus m}$ as $\varepsilon \downarrow 0$. (Here the $m$-fold concatenation of $t/m$ is indicated: i.e., for $s \in \mathbb{R}^n$, $s^{\oplus m}_r := s_{[(r-1)\mod n]+1}$ for $1 \le r \le mn$.) The inverse effective temperature corresponding to $t$ is $\beta(t) = t_\infty \lVert p \rVert \cdot \sqrt{  \lVert \gamma \rVert^2 + 1 }$; similarly, $\beta((t/m)^{\oplus m}) = t_\infty \lVert p \rVert \cdot \sqrt{ m  \lVert \gamma \rVert^2 + 1 }/\sqrt{m}$. \footnote{Compare this with the behavior for the Shannon entropy: $S((t/m)^{\oplus m}) = S(t) + \log m$.} Therefore (unless $p$ is approximately uniform, in which case $\gamma \approx 0$), $\beta((t/m)^{\oplus m}) \approx \beta(t)$. 

This shows that the effective temperature of the cat flow w/r/t the Markov section $\mathcal{R}'$ is qualitatively similar to the effective temperature for the cat map w/r/t the Markov partition $\mathcal{R}$. The adaptation of results concerning hyperbolic toral automorphisms to the corresponding flows is straightforward.

Unlike the cat map, however, the cat flow is not mixing due to the so-called Anosov alternative, which states that an Anosov flow is either a suspension of an Anosov diffeomorphism with constant ceiling or every strong stable and unstable manifold is everywhere dense, in which case it is mixing \cite{Chernov}. The cat flow is the prototypical example of such a suspension, and for this reason its utility as a model system for statistical physics is limited. 
Nevertheless, the cat flow is considered in magnetohydrodynamics as a \emph{kinematic fast dynamo}, i.e., an MHD system with prescribed velocity field supporting the exponential amplification of the magnetic field for small magnetic diffusivity \cite{AK}. The geodesic flow on a surface of negative curvature is also a kinematic fast dynamo, but it also represents the other, more physically relevant possibility under the Anosov alternative: we deal with it next.

\section{\label{sec:geodesic} The geodesic flow on a surface of constant negative curvature}

In this section we introduce the other reasonably accessible example of an Anosov flow, and one that is mixing (and moreover, motivated the theory of Anosov flows): namely, the geodesic flow on a compact Riemannian manifold of dimension $\ge 2$ with negative sectional curvature \cite{Anosov} (see also \cite{Klingenberg}). We then construct a map topologically conjugate to its Poincar\'e map and a corresponding Markov partition. 

The physical interpretation of the geodesic flow is given by the free-particle Hamiltonian \cite{CEG}
\begin{equation}
\label{eq:geodesichamiltonian}
H = \frac{1}{2m} \sum_{j,k} g^{jk}(x) p_j p_k,
\end{equation}
where as usual $(g^{jk})$ is the inverse of the matrix $(g_{jk})$ describing the Riemannian metric $ds^2 = \sum_{j,k} g_{jk} dx^j dx^k$ in local coordinates. Although the construction of symbolic dynamics for such geodesic flows dates to Hadamard \cite{Hadamard} and Morse \cite{Morse}, the fullest expression of the situation in the case of constant negative curvature is given in the complementary works of Bowen and Series \cite{BowenS} and Adler and Flatto \cite{AF}; we borrow liberally from the latter in the discussion below, which we have sought to make relatively self-contained at the expense of length. For a qualitative overview of the constructions involved, see \cite{Gutzwiller}.

Consider the Poincar\'e disk model of two-dimensional hyperbolic space, i.e., the complex unit disk $D$ endowed with the metric $ds^2 = dz d\bar{z}/(1-|z|^2)^2$. In this model geometry, the geodesics are circular arcs intersecting the boundary $S^1$ of $D$ at right angles. Useful coordinates on the unit tangent bundle are $u = (\xi, \eta, s)$, where $\xi$ and $\eta$ are respectively the terminal and initial ends of the geodesic determined by $u$, and where $s$ is the hyperbolic distance from the midpoint of that geodesic to the base point of $u$ in $D$. In this coordinate system the geodesic flow takes the form $(\xi, \eta, s) \rightarrow (\xi, \eta, s + t)$.

We have for $g \ge 2$ and $N = 8g-4$ the existence and (among others) the following properties of a bounded regular hyperbolic $N$-gon $F$ and of a corresponding discrete group $\Gamma$ of M\"obius transformations (i.e., maps of the form $z \mapsto \left(\begin{smallmatrix}
  a & b \\
  c & d
\end{smallmatrix} \right) \cdot z := \frac{az+b}{cz+d}$, for $ad-bc \ne 0$) that preserves $D$:
\begin{itemize}
\item the translates $\tau F$ for $\tau \in \Gamma$ tile $D$ and have pairwise disjoint interiors, so that $D/\Gamma$ is well defined;
\item there is a unique element of $\Gamma$ mapping a specific pair of edges of $F$ to each other, and gluing the paired edges of $F$ produces a compact surface of genus $g$ that can be identifed with $D/\Gamma$;
\item the geodesics extending the boundary $\partial F$ of $F$ are contained within the translates $\tau \partial F$ for $\tau \in \Gamma$ (this is called the ``extension condition'').
\end{itemize}
The Hamiltonian for the geodesic flow on $D/\Gamma$ takes the form (up to a constant)
\begin{equation}
\label{eq:geodesichamiltonianspecific}
H = (1-|z|^2)^2 \cdot p_z p_{\bar{z}} = (1-r^2)^2 \cdot p^2
\end{equation}
where Euclidean coordinates and momenta on $F$ are implicit in the RHS.

\begin{figure}[htbp]
\includegraphics[trim = 20mm 25mm 20mm 25mm, clip, width=8cm,keepaspectratio]{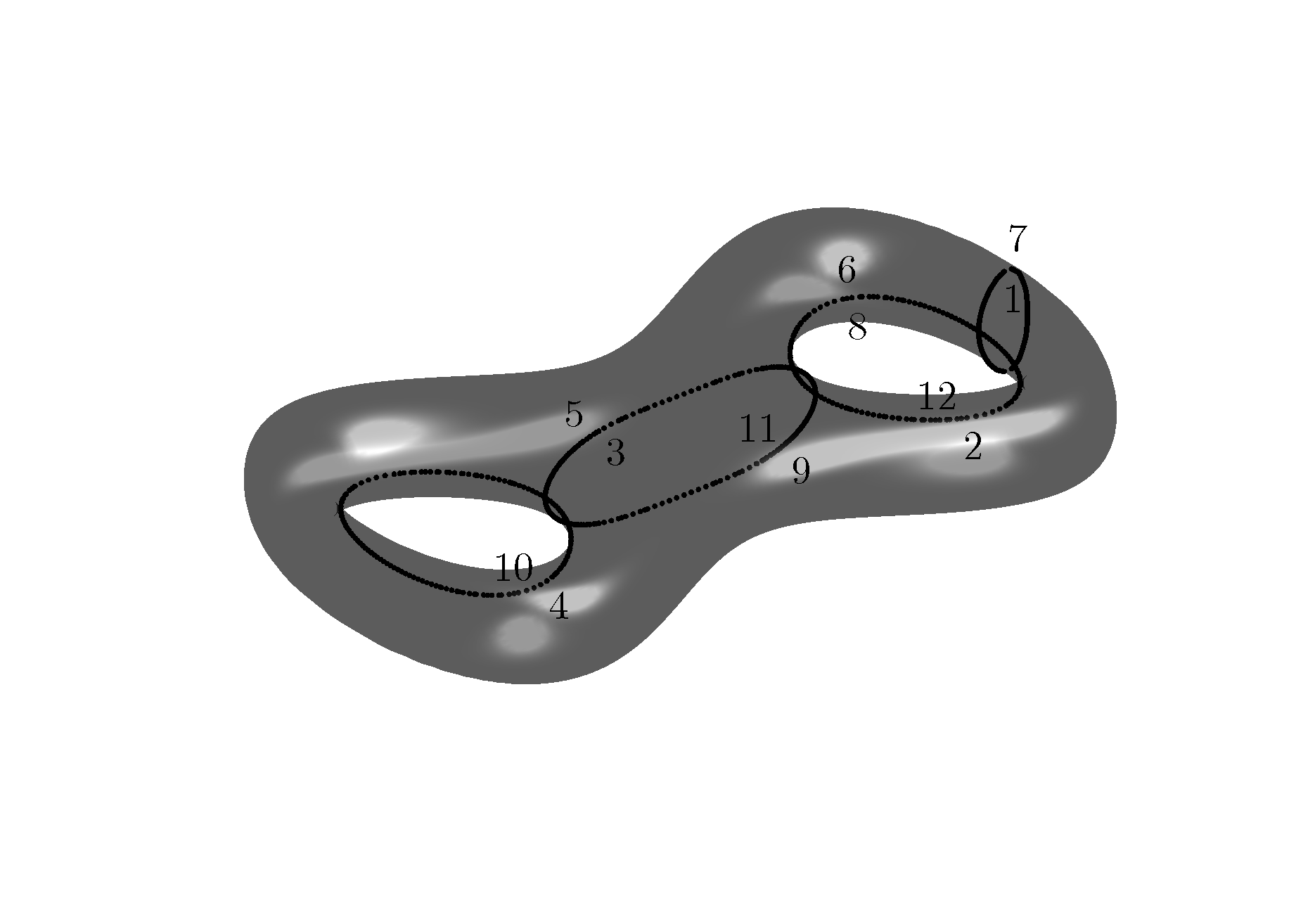}
\caption{ \label{fig:doubletorus} A surface with genus $g = 2$. For $g \ge 2$ there are surfaces of constant (or more generally global) negative curvature and whose geodesic flows are mixing (hence also ergodic). The construction of Adler and Flatto provides examples. For $g = 2$, the corresponding 12-gon can be recovered by cutting along the indicated paths. Incrementing $g$ by adding a handle also requires adding two more geodesic loops; one of the new loops and one of the old loops are separated into two arcs by their intersections with neighboring loops, and cutting along these arcs yields eight new edges. This construction underlies the pairing of edges of $F$, which is indicated explicitly here for $g=2$.}
\end{figure}

An explicit description of the $N$-gon $F$ is 
\begin{equation}
F = D \ \backslash \ \bigcup_{j=1}^{N} \left(\sqrt{a-1} \cdot D + \sqrt{a} e^{2\pi i(j-2g)/N}\right)
\end{equation}
with $a = \sec \frac{2\pi}{N}$. The edges inherit the implied index (in particular, we shall consider any indices deriving from edges of $F$ to be taken modulo $N$) and the pairing of edges $s_j$ turns out to be given in terms of
\begin{equation}
\sigma(j) := \begin{cases}
 (4g-j) \ \mbox{mod} \ N, & j \mbox{ odd} \\
 (2-j) \ \mbox{mod} \ N, & j \mbox{ even} 
\end{cases}
\end{equation}
by the existence of unique elements $T_j \in \Gamma$ satisfying $T_j(s_j) = s_{\sigma(j)}^{-1}$, where $s_j^{-1}$ is the orientation reversal of $s_j$ (see figure \ref{fig:doubletorus}).

The construction of a ``rectilinear'' map $T_R$ that is topologically conjugate to the ``curvilinear'' Poincar\'e map for the geodesic flow on $D/\Gamma$ is tractable and will be detailed below. In particular, an explicit Markov partition for $T_R$ will be constructed, and for this reason we shall restrict our attentions to this map. Given an edge $s_j$ of $\partial F$, extend it to a geodesic. Write $a_j^-$ and $a_j^+$ respectively for the initial and terminal ends of this geodesic. (NB. \cite{AF} uses the notation $a_j$ and $b_j$ instead.) A line of algebra shows that the $a_j^\pm$ are given in terms of
\begin{equation}
\theta_j^\pm := \frac{2\pi}{N}(j-2g) \pm \cos^{-1} \left( -\sqrt{\frac{a-1}{a}} \right)
\end{equation}
(where the principal branch of the inverse cosine is assumed) by 
\begin{equation}
a_j^\pm = \sqrt{a-1} \cdot e^{i\theta_j^\pm} + \sqrt{a} e^{2\pi i(j-2g)/N}.
\end{equation}

The sections for both the curvilinear and rectilinear Poincar\'e maps are specified in terms of the $a_j^\pm$, which occur in the (cyclic counterclockwise) order $a_1^-,a_0^+,a_2^-,a_1^+,\dots,a_N^-,a_{N-1}^+$. Because $T_j$ is a M\"obius transformation and as such maps circles to circles, it is easy to show that 
\begin{equation}
\label{eq:Tja}
T_j: \begin{cases}
\begin{array}{rcl}
a_j^\pm & \mapsto & a_{\sigma(j)}^\mp \\
a_{j-1}^\pm & \mapsto & a_{\sigma(j)+1}^\pm \\
a_{j+1}^\pm & \mapsto & a_{\sigma(j)-1}^\pm
\end{array}
\end{cases}
\end{equation}
Let $\ell_j := (a_j^-, a_j^+)$ and $u_j := (a_j^+, a_j^-)$. It follows that $T_j \times T_j$ sends (for example) $\ell_{j-1}$, $\ell_j$, $\ell_{j+1}$, $u_{j-1}$, $u_j$, $u_{j+1}$ respectively to $\ell_{\sigma(j)+1}$, $u_{\sigma(j)}$, $\ell_{\sigma(j)-1}$, $u_{\sigma(j)+1}$, $\ell_{\sigma(j)}$, $u_{\sigma(j)-1}$.

\begin{figure}[htbp]
\includegraphics[width=8cm,keepaspectratio]{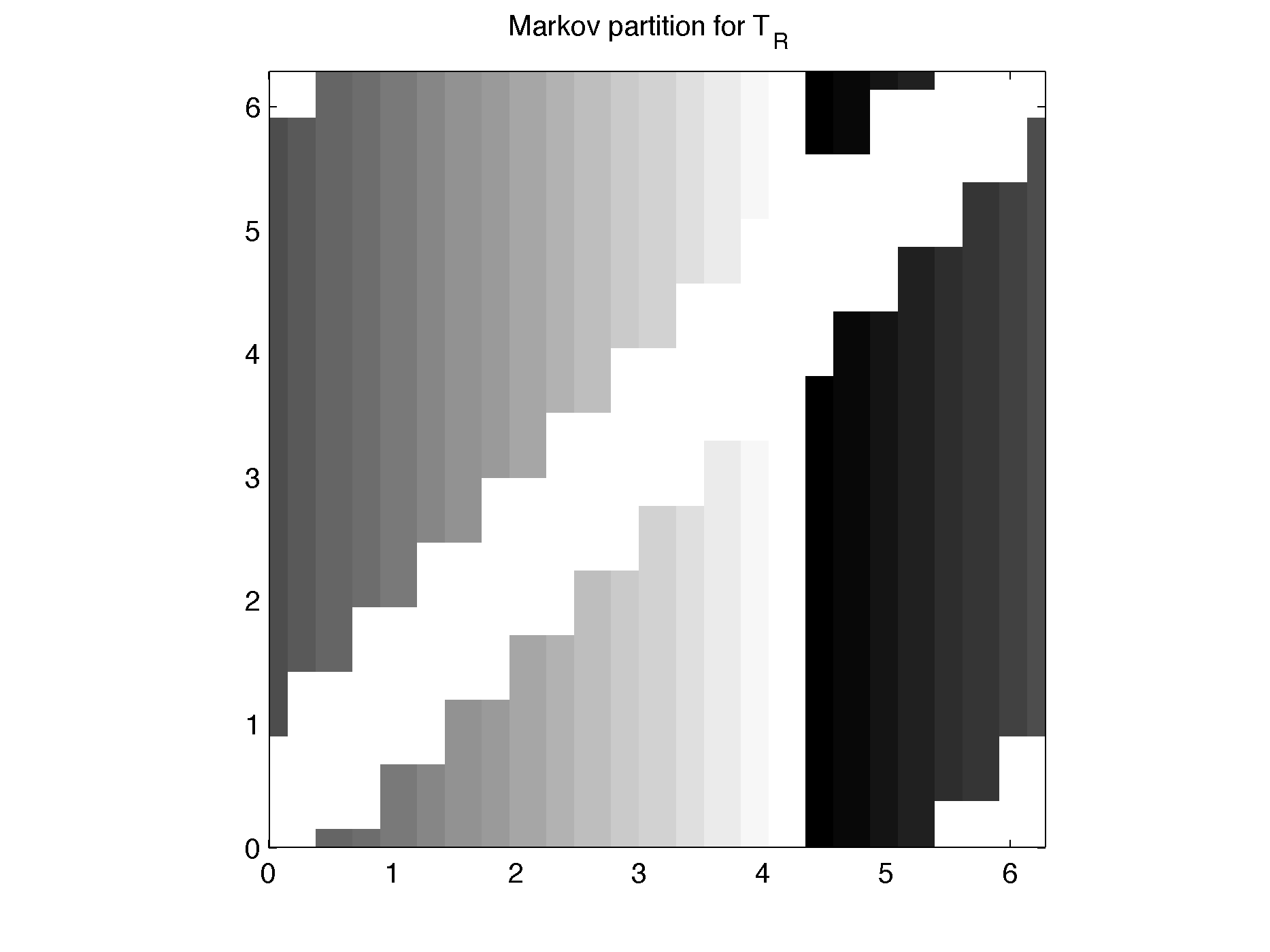}
\includegraphics[width=8cm,keepaspectratio]{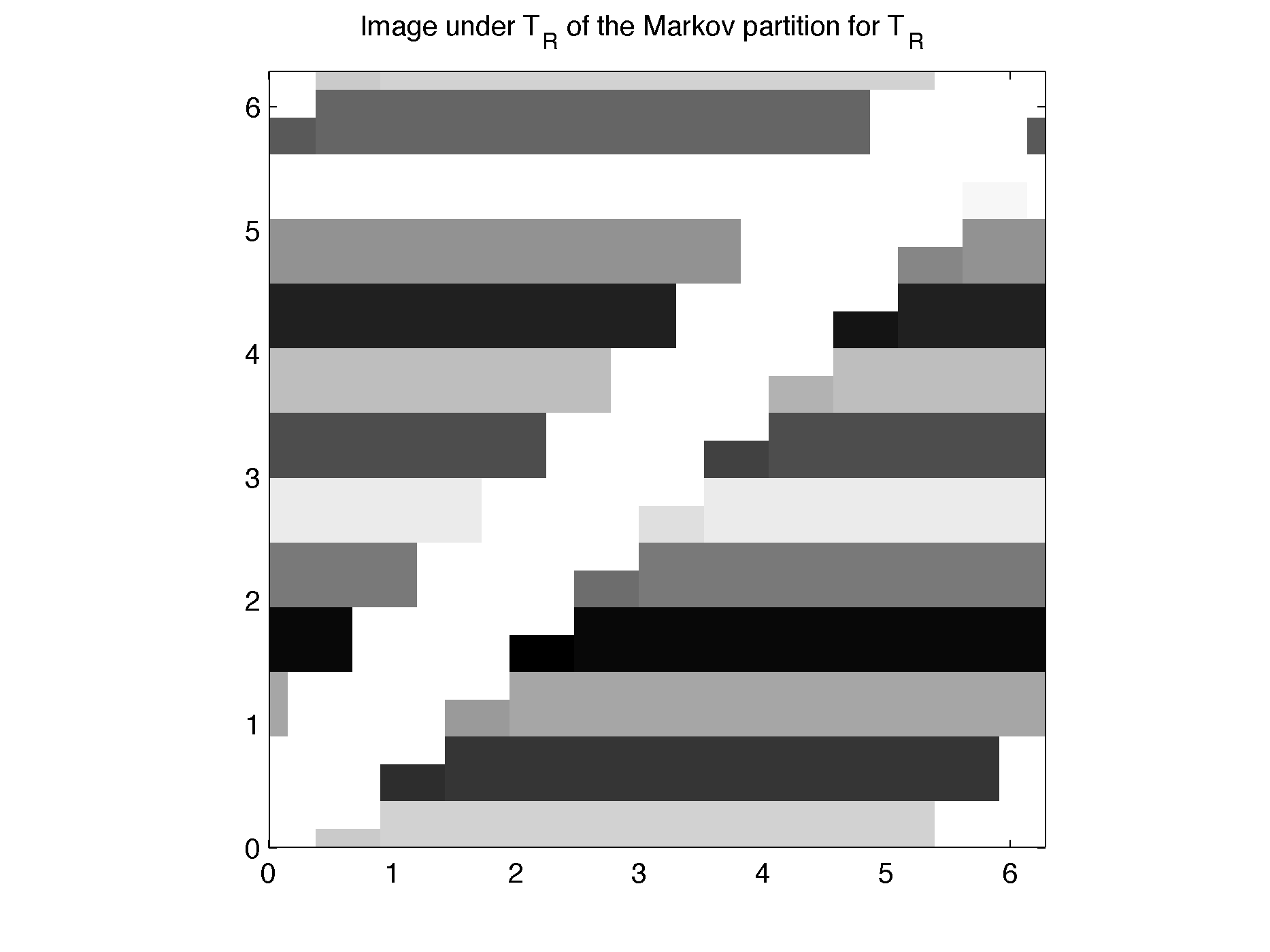}
\caption{ \label{fig:rectilinear1} (L) $\mathcal{R}$ for $g = 2$. (R) $T_R\mathcal{R}$. Rectangles are indicated by consistent shading.} 
\end{figure}

\begin{figure}[htbp]
\includegraphics[width=8cm,keepaspectratio]{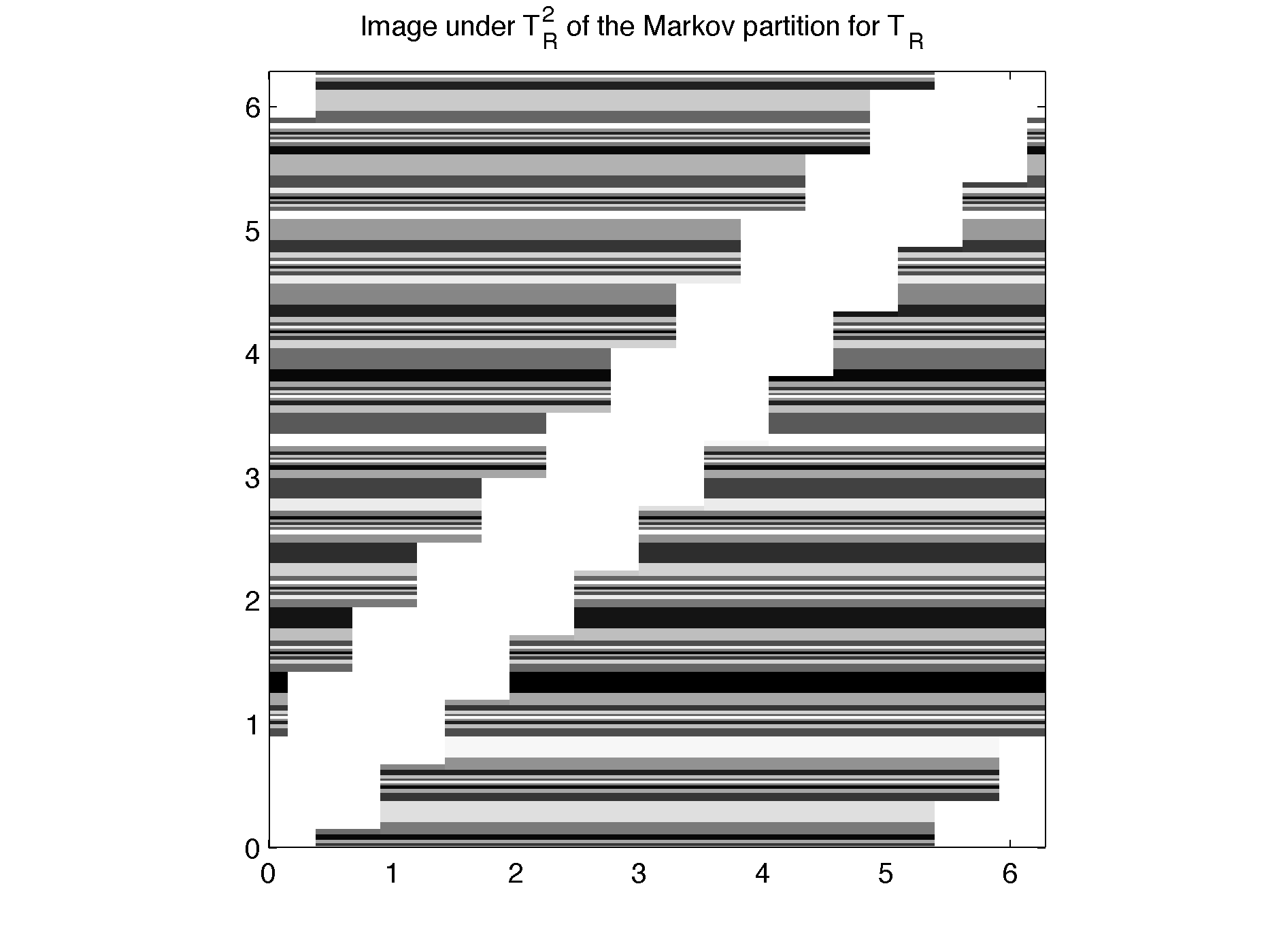}
\includegraphics[width=8cm,keepaspectratio]{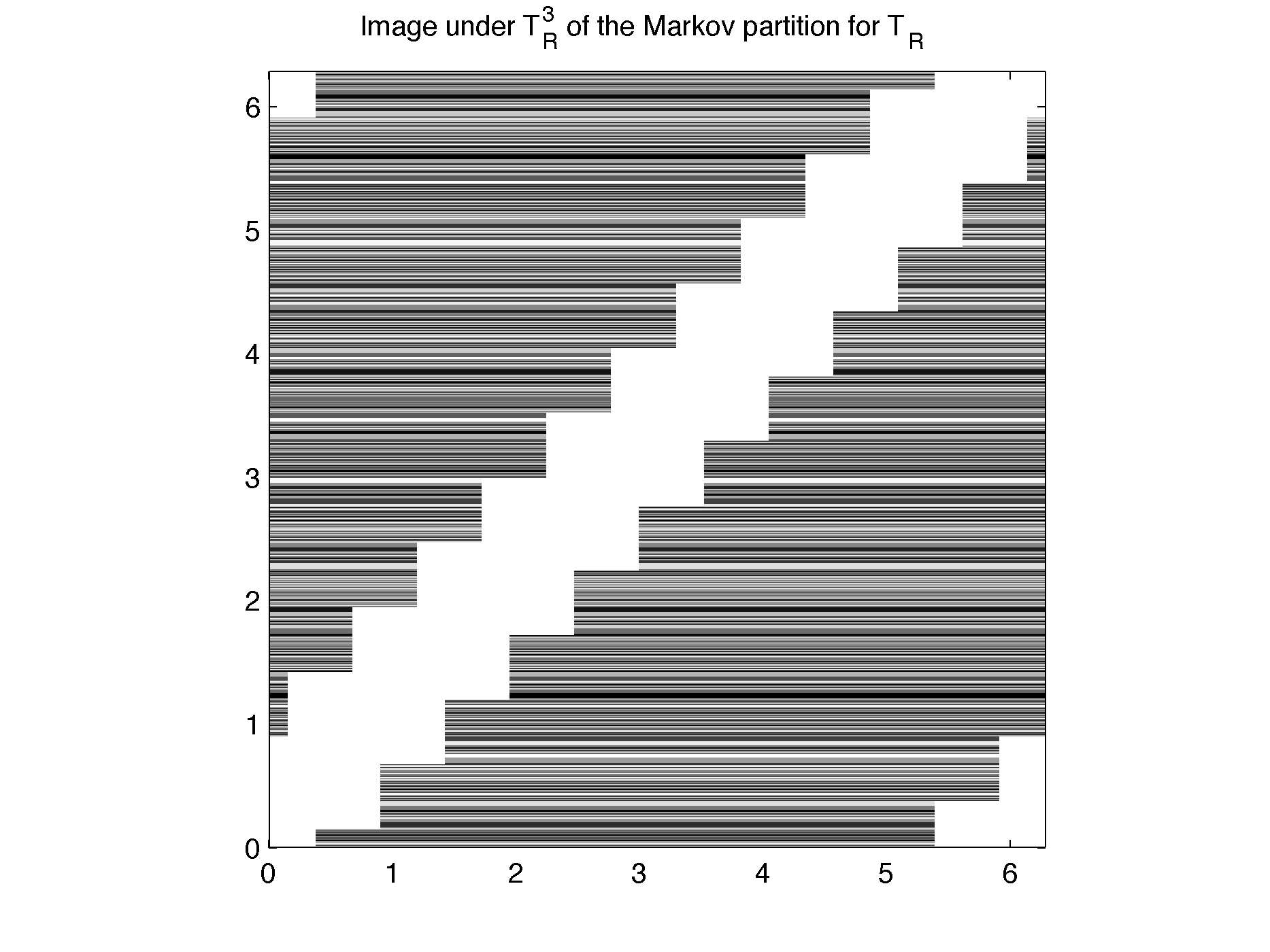}
\caption{ \label{fig:rectilinear2} (L) $T_R^2\mathcal{R}$. (R) $T_R^3\mathcal{R}$. Rectangles are indicated by consistent shading. Note that here the rectangles are not connected, though taking connected components yields refined Markov partitions.} 
\end{figure}

Write $\xi = e^{ix}$ and $\eta = e^{iy}$. Given two distinct points $(\xi,\eta)$, $(\xi',\eta')$ in $S^1 \times S^1$, let $\langle (\xi,\eta), (\xi',\eta') \rangle$ denote the path from $(\xi,\eta)$ to $(\xi',\eta')$ obtained by always keeping one argument fixed, first increasing the second argument (``up'') and subsequently the second argument (``right''), i.e.
\begin{equation}
\langle (\xi,\eta), (\xi',\eta') \rangle(t) := \begin{cases}
 (\xi,\eta e^{it}), & 0 \le t \le (y' - y) \ \mbox{mod} \ 2\pi \\
 (\xi e^{it},\eta'), & (y' - y) \ \mbox{mod} \ 2\pi \le t \le (y' - y) \ \mbox{mod} \ 2\pi +_\mathbb{R} (x' - x) \ \mbox{mod} \ 2\pi
\end{cases}
\end{equation}
where the $+_\mathbb{R}$ serves as an indication of the usual addition over $\mathbb{R}$. The domain $R$ of the rectilinear map $T_R$ is defined using the closed concatenation $\ell_R$ of paths $\langle \ell_j, \ell_{j+1}\rangle$ and the closed concatenation $u_R$ of paths $\langle u_j, u_{j+1}\rangle$. Specifically, let $R'$ be the region in $S^1 \times S^1$ bounded from below by $\ell_R$ and from above by $u_R$. Define $R := R' \ \backslash \left(\ell_R \cup \{u_1\} \cup \cdots \cup \{u_N\}\right)$ 
and let $R_j^*$ be the subset of pairs $(\xi,\eta)$ in $R$ such that $\xi$ is in the counterclockwise arc from $a_j^-$ to $a_{j+1}^-$, with the latter point excluded. Refine the partition $\{R_1^*,\dots,R_N^*\}$ of $R$ by dividing each $R_j^*$ into two rectangles, with the division along a segment of constant first coordinate (note that there is only one way to do this) and denote the result by $\mathcal{R}$. One of the main results of \cite{AF} is that $\mathcal{R}$ is a Markov partition for $T_R$ (see figures \ref{fig:rectilinear1} and \ref{fig:rectilinear2}): we now turn to the construction of this map, eliding its derivation.

It turns out that $T_R$ is given by $T_R(\xi,\eta) := (T_j(\xi),T_j(\eta))$ for $(\xi,\eta) \in R_j^*$. We can construct $T_j$ and hence $T_R$ as follows. Because a M\"obius transformation is determined by its action on (almost) any three points, using \eqref{eq:Tja} allows the explicit construction of $T_j$. Specifically, recall that the M\"obius transformation sending $z_1$, $z_2$, and $z_3$ respectively to 0, 1, and $\infty$ is $z \mapsto \mathfrak{H}(z_1,z_2,z_3) \cdot z$, where
\begin{equation}
\mathfrak{H}(z_1,z_2,z_3) := \begin{pmatrix}
  (z_2-z_3) & -z_1(z_2-z_3) \\
  (z_2-z_1) & -z_3(z_2-z_1) \\
\end{pmatrix}
\end{equation}
and as usual the action of $\left(\begin{smallmatrix}
  a & b \\
  c & d
\end{smallmatrix} \right) \in GL(2,\mathbb{C})$ on $z$ is given by $\left(\begin{smallmatrix}
  a & b \\
  c & d
\end{smallmatrix} \right) \cdot z = \frac{az+b}{cz+d}$. It follows that the M\"obius transformation sending $z_1$, $z_2$, and $z_3$ respectively to $z'_1$, $z'_2$, and $z'_3$ is 
\begin{equation}
z \mapsto [\mathfrak{H}^{-1}(z'_1,z'_2,z'_3) \mathfrak{H}(z_1,z_2,z_3)] \cdot z.
\end{equation}
In particular, substituting appropriate points immediately yields $T_j$ and completes the construction of $T_R$. 

\begin{figure}[htbp]
\includegraphics[width=8cm,keepaspectratio]{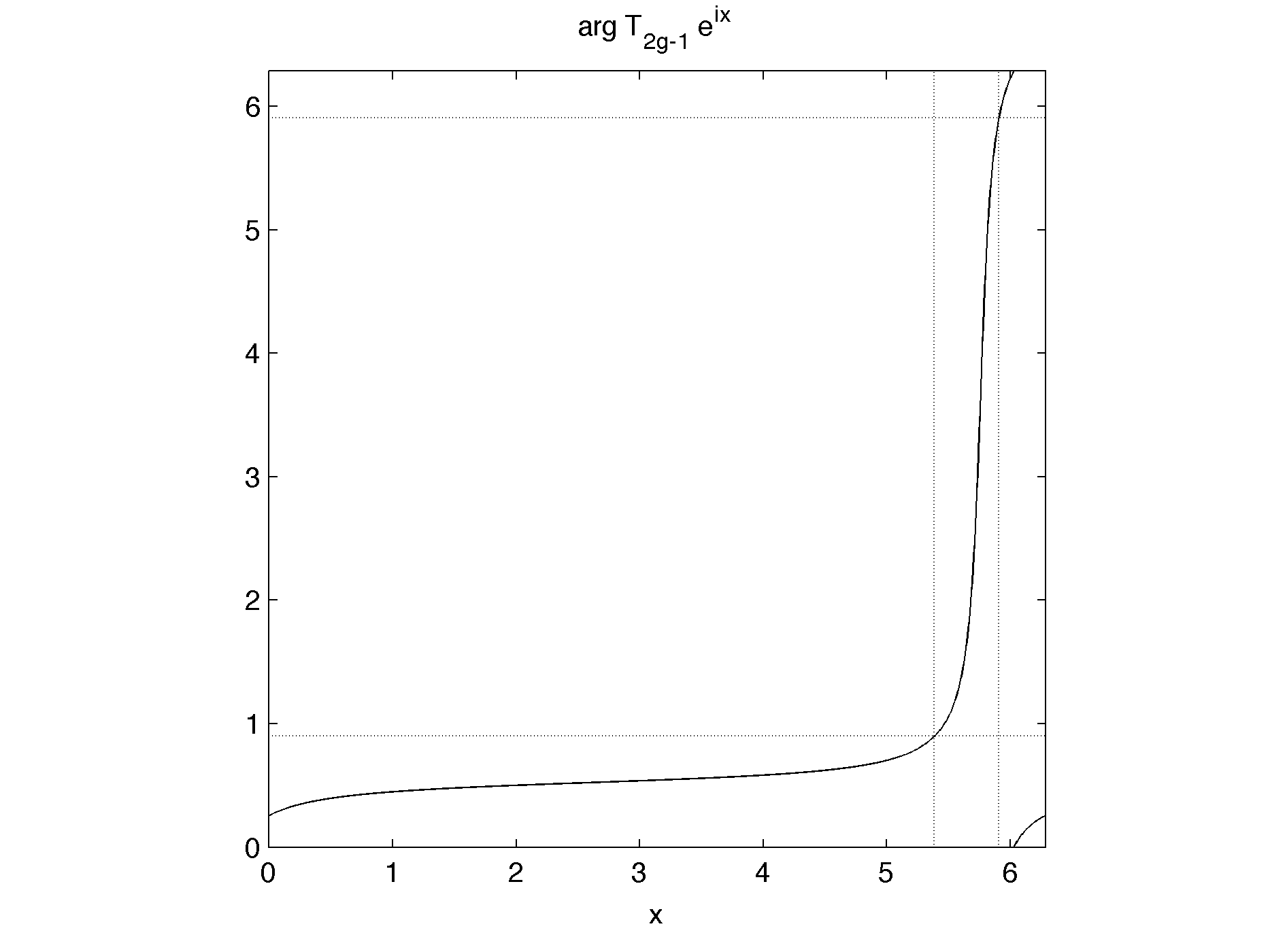}
\caption{ \label{fig:mobius0} The action of $T_{2g-1}$ on $S^1$ with $g=2$ is highly nonlinear. The restriction of its domain and range as a building block of $T_R$ on the first coordinate is indicated with dotted lines. The nonlinearity is also evidenced in the $T_R^m\mathcal{R}$ shown in figure \ref{fig:rectilinear2}.} 
\end{figure}

For example, if $g=2$ it can be shown (with some difficulty, or verified numerically with ease) that \begin{equation}
\label{eq:T2g1}
T_{2g-1}z = h \cdot z,
\end{equation}
where we exploit the fact that a constant multiple of a matrix in $GL(2,\mathbb{C})$ gives rise to the same M\"obius transformation to write
\begin{equation}
\label{eq:T2g1h}
h = \begin{pmatrix}
  e^{2\pi i/12} & -c \\
  c & -e^{-2\pi i/12} \\
\end{pmatrix}
\end{equation}
and here $c := (3/4)^{1/4}$. We thus obtain in this case the explicit formula
\begin{equation}
\label{eq:T2g1eix}
T_{2g-1}e^{ix} = -e^{ix} \cdot \frac{ce^{-ix} - e^{2\pi i/12}}{ce^{ix} - e^{-2\pi i/12}}.
\end{equation}
This is graphed in figure \ref{fig:mobius0}.

\section{\label{sec:geodesic2} SRB measures of Markov partitions for the geodesic flow}
 
In this section we outline a numerical calculation of the inverse effective temperature along similar lines as for the cat map. Up to normalization, the SRB measure of a rectangle (whose points satisfy $x > y$) is given by
\begin{equation}
\int_{y_1}^{y_2} \int_{x_1}^{x_2} \frac{dx \ dy}{\lvert e^{ix} - e^{iy}\rvert^2} = \log \sin \left ( \frac{x_2-y_2}{2} \right) - \log \sin \left ( \frac{x_2-y_1}{2} \right) - \log \sin \left ( \frac{x_1-y_2}{2} \right) + \log \sin \left ( \frac{x_1-y_1}{2} \right).
\end{equation}
Denoting this by $\mu([x_1,x_2] \times [y_1,y_2])$, we can see that $\mu([x_1,x_2] \times [y_1,y_2]) =\mu([x_1 + \Delta,x_2+ \Delta] \times [y_1+ \Delta,y_2+ \Delta])$. This partial translation invariance allows us to restrict our attention to subrectangles contained in two of the $2N$ rectangles in $\mathcal{R}$. The most convenient choice is the $(2g-1)$th pair, i.e., the two rectangles contained in $R_{2g-1}^*$ sharing an edge with $\xi = a_{2g-2}^+$ and that are always connected in the image of the usual map from the torus to $[0,2\pi]^2$. Write $Y :=(a_{2g}^+,a_{2g+1}^+,\dots,a_N^+,a_1^+,\dots,a_{2g-4}^+,a_{2g-2}^-,a_{2g-3}^+,a_{2g-1}^-)$. Now the $\eta$-coordinates of the subrectangular boundaries are essentially determined by suitable translations of the coordinates obtained by iterates of $T_{2g-1}$ on $Y$.

\begin{figure}[htbp]
\includegraphics[width=8cm,keepaspectratio]{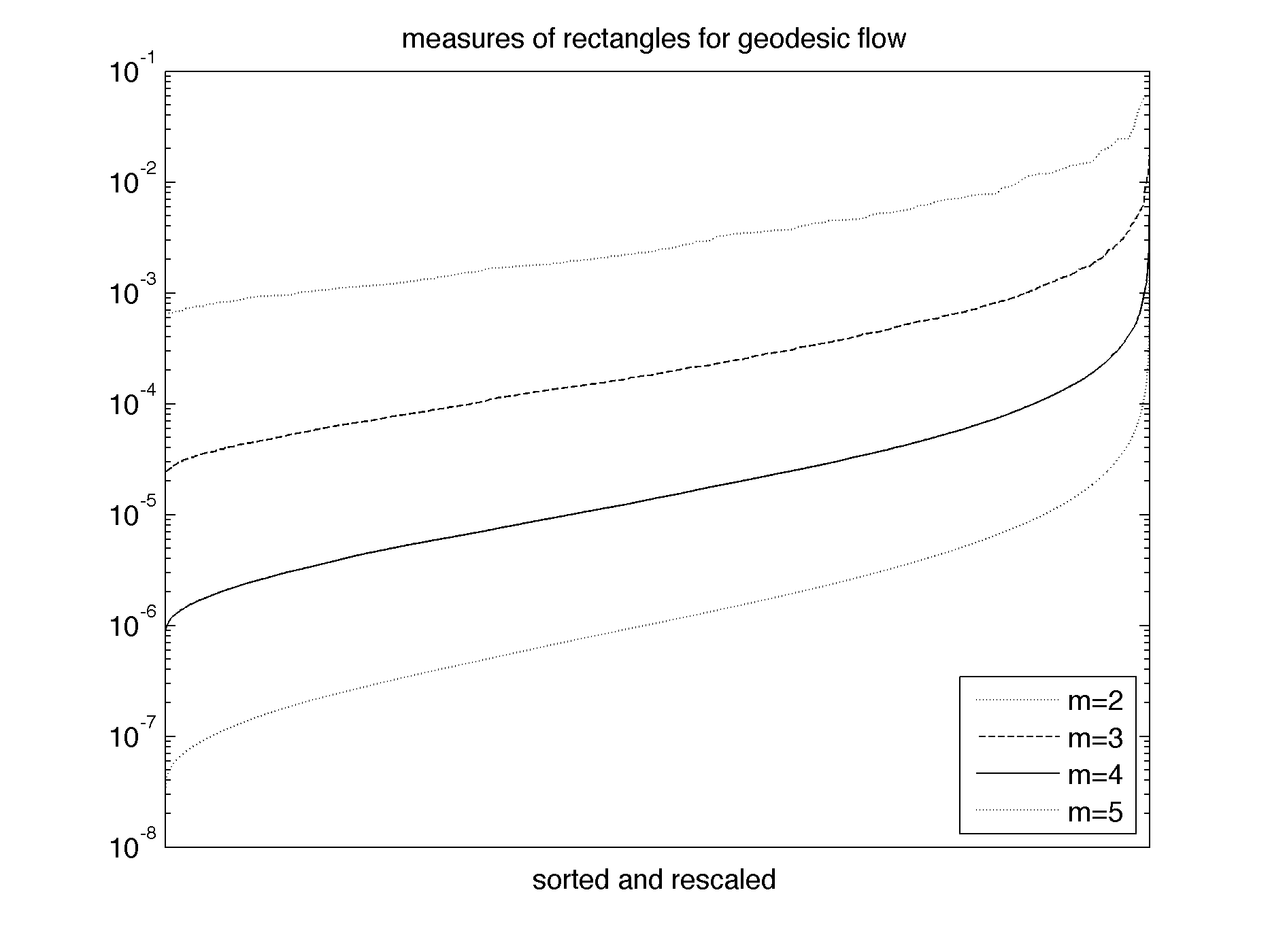}
\caption{ \label{fig:geodesicmeasures} Sorted measures of rectangles in successively refined partitions of the form $\mathcal{R}^\vee_m$.} 
\end{figure}

This fact facilitates a numerical calculation of the measures of partition refinements along the same lines as the previous ones for hyperbolic toral automorphisms. In the simplest case we consider refinements of the form $\mathcal{R}^\vee_m$. The results of this calculation indicate that $\beta/t_\infty$ diverges nearly exponentially as a function of the partition refinement. Although the vast growth in the number of rectangles precludes the use of refinements with more than a few iterates of $T_R$, for $T_A$ it suffices to consider only a few iterates in order to observe adequate numerical evidence of convergence, bolstering our view. The divergence can be ascribed to the nonlinearity of the $T_j$ (see figure \ref{fig:mobius0}), which produces rectangles in partition refinements with widely varying measures (see figure \ref{fig:geodesicmeasures} and recall section \ref{sec:nontriviality}). 
Proportionately fewer rectangles dominate the probabilities as $m$ increases, and the rectangles with comparatively large measures cause $\lVert \gamma \rVert$ to grow faster than $\lVert p \rVert$ decays. 

Because $T_R$ and $T_C$ have the same invariant measure and the topological conjugacy between them turns out to be built from isometries, the phenomenon described here applies to $T_C$ as well.

However, considering greedy refinements substantially changes this picture. A greedy refinement of $\mathcal{R}$ entails $N-1$ subsequent greedy refinements for the unaffected rectangle pairs $R^*_j$, and this pattern persists for later stages of greedy refinement as well. But each individual greedy refinement amounts to a boundary cutting across the images of multiple pairs $R^*_j$, so the resulting control over measures of rectangles is more significant than might otherwise be anticipated. 

In fact this point of view suggests that the cat map may actually be a sort of ``worst case'' for the controlling behavior of greedy refinements (at least for hyperbolic toral automorphisms in two dimensions), since the images of rectangles so obtained will only map across two rectangles in the unstable direction. It is tempting to speculate that may be related to the well-known fact that $\phi$ has the slowest converging continued fraction expansion of any irrational number.

\begin{figure}[htbp]
\includegraphics[width=8cm,keepaspectratio]{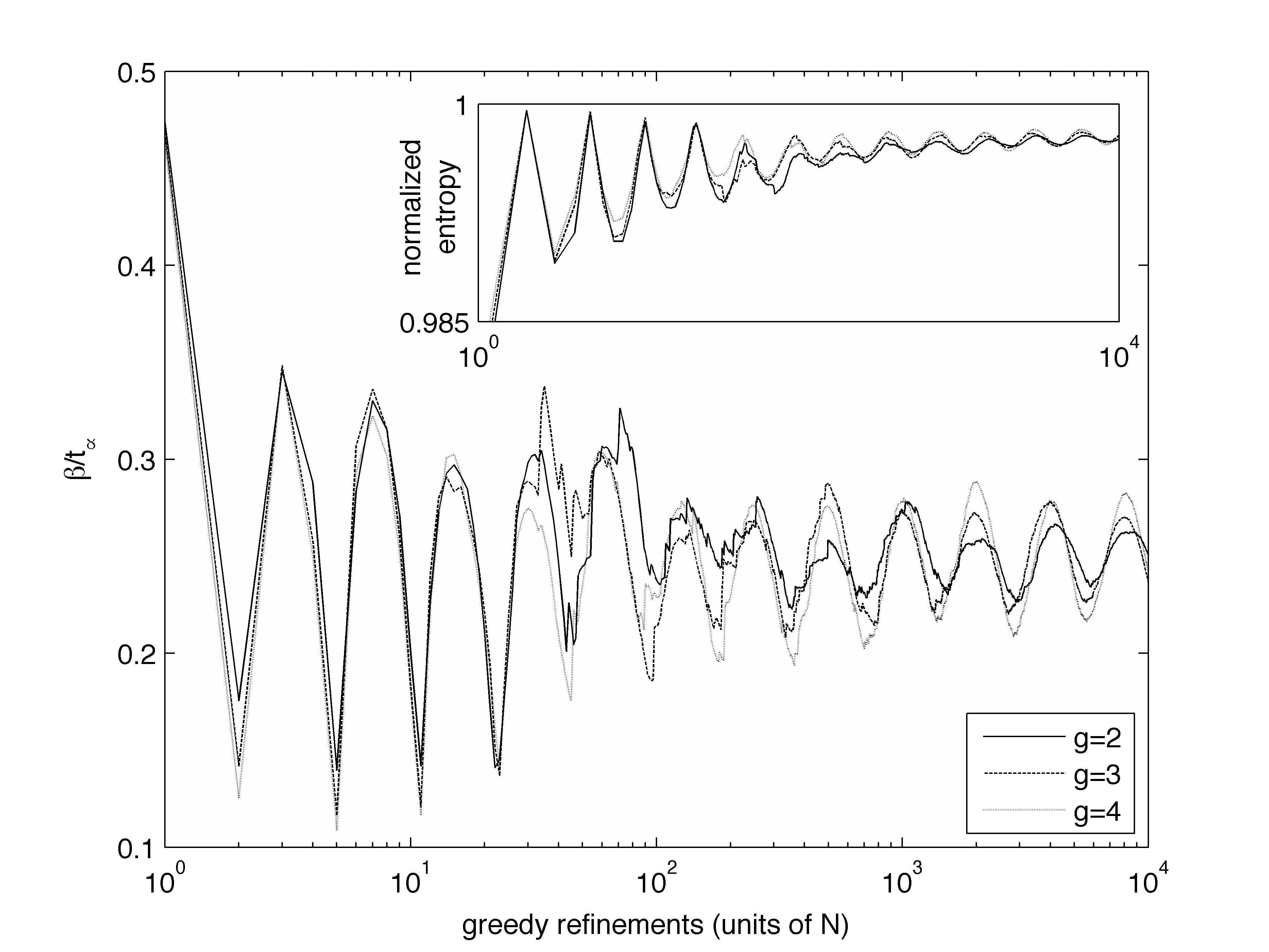}
\caption{ \label{fig:greedyg234} $\beta/t_\infty$ for $\mathcal{R}_R$ with $g = 2,3,4$ as a function of greedy refinements (in units of $N$). Note the logarithmic horizontal scale. This graph strongly suggests that $\liminf \beta/t_\infty$ is both nonzero and finite, and also independent of $g$. Inset: Normalized entropy. Note that convergence to 1 is visible but slow.} 
\end{figure}

Numerics clearly suggest that the limit inferior of $\beta/t_\infty$ over greedy refinements is positive and finite: see figure \ref{fig:greedyg234}. 
The computations are greatly simplified by the trivial observation that $\mu_{SRB}(R_j \cap T^{-1}R_k) = \mu_{SRB}(TR_j \cap R_k)$. Combining this with the other convenient facts mentioned in this section and using $T_{2g-1}^{-1}z = h^{-1} \cdot z$ facilitates the use of a simple data structure for the construction of greedy refinements and computation of the measures of their constituent rectangles \emph{in silico}.

\section{\label{sec:rugh} Comparison with the dynamical temperature}

The continuous cat map and cat flow do not allow for the self-consistent application of the dynamical temperature introduced by Rugh \cite{Rugh,Rugh2}. For a Hamiltonian system ergodic w/r/t the Liouville measure restricted to a nonsingular energy surface, the inverse dynamical temperature $\beta_{dyn}$ is given a.s. by the time average of (e.g.) $\Phi \equiv \nabla \cdot (\nabla H/\lVert \nabla H \rVert^2)$. For the continuous cat map from section \ref{sec:catflow} it is easy to see that $\Phi = -4H/5Kx^4$, but the requirements for the dynamical temperature construction to apply are not met. This breakdown is due to the fact that the constant-energy curves of $H$ are poorly behaved. In particular, the energy is only piecewise constant along trajectories, which are themselves nondifferentiable at points in $\pi(\partial([0,1]^2))$ and repelling from points in $\pi((0,1/2))$ (more generally, the cat map is not integrable). The additional fact that $\Phi(x) > 0 \iff x_2 >(\phi-1)x_1$, where $x$ is taken to be in $[0,1]^2 \subset \mathbb{R}^2$ only serves as a further complication. It is also easy to see that $\int_\varepsilon^1 \int_\varepsilon^1 \Phi$ (with $\Phi \equiv \nabla \cdot (\nabla H / \lVert \nabla H \rVert^2)$) diverges to $-\infty$ logarithmically w/r/t $\varepsilon$. Because the SRB measure for the cat map (or any hyperbolic toral automorphism) is just the pushforward of Lebesgue measure, it is clear that the Rugh formula $\beta_{dyn}(E(x)) = \lim_{t \uparrow \infty} t^{-1} \int_0^t \Phi(x(\tau)) \ d\tau$ for the inverse dynamical temperature cannot be sensibly interpreted in this context. 

In the case of Hamiltonian geodesic flow, experience suggests--and an explicit computation in the cases considered here shows--that $\Phi$ is inversely proportional to $H$, and so $\beta_{dyn}$ is too. It follows that $\beta_{dyn}$ and $\beta$ have the same scaling behavior \cite{Huntsman} provided that $t_\infty$ depends explicitly on the continuous evolution (as for example if $t_\infty$ is taken to be the mixing time). Although $\beta$ cannot be expected to recover the proper constant of proportionality \emph{ab initio}, it can be so adjusted and in other respects it will automatically reproduce the key behavior of the classical and dynamical inverse temperatures.

While the dynamical temperature has been evaluated in the context of a H\'enon-Heiles oscillator (HHO) \cite{Bannur} and a limited symbolic dynamical structure has also been identified for the HHO at the critical energy $E = 1/6$ \cite{AZ}, the prospects for evaluating the effective temperature in this context are presently poor for a number of technical reasons. 

Nevertheless, the preceding considerations highlight the comparative domains of utility (versus applicability) for the effective and dynamical temperatures. If a (well-behaved) Hamiltonian system is given, then the dynamical temperature is usually appropriate (see, e.g. \cite{YWL}), though the effective temperature will reproduce the same results (up to an overall choice of scale) provided it can be properly applied. If on the other hand a non-Hamiltonian dynamical evolution (such as is present in the canonical ensemble) or raw data in the form of symbolic dynamics are given, then the effective temperature is an appropriate generalization of the usual equilibrium temperature that also permits the identification of an effective energy function. 

It is particularly noteworthy that we have demonstrated the extraction of an effective temperature even in the microcanonical ensemble without any reliance on knowledge of the Hamiltonian beyond a constant energy surface, whereas the dynamical temperature requires this \cite{Rugh2}. Also, because the derivation of the dynamical temperature rests on the relation $\beta_{dyn} = dS/dE$ and there is no generally accepted physical definition of entropy for non-equilibrium steady states \cite{GC2,Klages}, it is far from clear how or if the dynamical temperature can be generalized to the nonequilibrium case.

\section{\label{sec:fdtemp} Comparison with the fluctuation-dissipation effective temperature}

The use of an effective temperature in glassy systems has a long history \cite{Tool,Nieuwenhuizen,LN} and has recently gained prominence through the \emph{fluctuation-dissipation (FD) temperature} in mean-field systems \cite{CKP}. In this section we discuss the nature of the FD temperature and its possible relationship with our effective temperature.

Recall that the connected correlation function of an observable $m$ is $C(t,t_w) := \langle m(t) m(t_w) \rangle - \langle m(t) \rangle \langle m(t_w) \rangle$ and the associated response and susceptibility are respectively given by
\begin{equation}
\label{eq:Rchi}
R(t,t_w) := \frac{\delta \langle m(t) \rangle}{\delta h(t_w)} \bigg |_{h = 0} \quad \mbox{and} \quad \chi(t,t_w) := \int_{t_w}^t R(t,t') \ dt',
\end{equation}
where as usual the field $h$ is conjugate to $m$ and $t_w$ is the ``waiting time'' of the system at which $h$ is first applied. In equilibrium the system is stationary and so we can write $C(t,t_w) \equiv C(t-t_w)$ and similarly for $R$ and $\chi$. In this context the \emph{fluctuation-dissipation theorem} (FDT) is a relation of the form 
\begin{equation}
\label{eq:fdt}
-\frac{\partial}{\partial t_w} \chi(t-t_w) = R(t-t_w) = \beta_o \frac{\partial}{\partial t_w} C(t-t_w),
\end{equation} 
where to avoid confusion we write $\beta_o$ to denote the true or ``ordinary'' inverse temperature.

The FDT is typically violated in nonequilibrium; however for mean-field systems the nature of the FDT violations permits the definition of a timescale-dependent FD temperature that is observable-independent and agrees with a thermometer tuned to the appropriate timescale. Towards this end, the \emph{fluctuation-dissipation ratio} $X$ is defined via 
\begin{equation}
\label{eq:fdtemp}
-\frac{\partial}{\partial t_w} \chi(t,t_w) = R(t,t_w) = \beta_o X(t,t_w) \frac{\partial}{\partial t_w} C(t,t_w).
\end{equation}
The enabling result of justifying the definition of the FD temperature for mean-field systems is that if $t$ and $t_w$ are large then $X(t,t_w) \approx X(C(t,t_w))$ \cite{CKP}. In this context the inverse FD temperature is defined by $\beta_{FD}(t,t_w) := \beta_o X(t,t_w)$. The functional dependence of the $\beta_{FD}$ so defined is considered to be an asset in the context of glassy dynamics for mean-field systems, since the well-known separation of timescales in glasses \cite{CugliandoloK} suggests the coexistence of multiple (effective) temperatures and the dependence on $t_w$ is expected due to nonstationarity or ``aging'' properties \cite{KurchanNature2007,BolthausenB}.

For systems other than mean-field glasses it has been found necessary to restrict consideration to an inverse FD temperature defined by $\beta_{FD,\infty} := X_\infty \beta_o$, where $X_\infty = \lim_{t_w \uparrow \infty} \lim_{t \uparrow \infty} X(t,t_w)$. This restriction emerged from the fact that $\beta_{FD}(t,t_w)$ exhibits dependence on observables for trap and one-dimensional Glauber-Ising models \cite{SFM}. In light of this result the domain of applicability of the restricted FD temperature has been extensively studied, and we provide a brief sketch of some representative results. Observable dependencies of $\beta_{FD,\infty}$ are still present in the one-dimensional Glauber-Ising model (although these dependencies disappear in two dimensions) and it is necessary in this case to restrict consideration to large spatial scales since at small scales the FDT still applies \cite{MBGS}. By the same token, while within the Gaussian approximation there is a well-defined $X_\infty$ for an $O(N)$ model, beyond the Gaussian approximation this ceases to be true \cite{CalabreseG}. In high dimensions the quenched Glauber-Ising model also exhibits two distinct values of $X_\infty$ depending on the initial magnetization, though there is no observable dependence for short-ranged interactions \cite{GSPR}. The Fredrickson-Anderson and East models \cite{LMSBG} as well as the backgammon model \cite{GPR} display negative values of $X_\infty$, precluding an interpretation in terms of temperature. An exactly solvable lattice model in which energy is conserved also admits a definition of temperature along the usual lines (viz. $\beta_o = dS/dE$) that differs from $\beta_{FD}$ \cite{BDD1,BDD2}. Finally, recent work indicates observable-dependence in a class of stationary nonequilibrium systems even in mean-field (but excluding any aging systems). In this event a signature of the breakdown is a difference between the Shannon entropies of the nonequilibrium state and of the equilibrium state with the same average energy \cite{MartensBD1,MartensBD2}. While it has been suggested that at least some defects in applicability for $\beta_{FD}$ might be remedied by considering nonlinear generalizations of the FDT \cite{LCZ,CLSZ,CasasVJ}, it is nevertheless apparent that the domain of applicability of $\beta_{FD}$ as an effective temperature is limited \cite{ZBCK}, even as its interpretation remains to be fully realized. 

Although $\beta_{FD}$ carries a clear physical significance and interpretation for mean-field glasses, these systems are fundamentally ill-suited for direct characterization via a single inverse effective temperature $\beta$. On a basic level this can be ascribed to nonstationarity and separation of timescales, mandating consideration of a parametrized family of inverse effective temperatures. Non-mean-field glasses exhibit not only these features but also spatial and temporal dynamical heterogeneity, which serve as additional complications \cite{Edinger,AppignanesiRF}. Moreover, it has been observed that the long-time dynamics of spin glass models beyond mean-field are invariant under generic time reparametrization (and a weaker class of reparametrization still holds for critical coarsening systems \cite{CorberiC}), which further indicates the difficulties in applying our framework directly to glassy dynamics in general \cite{CKCC,Castillo}. 

Finally, it is not at all obvious how to obtain a description of the dynamics of such systems that is appropriate for our framework. It is worth explicitly recalling the main theme of this paper: namely that even in the case of Anosov systems, where as we have seen a natural class of discretizations is provided, it still appears necessary (but also sufficient) to put in significant effort in order to obtain a physically reasonable value of the effective temperature.

This is not to say that $\beta_{FD}$ and $\beta$ cannot be related, but only that the existence or nature of any such relationship is not clear. The existence of such a relationship seems plausible since for an oscillator coupled to a nonequilibrium environment, accounting for a frequency-dependent $\beta_{FD}$ in the definition of entropy production rate causes the (Gallavotti-Cohen) fluctuation theorem to hold \cite{ZBCK}. Meanwhile based on the evidence in the present paper it is not unreasonable to speculate that a parametrized family of inverse effective temperatures $\beta$ might be appropriate for systems with multiple timescales, and that an appropriate coarse-graining (or suitable measure on dynamical modes, etc.) can in principle be found. In a similar vein we mention the possibility of applying superstatistics for systems with multiple timescales \cite{BeckC1,BeckC2,Beck}. 

Finally, we note that it is possible to take $t_\infty$ to be (very nearly) inversely proportional to the largest energy scales for a single Glauber-Ising spin (at high temperatures, this is dictated by the environment; at low temperatures, by the product of the magnetic moment and applied field) in such a way that $\beta = \beta_{FD} \equiv \beta_o$. This is detailed in appendix \ref{sec:glauber}. Because mixing and relaxation times are both broadly interchangeable with inverse spectral gaps, such a choice of $t_\infty$ is consistent with and provides additional support for the particulars of our approach to Anosov systems.

We conclude this section by mentioning that a coarse-graining scheme for glass-forming systems in which quasi-species are defined via the average local coordination statistics of particles superficially appears to indicate a possible avenue for applying our effective temperature to glasses in an altogether different manner. In this scheme the relative concentration of quasi-species are determined by a generalized Gibbs distribution and a relaxation time naturally enters the phenomenological framework \cite{BoueLPZ}. However, this relaxation time scales superexponentially (versus linearly) with $\beta_o$ and so using it for $t_\infty$ would eventually lead to $\beta_o < \beta$, while an effective temperature for glasses should satisfy the reverse inequality. Indeed, an effective temperature appropriate to this scheme has already been developed \cite{BoueHPRZ}. It would be of interest to see if the timescale obtained by inverting our framework while forcing $\beta^{-1}$ to equal this particular effective temperature turns out to carry physical significance.

\section{\label{sec:conclusion} Conclusion}

The relevance of the cat map and the geodesic flow on a surface of constant negative curvature for statistical physics derives from their privileged status as the prototypical Anosov diffeomorphism and flow, respectively. The Markovian symbolic dynamics exhibited by Anosov systems not only highlights their chaotic properties but also their correspondence with spin systems. Moreoever, the chaotic hypothesis maintains that Anosov systems are essential to the dynamics of nonequilibrium steady states.

While of course Markov structures (i.e., partitions or sections) are not unique, it is nevertheless evident that phenomena which hold for \emph{any} Markov structure of an Anosov system are likely to be of relevance to statistical physics. One remarkable phenomenon in this vein is the observed limiting behavior of $\beta/t_\infty$ as calculated on greedy partitions for both two-dimensional hyperbolic toral automorphisms and the geodesic flow. While any proof of these phenomena would seem to require the development of new and nontrivial mathematics, the observed behavior nevertheless appears to be generic.

Of particular importance is the implications that the evidence of this limiting behavior has for a proposed general theory of nonequilibrium steady states. As we have argued in this paper, it is plausible to extend the chaotic hypothesis with a conjectured generalization of the variational principle in which a suitable minimization of the effective free energy uniquely determines the inverse effective temperature. Moreover, such results justify a proposal for a comprehensive framework for nonequilibrium statistical physics that simultaneously incorporates and extends the formalism originally introduced by Ruelle and subsequently refined by Gallavotti, Cohen and others. The framework has as its goal a broad theory of nonequilibrium statistical physics that is truly intrinsic: i.e., that provides information about physical observables simply in terms of raw temporal information about the dynamics.

One reason to consider a proposal of the sort described here, in which the concept of (effective) temperature plays such a central role, is because there is no generally accepted physical definition of entropy for non-equilibrium steady states. It is hoped that the present work will serve to elicit fruitful investigations into the fundamental nature of stationary physical systems far from equilibrium.

\appendix

\section{\label{sec:lemmaapp} Proof of the lemma from section \ref{sec:greedy}}

W/l/o/g, let $n = \lvert \mathcal{R} \rvert$: it suffices to show that $\mathcal{R}' := \mathcal{R} \vee_T R_n$ is Markov. Let $\{ \ell_1,\dots,\ell_L\} := \{ \ell: a(\mathcal{R})_{n\ell} \ne 0 \}$ and (again w/l/o/g) set $R'_j := R_j$ for $1 \le j < n$, and $R'_{n+r-1} := T^{-1}(TR_n \cap R_{\ell_r}) = R_n \cap T^{-1}R_{\ell_r}$ for $1 \le r \le L$. For the proof we may disregard boundaries of rectangles and we shall also employ the convention that $1 \le k < n$ and $1 \le s \le L$. In the event that $x \in R_j \cap T^{-1}R_k$ for some $j$ and $k$ the Markov conditions are trivially satisfied. There are three remaining cases which we shall address individually. 

\emph{Case 1:} $x \in R_j \cap T^{-1}R'_{n+s-1}$ for some $j$, $s$. We must show that i) $TW^s(x,R_j) \subset W^s(Tx, R'_{n+s-1})$ and ii) $W^u(Tx,R'_{n+s-1}) \subset TW^u(x,R_j)$. To see i), note first that $W^s(Tx, R'_{n+s-1}) = W^s(Tx,R_n) \cap T^{-1}R_{\ell_s}$. Since $\mathcal{R}$ is Markov, we have that $TW^s(x,R_j) \subset W^s(Tx,R_n)$, so for i) we need only show that $TW^s(x,R_j) \subset T^{-1}R_{\ell_s}$. Since $T^{-1} \mathcal{R}$ is Markov, we have that $TW^s(x,T^{-1}R_n) \subset W^s(Tx,T^{-1}R_{\ell_s}) \subset T^{-1}R_{\ell_s}$, so we need only show that $W^s(x,R_j) \subset W^s(x,T^{-1}R_n)$. But this follows directly since $\mathcal{R}$ is Markov, establishing i). To see ii), note that $W^u(Tx,R'_{n+s-1}) = W^u(Tx,R_n) \cap T^{-1}R_{\ell_s}$ and $W^u(Tx,R_n) \subset TW^u(x,R_j)$ since $\mathcal{R}$ is Markov. This establishes ii) and hence Case 1.

\emph{Case 2:} $x \in R'_{n+r-1} \cap T^{-1}R_k$ for some $r$, $k$. In this event, since $\mathcal{R}$ is a partition, it follows that $\ell_r = k$. So we must show that iii) $TW^s(x,R'_{n+r-1}) \subset W^s(Tx,R_{\ell_r})$ and iv) $W^u(Tx,R_{\ell_r}) \subset TW^u(x,R'_{n+r-1})$. To see iii), note that $TW^s(x,R'_{n+r-1}) = TW^s(x,R_n) \cap R_{\ell_r}$. Since $\mathcal{R}$ is Markov, $TW^s(x,R_n) \subset W^s(Tx,R_{\ell_r})$, from which iii) follows. To see iv), note that $TW^u(x,R'_{n+r-1}) = TW^u(x,R_n) \cap R_{\ell_r}$, and since $\mathcal{R}$ is Markov $W^u(Tx,R_{\ell_r}) \subset TW^u(x,R_n)$. This establishes iv) and hence Case 2.

\emph{Case 3:} $x \in R'_{n+r-1} \cap T^{-1}R'_{n+s-1}$ for some $r$, $s$. We must show that v) $TW^s(x,R'_{n+r-1}) \subset W^s(Tx,R'_{n+s-1})$ and vi) $W^u(Tx,R'_{n+s-1}) \subset TW^u(x,R'_{n+r-1})$. To see v), note that $TW^s(x,R'_{n+r-1}) = TW^s(x,T^{-1}R_{\ell_r}) \cap TR_n$ and $W^s(Tx,R'_{n+s-1}) = W^s(Tx,T^{-1}R_{\ell_s}) \cap R_n$. Since $T^{-1}\mathcal{R}$ is Markov and $x \in T^{-1}R_{\ell_r} \cap T^{-1}(T^{-1}R_{\ell_s})$, we have that $TW^s(x,T^{-1}R_{\ell_r}) \subset W^s(Tx,T^{-1}R_{\ell_s})$, so we need only show that $W^s(Tx,T^{-1}R_{\ell_s}) \cap TR_n \subset W^s(Tx,T^{-1}R_{\ell_s}) \cap R_n$. But this follows since $\mathcal{R}$ is Markov and $Tx \in TR_n \cap R_n$. 

To see vi), note that $W^u(Tx,R'_{n+s-1}) = W^u(Tx,T^{-1}R_{\ell_s}) \cap R_n$ and $TW^u(x,R'_{n+r-1}) = TW^u(x,T^{-1}R_{\ell_r}) \cap TR_n$. Since $T^{-1}\mathcal{R}$ is Markov and $x \in T^{-1}R_{\ell_r} \cap T^{-1}(T^{-1}R_{\ell_s})$, we have that $W^u(Tx,T^{-1}R_{\ell_s}) \subset TW^u(x,T^{-1}R_{\ell_r})$. Since $x \in R_n \cap T^{-1}R_n$ and $\mathcal{R}$ is Markov, $W^u(Tx,R_n) \subset TW^u(x,R_n)$. It follows that $W^u(Tx,T^{-1}R_{\ell_s}) \cap W^u(Tx,R_n) \subset TW^u(x,T^{-1}R_{\ell_r}) \cap TW^u(x,R_n)$, from which vi), Case 3 and the lemma follow in turn. $\Box$

\section{\label{sec:variational} Variational principle}


A subset $E \subset M$ is called $(k, \varepsilon)$-separated iff for every $x \ne y$ there exists $j \in \{0,\dots,k-1\}$ s.t. $d(T^jx,T^jy) \ge \varepsilon$, in which case we write $E \equiv E(T,k,\varepsilon)$. If $T$ is hyperbolic or expanding, a natural way to try to build maximal $(k, \varepsilon)$-separated sets is to consider points spaced at distances $\approx \varepsilon$ on a local unstable manifold. We will use this intuition to provide a self-contained heuristic sketch of the so-called \emph{variational principle} as it applies to Anosov systems. For a more detailed and rigorous discussion that is also more generally applicable, see \cite{Bowen2}.

For $\phi$ nice define $\phi_{(\Sigma),k}(x) := \sum_{j=1}^{k-1} \phi(T^jx)$ and consider $z(T,\phi,E) := \sum_{x \in E} \exp[\phi_{(\Sigma),k}(x)]$. Write $z(T,\phi,k,\varepsilon) := \sup_{E(T,k,\varepsilon)} z(T,\phi,E(T,k,\varepsilon))$, and let $E^*(T,k,\varepsilon)$ be a saturating set for which the supremum is attained. Under the reasonable assumption that $E^*(T,k,\varepsilon)$ can effectively be given by the construction described above, then $z(T,\phi,E^*(T,k,\varepsilon)) = z(T,\phi,k,\varepsilon)$ should be approximately equal to $\lvert E^*(T,k,\varepsilon) \rvert \cdot \exp \left[ k \int_M \phi \ d\mu_* \right]$, where $\mu_*$ is a $T$-invariant probability measure on $M$ that has nice regularity properties on unstable manifolds.

We consequently expect that 
\begin{equation}
\label{eq:varprin0}
\frac{1}{k} \log z(T,\phi,k,\varepsilon) \approx \frac{1}{k} \log \lvert E^*(T,k,\varepsilon) \rvert + \int_M \phi \ d\mu_*.
\end{equation}
It can be shown that the \emph{topological entropy}
\begin{equation}
\label{eq:topent}
h(T) := \lim_{\varepsilon \downarrow 0} h(\varepsilon,T) \equiv \lim_{\varepsilon \downarrow 0} \limsup_{k \uparrow \infty} \frac{1}{k} \log \left ( \sup_{E(T,k,\varepsilon)} \lvert E(T,k,\varepsilon) \rvert \right)
\end{equation}
is well-defined 
\footnote{
By a result of Margulis, $h(T)$ governs the growth of the number $N(t)$ of periodic orbits of period $\le t$ when $T$ is Anosov through the relationship $htN(t)/e^{ht} \rightarrow 1$ \cite{MS}. (The idea is that an experimenter can distinguish $\sup_{E(T,k,\varepsilon)} \lvert E(T,k,\varepsilon) \rvert$ trajectories after $k$ timesteps of evolution under $T$, so that $h(\varepsilon,T)$ describes the growth of distinguishable trajectories over long times.) It can also be shown that $h(T)$ is the supremum of metric or Kolmogorov-Sinai entropies over the $T$-invariant Borel measures; by the Pesin-Ruelle inequality, the supremum is attained by a SRB measure.
}
and \eqref{eq:varprin0} and \eqref{eq:topent} combine to illustrate why the \emph{topological pressure} $P(T,\phi)$ (which is really a generalized free energy density) satisfies
\begin{equation}
\label{eq:varprin1}
P(T,\phi) := \lim_{\varepsilon \downarrow 0} \limsup_{k \uparrow \infty} \frac{1}{k} \log z(T,\phi,k,\varepsilon) = h(T) + \int_M \phi \ d\mu_*,
\end{equation}
for a suitable $T$-invariant probability measure $\mu_*$, called an ``equilibrium state'' for $\phi$ w/r/t $T$ (note that in the present context an equilibrium state can actually describe a physical nonequilibrium steady state). As we have just sketched, SRB measures--which can be defined as precisely the $T$-invariant measures with the strongest regularity properties on unstable manifolds--are equilibrium states.

Finally, it can be shown that 
\begin{equation}
\label{eq:varprin2}
P(T,\phi) \ge h_\mu(T) + \int_M \phi \ d\mu
\end{equation}
for a generic $T$-invariant Borel probability measure $\mu$, where $h_\mu(T)$ denotes the \emph{metric} or \emph{Kolmogorov-Sinai} entropy $\sup_\mathcal{X} \lim_{k \uparrow \infty} k^{-1} S_\mu(\lor_{j=0}^{k-1} T^{-j}\mathcal{X})$, and where $S_\mu(\mathcal{X}) := -\sum_{X \in \mathcal{X}} \mu(X) \log \mu(X)$ is the (Shannon) entropy of $\mathcal{X}$ w/r/t $\mu$. The equations \eqref{eq:varprin1} and \eqref{eq:varprin2} comprise the variational principle. This principle highlights not only the relationship between topological and metric entropies, but also characterizes equilibrium states or (equivalently for our purposes, by the symbolic dynamics correspondence) Gibbs measures uniquely determined by well-behaved potentials on subshifts of finite type--i.e., by physically reasonable interactions on one-dimensional spin systems.

\section{\label{sec:glauber} A single Glauber-Ising spin}

Consider a single Glauber-Ising spin $\sigma$. Its dynamics are determined by an overall inverse timescale $a$ and $b := \tanh (\beta_o \mu h)$, where $\mu$ is the magnetic moment and $h$ is the magnetic field \cite{Glauber}. The conjugate to $\sigma$ is $-\mu h$ and the dynamics corresponding to $\sigma = (-1,1)^*$ are given by the stochastic generator matrix 
\begin{equation}
Q := \frac{a}{2} \begin{pmatrix}
  -(1+b) & 1+b \\
  1-b & -(1-b)
\end{pmatrix}
\end{equation}
with corresponding invariant distribution $p = \frac{1}{2}(1-b,1+b)$.

In general $b$ and $Q$ inherit any functional time-dependence of $h$, requiring that we consider the Markov propagator $U(t_w,t) := U^{-1}(t_w)U(t)$, where
\begin{equation}
U(t) := \mathcal{O}^* \exp \int_0^t Q(s) \ ds,
\end{equation}
and $\mathcal{O}^*$ indicates the formal adjoint or \emph{reverse ordering operator}. Thus (e.g.) the distribution $p(s)$ is propagated as $p(t) = p(t_w)U(t_w,t)$. In the situation of interest however $h$ merely changes from one constant value $h$ on $(-\infty, t_w)$ to another constant value $h + \delta h$ on $[t_w,\infty)$, so $U(t_w,t)$ is just a matrix exponential. A routine calculation using the identities $\langle f(t) f(t_w) \rangle = \sum_{j,k} p_j(t_w) f_j U_{jk}(t_w,t) f_k$, $\langle f(t_w) \rangle = \sum_j p_j(t_w) f_j$ and $\langle f(t) \rangle = \sum_{j,k} p_j(t_w) U_{jk}(t_w,t) f_k$ for time-independent $f$ leads to
\begin{equation}
\langle \sigma(t) \sigma(t_w) \rangle - \langle \sigma(t) \rangle \langle \sigma(t_w) \rangle \equiv C(t-t_w) = (1-b^2)e^{-a(t-t_w)}.
\end{equation}

Moreover, 
\begin{equation}
\langle \sigma(t) \rangle = \langle \sigma(t_w) \rangle + \int_{t_w}^t ae^{-a(t-t_w)} b(t') \ dt'
\end{equation}
and in the present context this simplifies to $\langle \sigma(t) \rangle = (b + \delta b) - \delta b \cdot e^{-a(t-t_w)}$. Now $db/d(-\mu h) |_0 = -\beta_o$ and so $\delta \langle \sigma(t) \rangle/\delta(-\mu h(t_w)) |_0 = \beta_o e^{-a(t-t_w)}$. When taking partial derivatives w/r/t $t_w$ or $t$ it is appropriate to set $a \equiv 1$ (or to rescale the time variables by $a$). A line of algebra now shows that the equilibrium FDT holds for any $h$, not just $h = 0$, i.e. $\beta_{FD} \equiv \beta_o$. This is in some sense a peculiarity of the single-spin system: there is only one nontrivial observable, so any number of conjugate fields can be amalgamated.

Meanwhile one has $\lVert p \rVert^2 = (1+b^2)/2$ and $\lVert \gamma \Vert^2 = 2(\beta_o \mu h)^2$, so that 
\begin{equation}
\beta^2 = t_\infty^2 \cdot \frac{1+b^2}{2}(2[\beta_o \mu h]^2 +1),
\end{equation}
and enforcing $\beta = \beta_o$ leads to 
\begin{equation}
t_\infty = \left(\frac{2\beta_o^2}{(1+b^2)(2[\beta_o \mu h]^2 +1)} \right)^{1/2}.
\end{equation}
The question now is whether or not this is a physically reasonable characteristic timescale. We have that $t_\infty \approx \sqrt{2} \beta_o$ for $\beta_o$ small and $t_\infty \approx 1/\sqrt{2} \mu h$ for $\beta_o$ large. In both regimes $t_\infty$ is asymptotically inversely proportional to the largest energy scale, and indeed by equipartition the constant of proportionality can be chosen to be the same in both cases. On this basis $t_\infty$ can be regarded as a physically reasonable characteristic timescale similar to a relaxation (or mixing) time. \footnote{An alternative choice for $t_\infty$ motivated by recurrence considerations has been considered elsewhere, namely $t_\infty = Q_{-+}^{-1} + Q_{+-}^{-1} = 4a^{-1}(1-b^2)^{-1} = 4a^{-1}\cosh^2(\beta_o \mu h)$ \cite{Ford2}. (As with $\beta_{FD}$, it would be appropriate here to require $a \equiv 1$.) It is worth noting that here as in general there is a vast gap between recurrence and mixing or relaxation times.} It would be of interest to determine if a relationship of this sort applies to other systems.

\begin{acknowledgements}
It is a pleasure to thank a referee for suggestions that led to this paper, MathOverflow participants for helpful discussions and David Ford for suggesting supplementary references.
\end{acknowledgements}

\bibliography{anosovbib}

\end{document}